\PassOptionsToPackage{dvipsnames}{xcolor}
\PassOptionsToPackage{table}{xcolor}
\documentclass[journal]{IEEEtran}

\usepackage[utf8]{inputenc}

\usepackage{amsmath,amssymb,amsfonts}
\usepackage{array}
\usepackage[caption=false,font=normalsize,labelfont=sf,textfont=sf]{subfig}
\usepackage{textcomp}
\usepackage{stfloats}
\usepackage{url}
\usepackage{verbatim}
\usepackage{cite}

\usepackage{algorithm}
\usepackage[noend]{algpseudocode}
\usepackage{siunitx}
\usepackage{tikz}
\usepackage{hyperref}% hypertext links
    \hypersetup{colorlinks=true,linkcolor=black,citecolor=black,urlcolor=black}
\usepackage{amsthm}
\usepackage{listings}

\usepackage{algorithm}
\usepackage{cleveref}% automatic names for labels
    \crefname{subsection}{\S}{\S}
    \crefname{subsubsection}{\S}{\S}
    \crefname{paragraph}{\S}{\S}
    \crefname{algorithm}{algorithm}{algorithms}
\usepackage{ifthen}% branching in functions
\usepackage{xspace}% smart spacing after name commands
\usepackage{listings}

\usepackage{graphicx}

\usepackage{ragged2e}

\usepackage{multirow}
\usepackage{graphicx}
\usepackage[normalem]{ulem}
\useunder{\uline}{\ul}{}

\usepackage{comment}

\usepackage{amsmath,amssymb,amsfonts}
\usepackage{graphicx}

\definecolor{codegreen}{rgb}{0,0.6,0}
\definecolor{codegray}{rgb}{0.5,0.5,0.5}
\definecolor{codepurple}{rgb}{0.58,0,0.82}
\definecolor{backcolour}{rgb}{0.95,0.95,0.92}
\lstdefinestyle{mystyle}{
    commentstyle=\color{codegreen},
    keywordstyle=\color{magenta},
    numberstyle=\tiny\color{black},
    stringstyle=\color{codepurple},
    basicstyle=\ttfamily\footnotesize,
    breakatwhitespace=false,         
    breaklines=true,                 
    captionpos=b,                    
    keepspaces=true,                 
    numbers=left,                    
    showspaces=false,                
    showstringspaces=false,
    showtabs=false,                  
    tabsize=2,
    frame=single,
    framesep=4pt,
    boxpos=c
}

\lstdefinelanguage{SMRL}{%
    language = Java,
    morekeywords = {MR, var},
}

\usepackage[backgroundcolor=white,bordercolor=blue,linecolor=blue]{todonotes}

\usepackage{pdfpages}

\newcommand{\aim}{\text{AIM}\xspace}
\newcommand{\aimLong}{Automated Input Minimizer\xspace}
\newcommand{\kmeans}{\text{K-means}\xspace}
\newcommand{\dbscan}{\text{DBSCAN}\xspace}
\newcommand{\hdbscan}{\text{HDBSCAN}\xspace}
\newcommand{\nsgaTwo}{\text{NSGA-II}\xspace}
\newcommand{\speaTwo}{\text{SPEA2}\xspace}
\newcommand{\nsgaThree}{\text{NSGA-III}\xspace}
\newcommand{\mosa}{\text{MOSA}\xspace}

\newcommand{\mst}{\text{MST}\xspace}

\newcommand{\mstWi}{\text{MST-wi}\xspace}
\newcommand{\impro}{\text{IMPRO}\xspace}
\newcommand{\improLong}{Input set Minimization Problem Reduction Operator\xspace}
\newcommand{\geneticAlgo}{\text{MOCCO}\xspace}
\newcommand{\geneticAlgoLong}{Many-Objective Coverage and Cost Optimizer\xspace}
\newcommand{\jenkins}{\text{Jenkins}\xspace}
\newcommand{\joomla}{\text{Joomla}\xspace}
\newcommand{\crawljax}{\text{Crawljax}\xspace}

\newcommand{\configLev}{\text{L}}
\newcommand{\configBag}{\text{B}}
\newcommand{\configKmeans}{\text{K}}
\newcommand{\configDbscan}{\text{D}}
\newcommand{\configHdbscan}{\text{H}}
\newcommand{\configSep}{}
\newcommand{\LevKmeansKmeans}{\text{\configLev\configSep\configKmeans\configSep\configKmeans}\xspace}
\newcommand{\LevKmeansDbscan}{\text{\configLev\configSep\configKmeans\configSep\configDbscan}\xspace}
\newcommand{\LevKmeansHdbscan}{\text{\configLev\configSep\configKmeans\configSep\configHdbscan}\xspace}
\newcommand{\LevDbscanKmeans}{\text{\configLev\configSep\configDbscan\configSep\configKmeans}\xspace}
\newcommand{\LevDbscanDbscan}{\text{\configLev\configSep\configDbscan\configSep\configDbscan}\xspace}
\newcommand{\LevDbscanHdbscan}{\text{\configLev\configSep\configDbscan\configSep\configHdbscan}\xspace}
\newcommand{\LevHdbscanKmeans}{\text{\configLev\configSep\configHdbscan\configSep\configKmeans}\xspace}
\newcommand{\LevHdbscanDbscan}{\text{\configLev\configSep\configHdbscan\configSep\configDbscan}\xspace}
\newcommand{\LevHdbscanHdbscan}{\text{\configLev\configSep\configHdbscan\configSep\configHdbscan}\xspace}
\newcommand{\BagKmeansKmeans}{\text{\configBag\configSep\configKmeans\configSep\configKmeans}\xspace}
\newcommand{\BagKmeansDbscan}{\text{\configBag\configSep\configKmeans\configSep\configDbscan}\xspace}
\newcommand{\BagKmeansHdbscan}{\text{\configBag\configSep\configKmeans\configSep\configHdbscan}\xspace}
\newcommand{\BagDbscanKmeans}{\text{\configBag\configSep\configDbscan\configSep\configKmeans}\xspace}
\newcommand{\BagDbscanDbscan}{\text{\configBag\configSep\configDbscan\configSep\configDbscan}\xspace}
\newcommand{\BagDbscanHdbscan}{\text{\configBag\configSep\configDbscan\configSep\configHdbscan}\xspace}
\newcommand{\BagHdbscanKmeans}{\text{\configBag\configSep\configHdbscan\configSep\configKmeans}\xspace}
\newcommand{\BagHdbscanDbscan}{\text{\configBag\configSep\configHdbscan\configSep\configDbscan}\xspace}
\newcommand{\BagHdbscanHdbscan}{\text{\configBag\configSep\configHdbscan\configSep\configHdbscan}\xspace}

\newcommand{\Rt}{\text{R}\xspace}
\newcommand{\Art}{\text{A}}
\newcommand{\ArtKmeans}{\text{\Art\configSep\configKmeans}\xspace}
\newcommand{\ArtDbscan}{\text{\Art\configSep\configDbscan}\xspace}
\newcommand{\ArtHdbscan}{\text{\Art\configSep\configHdbscan}\xspace}

\newcommand{\algoRandom}{\text{Random}\xspace}
\newcommand{\algoGreedy}{\text{Greedy}\xspace}

\newcommand{\quotes}[1]{\text{``{#1}''}}

\newcommand{\todoColorDefault}{Black}% is the default color
\newcommand{\todoColor}{\todoColorDefault}% set initial value
\newcommand{\todoSetColor}[1]{% You can add new users in this macro
    \ifthenelse{\equal{#1}{LB}}{
        \renewcommand{\todoColor}{Red}}{}%
    \ifthenelse{\equal{#1}{FP}}{
        \renewcommand{\todoColor}{Maroon}}{}%
    \ifthenelse{\equal{#1}{YM}}{
        \renewcommand{\todoColor}{BurntOrange}}{}%
    \ifthenelse{\equal{#1}{NB}}{
        \renewcommand{\todoColor}{RedViolet}}{}%
}

\makeatletter
\newcommand{\oset}[3][0ex]{%
  \mathrel{\mathop{#3}\limits^{
    \vbox to#1{\kern-3.5\ex@
    \hbox{$\scriptstyle#2$}\vss}}}}
\makeatother
\DeclareMathOperator{\eqdef}{\oset{\text{\tiny def}}{=}}

\newcommand{\knowing}{\ |\ }
\newcommand{\set}[2]{\{{#1}\ifthenelse{\equal{\unexpanded{#2}}{}}{}{\knowing {#2}}\}}
\newcommand{\interval}[2]{[{#1}\ifthenelse{\equal{\unexpanded{#2}}{}}{}{,{#2}}]}
\newcommand{\card}[1]{\left|{#1}\right|}
\newcommand{\dataList}[1]{[\ifthenelse{\equal{\unexpanded{#1}}{}}{\,}{#1}]}

\newcommand{\getIntToStringA}{wrongInput}
\newcommand{\setIntToStringA}[1]{%
    \ifthenelse{\equal{#1}{1}}{\renewcommand{\getIntToStringA}{One}}{}%
    \ifthenelse{\equal{#1}{2}}{\renewcommand{\getIntToStringA}{Two}}{}%
    \ifthenelse{\equal{#1}{3}}{\renewcommand{\getIntToStringA}{Three}}{}%
    \ifthenelse{\equal{#1}{4}}{\renewcommand{\getIntToStringA}{Four}}{}%
    \ifthenelse{\equal{#1}{5}}{\renewcommand{\getIntToStringA}{Five}}{}%
    \ifthenelse{\equal{#1}{6}}{\renewcommand{\getIntToStringA}{Six}}{}%
    \ifthenelse{\equal{#1}{7}}{\renewcommand{\getIntToStringA}{Seven}}{}%
    \ifthenelse{\equal{#1}{8}}{\renewcommand{\getIntToStringA}{Eight}}{}%
    \ifthenelse{\equal{#1}{9}}{\renewcommand{\getIntToStringA}{Nine}}{}%
    \ifthenelse{\equal{#1}{10}}{\renewcommand{\getIntToStringA}{Ten}}{}%
}
\newcommand{\getIntToStringB}{wrongInput}
\newcommand{\setIntToStringB}[1]{%
    \ifthenelse{\equal{#1}{1}}{\renewcommand{\getIntToStringB}{One}}{}%
    \ifthenelse{\equal{#1}{2}}{\renewcommand{\getIntToStringB}{Two}}{}%
    \ifthenelse{\equal{#1}{3}}{\renewcommand{\getIntToStringB}{Three}}{}%
    \ifthenelse{\equal{#1}{4}}{\renewcommand{\getIntToStringB}{Four}}{}%
    \ifthenelse{\equal{#1}{5}}{\renewcommand{\getIntToStringB}{Five}}{}%
    \ifthenelse{\equal{#1}{6}}{\renewcommand{\getIntToStringB}{Six}}{}%
    \ifthenelse{\equal{#1}{7}}{\renewcommand{\getIntToStringB}{Seven}}{}%
    \ifthenelse{\equal{#1}{8}}{\renewcommand{\getIntToStringB}{Eight}}{}%
    \ifthenelse{\equal{#1}{9}}{\renewcommand{\getIntToStringB}{Nine}}{}%
    \ifthenelse{\equal{#1}{10}}{\renewcommand{\getIntToStringB}{Ten}}{}%
}
\newcommand{\nameAB}[3]{%
    \begingroup\setIntToStringA{#2}\setIntToStringB{#3}%
    \expandafter\endgroup
    \csname #1\getIntToStringA\getIntToStringB\endcsname
}

\newcommand{\normalization}[1]{\omega({#1})}

\DeclareMathOperator{\graph}{\mathcal{G}}

\newcommand{\mr}{\mathit{mr}}
\newcommand{\MRs}{\mathit{MRs}}
\newcommand{\costFunction}[1]{\mathit{cost}({#1})}

\newcommand{\dataUrl}{\mathit{url}}
\newcommand{\urlAct}[1]
{\mathit{url}({#1})} 
\newcommand{\distURL}[2]
{\mathit{urlDist}({#1}, {#2})} 
\newcommand{\lca}[2]
{\mathit{LCA}({#1}, {#2})}
\newcommand{\paramsAct}[1]
{\mathit{res}({#1})} 
\newcommand{\distParam}[2]
{\mathit{paramDist}({#1}, {#2})} 

\newcommand{\params}{\mathit{resids}}
\newcommand{\typeOf}[1]
{\mathit{type}({#1})}

\newcommand{\listInd}[2]
{#1^{[{#2}]}} 

\newcommand{\actionDist}[2]
{\mathit{actionDist}({#1}, {#2})}

\newcommand{\distParamVal}[2]
{\mathit{paramValDist}({#1}, {#2})}

\newcommand{\editDistance}[2]
{\mathit{LevenshteinDist}({#1}, {#2})} 

\newcommand{\numberOfExecutedActions}[2]
{\mathit{nbActions}({#1}, {#2})}

\newcommand{\abs}[1]{|{#1}|}

\newcommand{\bigO}[1]{\mathcal{O}({#1})}
\newcommand{\action}{\mathit{act}}
\newcommand{\sequence}{\mathit{in}}
 
\newcommand{\outputClass}{\mathit{outCl}}
\newcommand{\actionClass}{\mathit{actSet}}
\newcommand{\actionSublass}{\mathit{bl}}
\newcommand{\TestSuite}{\mathit{I}}
\newcommand{\labelInit}{\mathit{init}}
\newcommand{\TestSuiteInit}{\TestSuite_{\labelInit}}

\newcommand{\TestSuiteFinal}{\TestSuite_{\mathit{final}}}
 
\newcommand{\length}[1]{\mathit{len}({#1})} 
\newcommand{\getAction}[2]{\mathit{action}({#1}, {#2})} 
\newcommand{\getOutput}[2]{\mathit{output}({#1}, {#2})} 
\newcommand{\getOutputClass}[2]{\mathit{OutputClass}({#1},{#2})} 
\newcommand{\getActionClass}[1]{\mathit{ActionSet}({#1})} 
\newcommand{\getActionSubclass}[2]{\mathit{ActionSubclass}({#1},{#2})} 
\newcommand{\getSubclassFromAct}[2]{\mathit{Subclass}({#1}, {#2})} 
\newcommand{\getSubclasses}[1]{\mathit{Cover}({#1})}

\newcommand{\getSequences}[1]{\mathit{Inputs}({#1})}

\newcommand{\superposition}[2]{\mathit{superpos}({#1}, {#2})}
\newcommand{\redundancy}[2]{\mathit{redundancy}({#1}, {#2})}
\newcommand{\Redundant}[1]{\mathit{Redundant}({#1})}
\DeclareMathOperator{\overlap}{\sqcap}
\newcommand{\OverlapGraph}[1]{\graph_{\overlap}({#1})}

\newcommand{\RedundantComponents}[1]{\mathit{Comps}({#1})}
\newcommand{\RedComp}{\mathit{C}}

\newcommand{\ReducOrders}[1]{\mathit{ValidOrders}({#1})}

\newcommand{\gain}[1]{\mathit{gain}({#1})}
\newcommand{\gainExhaust}[1]{\gain{#1}}

\newcommand{\potential}[2]{\mathit{potential}({#1}, {#2})}
\newcommand{\nbSubClasses}{n}
\newcommand{\objFun}[2]{f_{#1}({#2})}
\newcommand{\fitnessVector}[1]{\mathit{F}({#1})}
\newcommand{\fitnessVectorComponent}[1]{\mathit{F}_{\RedComp}({#1})}
\DeclareMathOperator*{\minimize}{\text{minimize}}
\DeclareMathOperator{\dominates}{\succ}
\DeclareMathOperator{\notDominates}{\not\hspace{-0.1cm}\dominates}

\newcommand{\SearchSpace}{\TestSuite_{\mathit{search}}}
\newcommand{\RedundantInputs}{\TestSuite_{\mathit{redund}}}
\newcommand{\NecessaryInputs}{\TestSuite_{\mathit{necess}}}
\newcommand{\NecessaryInputsNew}{\NecessaryInputs^{\mathit{new}}}
\newcommand{\Objectives}{\mathit{Coverage}_{\mathit{obj}}}

\newcommand{\localDom}{\sqsubseteq}

\newcommand{\sizePop}{n_\mathit{size}}
\newcommand{\nbGens}{n_\mathit{gens}}
\newcommand{\timeBudget}{\mathit{time}_\mathit{budget}}
\newcommand{\timeStart}{\mathit{time}_\mathit{start}}
\newcommand{\getTime}{\mathit{getTime}()}
\newcommand{\mutation}[1]{\mathit{mutate}({#1})}
\newcommand{\reduction}[1]{\mathit{reduce}({#1})}
\newcommand{\RoofersVar}{\mathit{Roofers}}
\newcommand{\MisersVar}{\mathit{Misers}}
\newcommand{\initRoofers}[1]{\mathit{initRoofers}({#1})}
\newcommand{\Roofers}[1]{\mathit{Roofers}({#1})}
\newcommand{\Misers}[1]{\mathit{Misers}({#1})}
\newcommand{\halfObj}{O}
\newcommand{\halfSeqs}{S}
\newcommand{\select}[2]{\mathit{select}({#1}, {#2})}
\newcommand{\symbProba}{\text{P}}
\newcommand{\distrib}[1]{\ifthenelse{\equal{\unexpanded{#1}}{}}{\symbProba}{\symbProba_{#1}}}
\newcommand{\proba}[2]{\distrib{#1}({#2})}

\newcommand{\minpts}{n}

\newcommand{\aimChar}[1]{{}\lq{#1}\rq}
\newcommand{\aimStr}[1]{{}\quotes{{#1}}}

\newcommand{\complexity}[1]{\mathit{O}({#1})}

\newcommand{\exposure}[1]{\mathit{exposure}({#1})}
\newcommand{\occurrence}[1]{\mathit{occurrence}({#1})}

\newcommand{\pValue}{p}
\newcommand{\vda}{A}
\newcommand{\effectSize}{E}
\newcommand{\vdaLong}{A_{12}}
\newcommand{\cliffsDelta}{\delta}
\newcommand{\configRandomVar}{M}
\newcommand{\tPlus}{R^{+}}
\newcommand{\tMinus}{R^{-}}

\newcommand{\TSE}[2]{{#2}}

\begin{document}

\title{AIM: Automated Input Set Minimization\\for Metamorphic Security Testing}
\markboth{IEEE Transactions on Software Engineering,~Vol.~\textbf{XX}, No.~\textbf{XX}, \textbf{Month}~2024}%
{Bayati Chaleshtari \MakeLowercase{\textit{et al.}}: AIM: Automated Input Set Minimization for Metamorphic Security Testing}

\author{%
Nazanin~Bayati~Chaleshtari,
Yoann~Marquer,
Fabrizio~Pastore,~\IEEEmembership{Member,~IEEE,}
and~Lionel~C.~Briand,~\IEEEmembership{Fellow,~IEEE}%
\IEEEcompsocitemizethanks{\IEEEcompsocthanksitem N. Bayati~Chaleshtari and L. Briand are with the School of Electrical and Computer Engineering of University of Ottawa, Canada,
Y. Marquer and F. Pastore are with the Interdisciplinary Centre for Security, Reliability, and Trust (SnT) of the University of Luxembourg, Luxembourg, and
L. Briand is also with Lero SFI Centre for Software Research and University of Limerick, Ireland. Part of this work was done when L. Briand was affiliated with the Interdisciplinary Centre for Security, Reliability, and Trust (SnT) of the University of Luxembourg.\protect\\
E-mail: n.bayati@uottawa.ca, yoann.marquer@uni.lu, fabrizio.pastore@uni.lu, lbriand@uottawa.ca
}%
\thanks{Manuscript received \textbf{Month DD}, 2024; revised \textbf{Month DD}, 2024.}
}

\maketitle

\begin{abstract}
Although the security testing of Web systems can be automated by generating crafted inputs, solutions to automate the test oracle, i.e., vulnerability detection, remain difficult to apply in practice. Specifically, though previous work has demonstrated the potential of metamorphic testing---security failures can be determined by metamorphic relations that turn valid inputs into malicious inputs---metamorphic relations are typically executed on a large set of inputs, which is time-consuming and thus makes metamorphic testing impractical.

We propose AIM, an approach that automatically selects inputs to reduce testing costs while preserving vulnerability detection capabilities.
AIM includes a clustering-based black-box approach, to identify similar inputs based on their security properties.
It also relies on a novel genetic algorithm to efficiently select diverse inputs while minimizing their total cost. 
Further, it contains a problem-reduction component to reduce the search space and speed up the minimization process.
We evaluated the effectiveness of AIM  on two well-known Web systems, Jenkins and Joomla, with documented vulnerabilities. 
We compared AIM's results with four baselines involving standard search approaches. Overall, AIM reduced metamorphic testing time by 84\% for Jenkins and 82\% for Joomla, while preserving the same level of vulnerability detection. Furthermore, AIM significantly outperformed all the considered baselines regarding vulnerability coverage.

\end{abstract}


\section{Introduction}
\label{sec:aim:intro}

%%%%%%%%%%%%%%%%
% \subsection{Context}
\label{sec:aim:intro:context}

Web systems, from social media platforms to e-commerce and banking systems, are a backbone of our society: they manage data that is at the heart of our social and business activities (e.g., public pictures, bank transactions), and, as such, should be  protected. To verify that Web systems are secure, engineers perform security testing, which consists of verifying that the software adheres to its security properties (e.g., confidentiality, availability, and integrity). Such testing is typically performed by simulating malicious users interacting with the system under test~\cite{Mai2028ISSRE,MAI2018165}. 

At a high-level, security testing does not differ from other software testing activities: it consists of providing inputs to the software under test and verifying that the software outputs are correct, based on specifications.
For such a verification, a \emph{test oracle}~\cite{How78} is required, i.e., a mechanism for determining whether a test case has passed or failed.
When test cases are manually derived, test oracles are defined by engineers and they generally consist of expressions comparing an observed output with the expected output, determined from software specifications.
In security testing, when a software output differs from the expected one, then a software vulnerability (i.e., a fault affecting the security properties of the software) has been discovered.

\TSE{1.1}{Deriving} test oracles for the software under test (SUT) is called the \emph{oracle problem}~\cite{TestOracleSurvey}, which entails distinguishing correct from incorrect outputs for all potential inputs.
Except for the verification of basic reliability properties---ensuring that the software provides a timely output and does not crash---the problem is not tractable without additional executable specifications (e.g., method post-conditions or detailed system models), which, unfortunately, are often unavailable.
%% The oracle problem is particularly acute in the context of security testing because security properties need to be verified by providing a large number of test inputs to all the system interfaces.
 % the point of interaction or communication between different components or systems, e.g. UI in a Web system
Further, since software vulnerabilities tend to be subtle, it is necessary to exercise each software interface with a large number of inputs (e.g., providing all the possible code injection strings to a Web form). When a large number of test inputs are needed, even in the presence of automated means to generate them (e.g., catalogs of code injection strings), testing becomes impractical if we lack solutions to automatically derive test oracles. 

%This problem may be tractable in traditional testing on simple systems, when a test oracle usually consists in comparing an observed output with the expected output.
%But the output of a system may not be known in advance, this output may not be easy to handle (for instance in the context of simulations or code generation~\cite{SFSRC16}), and verifying all outputs of a system may simply be infeasible due to the large number of potential inputs~\cite{MPGB20}.

Metamorphic testing was proposed to alleviate the oracle problem~\cite{SFSRC16} by testing not the input-output behavior of the system, but by comparing the outputs of multiple test executions~\cite{SFSRC16,BPGB23}.
It relies on metamorphic relations (MRs), which are specifications expressing how to derive a follow-up input from a source input and relations between the corresponding outputs.
%relations between outputs under certain input conditions, which are then used to derive a follow-up input from a source input to satisfy such relations.
Such an approach has shown to be useful for security testing, also referred to as metamorphic security testing (MST)~\cite{MPGB20,BPGB23}.
%which aims to identify security vulnerabilities that would result, for instance, in disclosing sensitive information or preventing the SUT to deliver the intended functionalities~\cite{MPGB20}.
MST consists in relying on MRs to modify source inputs to obtain follow-up inputs that mimic attacks and verify that known output properties captured by these MRs  hold (e.g., if the follow-up input differs from the source input in some way, then the output shall be different).
For instance, one may verify if URLs can be accessed by users who should not reach them through their user interface, thus enabling the detection of authorization vulnerabilities.

MST has been successfully applied to testing Web interfaces~\cite{MPGB20,BPGB23} in an approach called \mstWi~\cite{BPGB23}; in such context, source inputs are sequences of interactions with a Web system and can be easily derived using a Web crawler.
For example, a source input may consist of two actions: performing a login and then requesting a specific URL appearing in the returned Web page.
\mstWi integrates a catalog of 76 MRs enabling the identification of 101 vulnerability types.

The performance and scalability of metamorphic testing naturally depends on the number of source inputs to be processed. 
In the case of \mstWi, we demonstrated that scalability can be achieved through parallelism; however, such solution may not fit all development contexts (e.g., not all companies have an infrastructure enabling parallel execution of the software under test and its test cases).
Further, even when parallelization is possible, a reduction of the test execution time may provide tangible benefits, including earlier vulnerability detection.
In general, what is required is an approach to minimize the number of source inputs to be used during testing.
% However, such optimization remains an open research problem for MT.

%%%%%%%%%%%%%%%%
% \subsection{Contributions}
\label{sec:aim:intro:contributions}

In this work, we address the problem of minimizing source inputs used by MRs to make metamorphic testing more scalable, with a focus on Web systems, though many aspects are reusable to other domains.
We propose the \aimLong (\aim) approach, which aims at minimizing a set of source inputs (hereafter, the \emph{initial input set}), while preserving the capability of MRs to detect security vulnerabilities.
More in detail, this work includes the following contributions:
\begin{itemize}
    %\item We leverage the \mstWi framework to retrieve output data using a crawler and we update this framework to extract cost information about MRs without executing them.
   \item We propose \aim, an approach to minimize input sets for metamorphic testing
   %Web systems 
   while retaining inputs able to detect vulnerabilities. Note that many steps of \aim are not specific to Web systems while others would need to be tailored to other domains (e.g., desktop applications, embedded systems).
   %further, some steps can be reused to address input set minimization problems beyond metamorphic testing.
   %to perform metamorphic testing for different purposes (e.g., functional, robustness, security) and targeting any kind of software system (e.g., Web systems, desktop applications, embedded systems), we tailored it to address metamorphic security testing of Web systems.
   This approach includes the following novel components:
  \begin{itemize}
        \item An extension of the \mstWi framework to retrieve output data and extract cost information about MRs without executing them.
      \item A black-box approach leveraging clustering algorithms to partition the initial input set based on security-related characteristics in order to keep a small number of representative inputs.
      %\item A hyperparameter tuning method, based on the Silhouette score and Gini index, which automatically selects the optimal value of the hyperparameter(s) for the clustering algorithms.
        \item  \geneticAlgo (\geneticAlgoLong), a novel genetic algorithm specifically designed to fit our problem and which is able to efficiently select diverse inputs while minimizing their total cost.
      \item \impro (\improLong), an approach to reduce the search space to its minimal extent, and then divide it in smaller, easier to minimize,  independent parts.

  \end{itemize}

     \item We provide a prototype framework for \aim~\cite{AimReplicability}, integrating the above components and automating the process of input set minimization for Web systems.
       
    \item We report on an extensive empirical evaluation (about 800 hours of computation) aimed at assessing the effectiveness of \aim in terms of vulnerability detection and performance, considering 18 different \aim configurations and 
    5 search algorithms (including \geneticAlgo) for security testing, on the \jenkins and \joomla systems, which are the most used Web-based frameworks for development automation and context management.

     \item We also provide a proof of the correctness of the \aim approach in a separate appendix provided as supplementary material.
\end{itemize}

%%%%%%%%%%%%%%%%
% \subsection{Organization of the document}
\label{sec:aim:orgaDoc}

This paper is structured as follows.
We introduce background information necessary to state our problem and detail our approach (\Cref{sec:aim:background}).
We define the problem of minimizing the initial input set while retaining inputs capable of detecting distinct software vulnerabilities (\Cref{sec:aim:framework}).
We present an overview of \aim (\Cref{sec:aim:overview}) and then detail our core technical solutions (\Cref{sec:aim:system,sec:aim:doubleCLustering,sec:aim:search,sec:aim:genetic,sec:aim:postprocessing}).
We report on a large-scale empirical evaluation of \aim (\Cref{sec:aim:results}) and 
address the threats to the validity of the results (\Cref{sec:aim:results:threatsToValidity}).
We discuss and contrast related work (\Cref{sec:aim:related}) and draw conclusions (\Cref{sec:aim:conclusion}).
\section{Background}
\label{sec:aim:background}

In this section, we present the concepts required to define our approach.
We first provide a background on Metamorphic Testing (MT, \Cref{sec:aim:background:MT}), then we briefly describe \mstWi, our previous work on the application of MT to security (\Cref{sec:aim:background:MST}).
Next, we briefly describe three clustering algorithms: \kmeans (\Cref{sec:aim:background:kMeans}), \dbscan (\Cref{sec:aim:background:dbscan}), and \hdbscan(\Cref{sec:aim:background:hdbscan}).
Finally, we introduce optimization problems (\Cref{sec:aim:background:Optimization}).
%, as well as the biclustering approach (\Cref{sec:biclustering}), which presents similarities with our approach.

\subsection{Metamorphic Testing}
\label{sec:aim:background:MT}

In contrast to common testing practice, which compares for each input of the system the actual output against the expected output, MT examines the relationships between outputs obtained from multiple test executions.

MT is based on Metamorphic Relations (MRs), which are necessary properties of the SUT (system under test) in relation to multiple inputs and their expected outputs~\cite{CKL18}.
% In MT, a single test case run requires multiple executions of the system under test with distinct inputs~\cite{BPGB23}. 
The test result, either pass or failure, is determined by validating the outputs of various executions against the MR.

Formally, let $S$ be the SUT.
In the context of MT, inputs in the domain of S  are called \emph{source inputs}.
Moreover, we call \emph{source output} and we denote $S(x)$ the output obtained from a source input $x$.
An MR is the combination of:
\begin{itemize}
    \item A \emph{transformation function} $\theta$, taking values in source inputs and generating new inputs called \emph{follow-up inputs}.
    For each source input $x$, we call \emph{follow-up output} the output $S(\theta(x))$ of the follow-up input $\theta(x)$.
    \item An \emph{output relation} $R$ between source outputs and follow-up outputs.
\end{itemize}

The MR is \emph{executed} with a source input $x$ when the follow-up input $\theta(x)$ is generated, then the SUT is executed on both inputs to obtain outputs $S(x)$ and $S(\theta(x))$, and finally the relation $R(S(x), S(\theta(x)))$ is checked.
If this relation holds, then the MR is \emph{satisfied}, otherwise it is \emph{violated}.

%We now provide an example to further clarify the concepts presented above.

%\begin{ex}
For instance, consider a system implementing the cosine function. 
%we want to test a system which computes the cosines of the given input x. 
It might not be feasible to verify the $\cos(x)$ results for all possible values of $x$, except for special values of $x$, e.g., $cos(0) = 1$ or $cos(\frac{\pi}{2}) = 0$. However, the cosine function satisfies that, for each input $x$, $\cos(\pi - x)= -\cos(x)$. Based on this property, we can define an MR, where the source inputs are the possible angle values of $x$, the follow-up inputs are $y = \pi - x$, and the expected relation between source and follow-up outputs is $\cos(y)= -\cos(x)$.
The SUT is executed twice per source input, respectively with an angle $x$ and an angle $y = \pi - x$. The outputs of both executions are then validated against the output relation. If this relation is violated, then the SUT is faulty. 
%\end{ex}

\subsection{Metamorphic Security Testing}
\label{sec:aim:background:MST}

In previous work, we automated MT in the security domain by introducing a tool named MST-wi~\cite{BPGB23}. \mstWi enables software engineers to define MRs that capture the security properties of Web systems. MST-wi includes a data collection framework that crawls the Web system under test to automatically derive source inputs. Each source input is a sequence of interactions of the legitimate user with the Web system. Moreover, \mstWi includes a Domain Specific Language (DSL) to support writing MRs for security testing. Finally, MST-wi provides a testing framework that automatically performs security testing based on the defined MRs and the input data.

In MST, 
%source inputs are the input data obtained from the data collection framework. The 
follow-up inputs are generated by modifying source inputs, simulating actions an attacker might take to identify vulnerabilities in the SUT. These modifications can be done using 55 Web-specific functions enabling engineers to define complex security properties, e.g., \texttt{cannotReachThroughGUI}, \texttt{isSupervisorOf}, and \texttt{isError}. MRs capture security properties that hold when the SUT behaves in a safe way. 
If an MR, for any given source input, gets violated, then \mstWi detects a vulnerability in the SUT.
In that case, we say that the MR \emph{exercised} the vulnerability in the SUT.
MST-wi includes a catalog of 76 MRs, inspired by OWASP guidelines~\cite{OWASP} and vulnerability descriptions in the CWE database~\cite{CWE}, capturing a large variety of security properties for Web systems.

\begin{figure*}[t]
\centering
\begin{minipage}{0.9\linewidth}
\begin{lstlisting}[language=SMRL,escapechar=@]
MR CWE_668 {
{
   var sep = "/"; @\label{line:authMR:sep}@ 
   for (var par=0; par < 4; par++){  @\label{line:authMR:par}@ 
     for (Action action : Input(1).actions()){@\label{line:authMR:for}@
        var pos = action.getPosition();  @\label{line:authMR:pos}@ 
        var newUrl = action.urlPath+sep+RandomFilePath(); @\label{line:authMR:newUrl}@ 
        IMPLIES(
             !isAdmin(action.user) && @\label{line:authMR:isAdmin}@ 
             afterLogin(action) &&  @\label{line:authMR:afterLogin}@ 
             CREATE(Input(2), Input(1)) &&  @\label{line:authMR:create}@ 
             Input(2).actions().get(pos).setUrl(newUrl) &&  @\label{line:authMR:setUrl}@ 
             notTried(action.getUser(), newUrl)  @\label{line:authMR:notTried}@ 
              ,
             TRUE(              @\label{line:authMR:eval}@                       
                 Output(Input(2),pos).noFile()  ||
                 userCanRetrieveContent(action.getUser(), Output(Input(2),pos).file()) ||
                 different(Output(Input(1),pos), Output(Input(2),pos)))
        );//end-IMPLIES
     }//end-for
   sep=sep+"../";
   }//end-for
}//end-MR
}//end-package
\end{lstlisting}
\end{minipage}
\caption{MR for CWE 668: Exposure of resource to wrong sphere~\cite{cwe668}} %during metamorphic testing of OTG-AUTHZ-002.}
\label{fig:MR_668}
\end{figure*}
 % Output(Input(2),pos).noFile() ||

%\begin{ex}
%\label{ex:authMR}
We describe in \Cref{fig:MR_668} an MR written for CWE 668, which concerns unintended access rights~\cite{cwe668}.
% testing activity Testing Directory Traversal File Include 
This MR verifies that a file path passed in a URL should never enable a user to access data that is not already provided by the user interface.
% This MR verifies that requesting a file path should never enable a user to access data that is not provided by the user interface.
%A URL is a sequence of words separated by slashes ('/'). 
The first \texttt{for} loop iterates multiple times (\Cref{line:authMR:par}) to cover different system paths, e.g., \texttt{/} and \texttt{/../../}.
% , e.g., '/' and '/../'  
The second \texttt{for} loop iterates over all the actions of a source input (\Cref{line:authMR:for}).
% This includes a parameter $par$ that the user can send data to the server.
Each action in the sequence is identified by its position, i.e., the first action in a sequence has position 0.
The position of the current action is stored (\Cref{line:authMR:pos}) to be used to generate the corresponding follow-up action.
%The function \texttt{RandomFilePath()} returns a randomly selected system file path, which is not provided by the user interface, e.g., \texttt{config.xml}.
% We select paths of files in the Web system subfolder, ignoring images, and replacing symbolic links (e.g., ‘plugins’ is mapped to ‘plugin’ in Jenkins).
% like config.xml or credentials.xml
A new URL is defined by concatenating the URL of the current action and a randomly selected system file path, e.g., \texttt{config.xml}, (\Cref{line:authMR:newUrl}).
For instance, if the URL of the current action is \texttt{http://www.hostname.com}, the new URL can be \texttt{http://www.hostname.com/../../config.xml}.
The MR first checks if the user who is performing the action is admin (\Cref{line:authMR:isAdmin}), since an admin has direct access to the system file path, and hence will not exercise a vulnerability.
Then, the MR checks that the action is performed after a login (\Cref{line:authMR:afterLogin}), to ensure this action requires authentication.
Then, the MR generates a follow-up input, named Input(2), by copying the current sequence of actions Input(1) (\Cref{line:authMR:create}) and setting the URL of the current action to the new URL (\Cref{line:authMR:setUrl}).
% The MR first creates a copy of the current sequences of actions Input(1) as the follow-up input, named Input(2) (\Cref{line:authMR:create}).
% Then, in the follow-up input, the MR sets the URL of the current action to the new URL (\Cref{line:authMR:setUrl}).
To speed up the process, the MR verifies that the current user has not tried the same URL before (\Cref{line:authMR:notTried}).
The SUT is vulnerable if all the following conditions are violated:
1) the follow-up input does not access a file at the new URL, or 2) it accesses a file, but the user has the right to access it, 
%(based on the data collected by the crawler),
or 3) the source and follow-up inputs obtain different outputs, as the follow-up input tries to access a system file without access rights, while the source input is accessing the originally crawled URL.

This MR tests the initial set of source inputs, with different URLs and users, and transforms each one several times with different system file paths, leading to a combinatorial explosion.
% URLs multiple times with different users and system file paths, leading to a combinatorial explosion.
The more executed actions, the longer the execution time. The provided MR, with an input set of 160 source inputs on Jenkins, executed more than \num{200000} follow-up inputs in \num{17694} minutes (about 12 days) on a professional desktop PCs (Dell G7 7500, RAM 16Gb, Intel(R) Core(TM) i9-10885H CPU @ 2.40GHz).
Even when parallelization is possible, a reduction of the test execution time may provide tangible benefits, including earlier vulnerability detection.
This warrants an approach to minimize the initial set of source inputs, based on the cost of each input.

However, knowing the execution time of each source input would require to execute them on the considered MRs, hence defeating the purpose of input set minimization.
% Thus, we cannot rely on executing MRs and we need a surrogate metric to guide the minimization technique.
\TSE{1.2}{To avoid executing MRs but relying on the number of actions executed by an MR as a surrogate metric for execution time to guide the minimization technique, we computed the Spearman’s correlation coefficients between execution time and number of actions exercised by the MRs tested in a previous study~\cite{BPGB23}. It led to significant correlations (i.e., coefficients of correlation above 0.5 and p-value $\pValue$ below 0.05), thus confirming the feasibility of relying on the number of actions as a surrogate metric for execution time.}
% {Fortunately, based on the results in a previous study~\cite{BPGB23}, we have observed a significant correlation (i.e.,  coefficient of correlation above 0.5) between the execution time and the number of executed actions, considering the 76 MRs.
% Hence, the number of executed actions can be used as a surrogate metric for execution time.}

\begin{figure}[t]
\begin{center}
\includegraphics[width=1.0\linewidth]{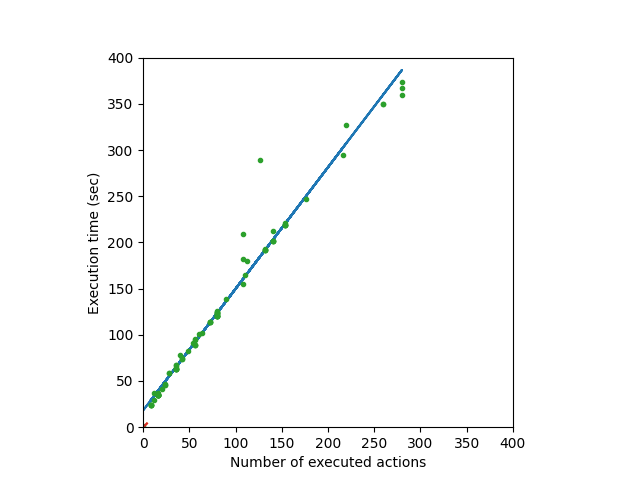}
\caption{Linear regression between the number of executed actions and the execution time of a metamorphic relation. Each input is represented by a green dot, while the blue line depicts the linear regression model.}
\label{fig:linearmodel}
\end{center}
\end{figure}

The example in \Cref{fig:linearmodel} depicts a typical linear correlation between the execution time of an MR and the number of executed actions.
It uses randomly selected source inputs and an MR written for CWE 863. 
% which has a shorter execution time than the one 
% presented in \Cref{fig:MR_668}. 
Each point represents the execution time (x-axis) and the number of executed actions (y-axis) for a given input. The linear regression is represented by the blue line and the corresponding  coefficient of determination is $97.8\%$, indicating a strong linear correlation.

% Since executing the MR provided in \Cref{fig:MR_668} is time consuming, the MR written for CWE\_863~\cite{CWE863} which has a shorter execution time is considered for illustration. 
% % We investigate here one MR written for CWE\_863~\cite{CWE863} and randomly selected source inputs.
% The selected MR is executed for each input and measured its execution time and the number of executed actions from source and follow-up inputs.
% Then, a linear regression is performed between execution time and the number of executed actions, represented by the blue line depicted in \Cref{fig:linearmodel}.
% The coefficient of determination is $97.8\%$, indicating a strong linear correlation between MR execution time and the number of executed actions.
% %\end{ex}

\subsection{Test Suite Minimization}
\label{sec:aim:background:testMinimization}

Test suites are prone to redundant test cases that, if not removed, can lead to a massive waste of time and resources~\cite{YH12}, thus warranting systematic and automated strategies to eliminate redundant test cases, that are referred to as test suite \emph{minimization}.

While test suite minimization techniques are very diverse~\cite{YH12}, most of them are white-box approaches aiming at minimizing the size of the test suite while maximizing code coverage.
For instance, several test minimization approaches used greedy heuristics to select test cases based on their code coverage~\cite{MB17,NH20}.
Black-box approaches include the FAST-R family of scalable approaches that leverages a representation of test source code (or command line inputs) in a vector-space model~\cite{CMVB19} and the ATM approach that is based on the abstract syntax tree of test source code~\cite{PGB23}.
%and a recent study uses word embedding to represent test cases specified in natural language~\cite{VPBB23}
Both use similarity metrics to cluster and then select test cases.

In the context of Web systems, both the system source code and test code are not available to determine similarity between source inputs, warranting a different black-box approach able to cluster and select source inputs for these systems.
Moreover, to make MST scalable, MR execution time should be minimized while preserving vulnerability detection, warranting an approach that minimizes the number of executed actions (\Cref{sec:aim:background:MST}) while covering all the input clusters.

\subsection{Clustering}
\label{sec:aim:background:Clustering}

Within the clustering steps in \Cref{sec:aim:doubleCLustering}, we rely on three well-known clustering algorithms: $\kmeans$ (\Cref{sec:aim:background:kMeans}), $\dbscan$ (\Cref{sec:aim:background:dbscan}), and $\hdbscan$ (\Cref{sec:aim:background:hdbscan}).
%, and we compare our approach with biclustering (\Cref{sec:biclustering}).

\subsubsection{\kmeans{}}
\label{sec:aim:background:kMeans}

K-means is a clustering algorithm which takes as input a set of data points and an integer $k$. 
K-means aims to assign data points to $k$ clusters by maximizing the similarity between individual data points within each cluster and the center of the cluster, called centroid.
The centroids are randomly initialized, then iteratively refined until a fixpoint is reached ~\cite{ASBZ17}.

\subsubsection{\dbscan{}}
\label{sec:aim:background:dbscan}

$\dbscan$ (Density-Based Spatial Clustering of Applications with Noise) is an algorithm that defines clusters using local density estimation.
This algorithm takes as input a dataset and two configuration parameters: the distance threshold $\epsilon$ and the minimum number of neighbors $\minpts$.

The distance threshold $\epsilon$ is used to determine the $\epsilon$-neighborhood of each data point, i.e.,  the set of data points that are at most $\epsilon$ distant from it.
There are three different types of data points in $\dbscan$, based on the number of neighbors in the  $\epsilon$-neighborhood of a data point:
\begin{description}
    \item[Core] If a data point has a number of neighbors above $\minpts$, it is then considered a core point.
   % For instance, in \Cref{fig:dbscan}, the points colored in red (e.g., A) are core points.
    \item[Border] If a data point has a number of neighbors below $\minpts$, but has a core point in its neighborhood, it is then considered a border point.
 %   For instance, in \Cref{fig:dbscan}, the points colored in yellow (B and C) are border points.
    \item[Noise] Any data point which is neither a core point nor a border point is considered noise.
%    For instance, in \Cref{fig:dbscan}, point N colored in blue is noise.
\end{description}

A cluster consists of the set of core points and border points that can be reached through their $\epsilon$-neighborhoods ~\cite{dbscan}. $\dbscan$ uses a single global $\epsilon$ value to determine the clusters.
But, if the clusters have varying densities, this could lead to suboptimal partitioning of the data.
$\hdbscan$ addresses this problem and we describe next.

\subsubsection{\hdbscan{}}
\label{sec:aim:background:hdbscan}

$\hdbscan$ (Hierarchical Density-Based Spatial Clustering of Applications with Noise) is an extension of $\dbscan$ (\Cref{sec:aim:background:dbscan}).
As opposed to $\dbscan$, $\hdbscan$ relies on different distance thresholds $\epsilon$ for each cluster, thus obtaining clusters of varying densities.

$\hdbscan$ first builds a hierarchy of clusters, based on various $\epsilon$ values selected in decreasing order.
Then, based on such a hierarchy, $\hdbscan$ selects as final clusters the most persistent ones, where cluster persistence represents how long a cluster remains the same without splitting when decreasing the value of $\epsilon$.
In $\hdbscan$, one has only to specify one parameter, which is the minimum number of individuals required to form a cluster, denoted by $n$~\cite{MHA17}. Clusters with less than $n$ individuals are considered noise and ignored.

\subsection{Many Objective Optimization}
\label{sec:aim:background:Optimization}

Engineers are often faced with problems requiring to fulfill multiple objectives at the same time, called \emph{multi-objective problems}.
For instance, multi-objective search algorithms were used in test suite minimization approaches to balance cost, effectiveness, and other objectives~\cite{WAG15,ZAY19}. %like feature pairwise coverage~\cite{WAG15} or transition coverage~\cite{ZAY19}.%.
%when minimizing a test suite (\Cref{sec:aim:background:testMinimization}), one may face tension between the number of test cases and the number of covered branches in the source code.
Multi-objective problems with at least (three or) four objectives are informally known as \emph{many-objective problems}~\cite{LLTY15}.
In both kind of problems,~\TSE{1.4.1}{one needs} a solution which is a good trade-off between the objectives.
Hence, we first introduce the Pareto front of a decision space (\Cref{sec:aim:background:paretoFront}).
Then, we describe genetic algorithms able to solve many-objective problems (\Cref{sec:aim:background:MaO}).

\subsubsection{Pareto Front}
\label{sec:aim:background:paretoFront}

Multi- and many-objective problems can be stated as minimizing several objective functions while taking values in a given decision space.
The goal of multi-objective optimization is to approximate the Pareto Front in the objective space~\cite{LLTY15}.

Formally, if $D$ is the decision space and $\objFun{1}{.}, \dots, \objFun{n}{.}$ are $n$ objective functions defined on $D$, then the \emph{fitness vector} of a decision vector $x \in D$ is
$\dataList{\objFun{1}{x},\dots, \objFun{n}{x}}$, hereafter denoted $\fitnessVector{x}$.
Moreover, a decision vector $x_1$ \emph{Pareto-dominates} a decision vector $x_2$ (hereafter denoted $x_1 \dominates x_2$) if 1) for each $1 \le i \le n$, we have $\objFun{i}{x_1} \le \objFun{i}{x_2}$, and 2) there exists $1 \le i \le n$ such that $\objFun{i}{x_1} < \objFun{i}{x_2}$.
If there exists no decision vector $x_1$ such that $x_1 \dominates x_2$, we say that the decision vector $x_2$ is \emph{non-dominated}.
The \emph{Pareto front} of $D$ is the set $\set{\fitnessVector{x_2}}{x_2 \in D \text{ and } \forall x_1 \in D: x_1 \notDominates x_2}$ of the fitness vectors of the non-dominated decision vectors.
Finally, a \emph{multi/many-objective problem} consists in:
$$\minimize_{x \in D} \fitnessVector{x} = \dataList{\objFun{1}{x},\dots, \objFun{n}{x}}$$
where the $\minimize$ notation means that we want to find or at least approximate the non-dominated decision vectors, hence the ones having a fitness vector in the Pareto front~\cite{LLTY15}.

\subsubsection{Solving Many-Objective Problems}
\label{sec:aim:background:MaO}

Multi-objective algorithms like \nsgaTwo~\cite{DPAM02} or \speaTwo~\cite{ZLT01, KHMW04} are not effective in solving many-objective problems~\cite{DJ14, PKT15} because of the following challenges:
\begin{enumerate}
    \item The proportion of non-dominated solutions becomes exponentially large with an increased number of objectives.
    This reduces the chances of the search being stuck at a local optimum and may lead to a better convergence rate~\cite{LLTY15}, but also slows down the search process considerably~\cite{DJ14}.
    \item With an increased number of objectives, diversity operators (e.g., based on crowding distance or clustering) become computationally expensive~\cite{DJ14}.
    \item If only a handful of solutions are to be found in a large-dimensional space, solutions are likely to be widely distant from each other.
    Hence, two distant parent solutions are likely to produce offspring solutions that are distant from them.
    In this situation, recombination operations may be inefficient and require crossover restriction or other schemes~\cite{DJ14}.
\end{enumerate}

To tackle these challenges, several many-objective algorithms have been successfully applied within the software engineering community, like \nsgaThree~\cite{DJ14, JD14} and \mosa~\cite{PKT15}.

\nsgaThree~\cite{DJ14, JD14} is based on \nsgaTwo~\cite{DPAM02} and addresses these challenges by assuming a set of supplied or predefined reference points.
Diversity (challenge 2) is ensured by starting the search in parallel from each of the reference points, assuming that largely spread starting points would lead to exploring all relevant parts of the Pareto front.
For each parallel search, parents share the same starting point, so they are assumed to be close enough so that recombination operations (challenge 3) are more meaningful.
Finally, instead of considering all solutions in the Pareto front, \nsgaThree focuses on individuals which are the closest to the largest number of reference points.
That way, \nsgaThree considers only a small proportion of the Pareto front (addressing challenge 1).

Another many-objective algorithm, \mosa~\cite{PKT15}, does not aim to identify a single individual achieving a  trade-off between objectives but a set of individuals, each satisfying one of the objectives. Such characteristic makes MOSA adequate for many software testing problems where it is sufficient to identify one test case (i.e., an individual) for each test objective (e.g., covering a specific branch or violating a safety requirement).
To deal with challenge 1, \mosa relies on a preference criterion amongst individuals in the Pareto front, by focusing on 1) extreme individuals (i.e., test cases having one or more objective scores equal to zero), and 2) in case of tie, the shortest test cases.
These best extreme individuals are stored in an archive during the search, and the archive obtained at the last generation is the final solution.
Challenges 2 and 3 are addressed by 
%preserving in the population, for each objective, the individual with the best fitness for such objective.
focusing the search, on each generation, on the objectives not yet covered by individuals in the archive.

%\subsubsection{Biclustering}
%\label{sec:biclustering}

%Biclustering is an approach taking as inputs a set $F$ of features (e.g., genes), a set $S$ of samples (e.g., conditions), and a $\card{F} \times \card{S}$ matrix $E$ where each coefficient $e_{fs}$ is a numerical value representing the expression of feature $f$ in sample $s$~\cite{BPP08}.
%For instance, on text mining based on a vector space model, $e_{fs}$ is the number of occurrences of the search word $f$ in the document $s$.

%Biclustering consists in partitioning features $F$ and samples $S$ at the same time, based on $E$, with the underlying assumption that a partition of $F$ (e.g., relevant groups of genes interfering with each other for the expression of physiological properties) should induce the partition of data observed for $S$ (e.g., some relevant conditions in which some genes can indeed work together).

%Biclustering is usually an NP-hard problem and biclustering techniques hence rely on meta-heuristics, usually minimizing a cost function for the obtain clusters, with a variety of meta-heuristics and associated cost functions proposed in the literature~\cite{PGAR15}.
% YM: and there is (apparently) no consensus on how to compare these techniques...

\section{Problem Definition}
\label{sec:aim:framework}

As the time required to execute a set of considered MRs may be large (\Cref{sec:aim:background:MST}), we aim to minimize the set of source inputs (hereafter, \emph{input set}) to be used when applying MST to a Web system, given a set of MRs.
In our context, each input is a sequence of actions used to communicate with the Web system and each action leads to a different Web page.
%Further, the initial input set to be minimized consists of inputs collected through crawling (\Cref{sec:aim:background:MST}).

%, such that we do not reduce its vulnerability detection capability.
%To achieve such an objective, we need to recall that MST aims to exercise the SUT by altering valid source input sequences through either changing the order of input actions or changing their values.
To ensure that a \emph{minimized input set} can exercise the same vulnerabilities as the original one, intuitively, we should ensure that they belong to the same \emph{input blocks}.
Indeed, in software testing, after identifying an important characteristic to consider for the inputs, one can partition the input space in blocks, i.e., pairwise disjoint sets of inputs, such that inputs in the same block exercise the SUT in a similar way~\cite{AO16}.
% For instance, consider a scenario where a code injection command is successful only when
% in the case of a code injection command that can succeed only if
% injected after an empty string, we want the empty strings to be taken into account in the same input block, given that they exercise the SUT in a similar way.
As the manual identification of relevant input blocks for a large system is extremely costly, we rely on clustering for that purpose (\Cref{sec:aim:doubleCLustering}).
%Moreover, since we consider several characteristics, based on input parameters and the outputs of the SUT, we obtain several partitions. Thus,
Since an input is a sequence of actions, it can exercise several input blocks.
In the rest of the paper, we rely on the notion of \emph{input coverage}, indicating the input blocks an input belongs to.

%For the problem we consider, we first assume we know, for each input in the initial input set, the cost and coverage of this input.
%Then, we state and motivate our goals (\Cref{sec:aim:goals}): identify a subset of the initial input set that minimizes total cost while maintaining the same coverage as the initial input set.
%We explain why we consider each input block to be covered as an individual objective (\Cref{sec:aim:mao}), hence facing a many-objective optimization problem.
%Then, we define the objective functions we consider for each individual objective (\Cref{sec:aim:problemDefinition}).
%Finally, we describe the solutions to our many-objective problem (\Cref{sec:aim:solutions}).

%%%%%%%%%%%%%%%%
\subsection{Assumptions and Goals}
\label{sec:aim:goals}
\label{defi:coveringSeqs}

We assume we know, for each input in the initial input set, 1) its cost and 2) its coverage.

1) Because we want to make \mst scalable, the cost $\costFunction{\sequence}$ of an input $\sequence$ corresponds to the execution time required to verify if the considered MRs are satisfied with this input.
Because we aim to reduce this execution time without having to execute the MRs,  as it would defeat the purpose of input set minimization, we use the number $\numberOfExecutedActions{\mr}{\sequence}$ of actions to be executed by an MR $\mr$ using input $\sequence$ as a surrogate metric for its execution time (see \Cref{sec:aim:background:MST}).
We thus define the cost of an input as follows:
\label{defi:cost}
$$\costFunction{\sequence} \eqdef \sum_{\mr \in \MRs} \numberOfExecutedActions{\mr}{\sequence}$$
When $\costFunction{\sequence} = 0$, input $\sequence$ was not exercised by any MR due to the preconditions in these MRs.
Hence, $\sequence$ is not useful for MST and can be removed without loss from the initial input set.
Finally, the total cost of an input set $\TestSuite$ is $\costFunction{\TestSuite}\eqdef \sum_{\sequence \in \TestSuite} \costFunction{\sequence}$.

2) To minimize the cost of metamorphic testing, we remove unnecessary inputs from the initial input set, but we want to preserve all the inputs able to exercise distinct vulnerabilities.
Hence, we consider, for each initial input $\sequence$, its coverage $\getSubclasses{\sequence}$.
In our study, $\getSubclasses{\sequence}$ is the set of input blocks the input $\sequence$ belongs to, and we determine these input blocks in \Cref{sec:aim:doubleCLustering} using \emph{double-clustering}.
%In short, each input is a sequence of actions used to communicate with the Web system and each action leads to a different Web page.
%Since we focus on security vulnerabilities in Web systems we consider as input characteristics 1) the system outputs since they characterize system states,
%and 2) action parameters (e.g., the URL, values belonging to a submitted form, or the method of sending a request to the server)
%since vulnerabilities might be detected through specific combinations of parameter values.
%We first cluster the system outputs (i.e., textual content extracted from Web pages) to obtain \emph{output classes}.
%Then, for each output class, we use action parameters to cluster the actions producing outputs in the class, obtaining \emph{action subclasses}.
%Finally, we define the coverage of each input (i.e., each sequence of actions) as the subclasses actions in the sequence belong to.
%We considered these input characteristics since we focus on security vulnerabilities in Web systems.
For now, we assume that the coverage of an input is known.
The total coverage of an input set $\TestSuite$ is
$\getSubclasses{\TestSuite} \eqdef \cup_{\sequence \in \TestSuite} \getSubclasses{\sequence}$.

We can now state our goals.
We want to obtain a subset $\TestSuiteFinal \subseteq \TestSuiteInit$ of the initial input set such that 1)  $\TestSuiteFinal$ does not reduce total input coverage, i.e., $\getSubclasses{\TestSuiteFinal} = \getSubclasses{\TestSuiteInit}$  and 2) $\TestSuiteFinal$ has minimal cost, i.e., $\costFunction{\TestSuiteFinal} = \min \{\costFunction{\TestSuite}$ $|$ $\TestSuite \subseteq \TestSuiteInit \land \getSubclasses{\TestSuite} = \getSubclasses{\TestSuiteInit}\}$.
Note that a solution $\TestSuiteFinal$ may not be necessarily unique.

%%%%%%%%%%%%%%%%
\subsection{A Many-Objective Problem}
\label{sec:aim:mao}

To minimize the initial input set, we focus on the \emph{selection} of inputs that belong to the same input blocks as the initial input set.
A potential solution to our problem is an input set $\TestSuite \subseteq \TestSuiteInit$.
Obtaining a solution $\TestSuite$ able to reach full input coverage is straightforward since, for each block $\actionSublass$, one can simply select an input in $\getSequences{\actionSublass} \eqdef \set{\sequence \in \TestSuiteInit}{\actionSublass \in \getSubclasses{\sequence}}$.
The hard part of our problem is to determine a combination of inputs able to reach full input coverage at a minimal cost.
Hence, we have to consider an input set as a whole and not focus on individual inputs.

This is similar to the \emph{whole suite} approach~\cite{FA13} targeting white-box tes\-ting.
They use as objective the total number of covered branches.
But, in our context, counting the number of uncovered blocks would consider as equivalent input sets that miss the same number of blocks, without taking into account that it may be easier to cover some blocks than others (e.g., some blocks may be covered by many inputs, but some only by a few) or that a block may be covered by inputs with different costs.
Thus, to obtain a combination of inputs that minimizes cost while preserving input coverage, we have to investigate how input sets cover each input block.
 
Hence, we are interested in covering each input block as an individual objective, in a way similar to the coverage of each code branch for white-box testing~\cite{PKT15}.
Because the total number of blocks to be covered is typically large ($\ge 4$), we deal with a \emph{many-objective problem}~\cite{LLTY15}.
This can be an advantage, because a many-objective reformulation of complex problems can reduce the probability of being trapped in local optima and may lead to a better convergence rate~\cite{PKT15}.
But this raises several challenges (\Cref{sec:aim:background:MaO}) that we tackle while presenting our search algorithm (\Cref{sec:aim:genetic}).

%%%%%%%%%%%%%%%%
\subsection{Objective Functions}
\label{defi:superposition}
\label{defi:redundancy}
\label{defi:redundantSeq}
\label{defi:reduction}
\label{defi:validSteps}
\label{defi:gainExhaust}
\label{sec:aim:reduction}
\label{sec:aim:gain}
\label{sec:aim:problemDefinition}

To provide effective guidance to a search algorithm, we need to quantify when an input set is closer to the objective of covering a particular block $\actionSublass$ than another input set.
In other words, if $\TestSuite_1$ and $\TestSuite_2$ are two input sets which do not cover $\actionSublass$ but have the same cost, we need to determine which one is more desirable to achieve the goals introduced in \Cref{sec:aim:goals} by defining appropriate objective functions.

%In this example, $\TestSuite_1$ and $\TestSuite_2$ would only differ with respect to their particular combination of inputs and what could happen when extending one or the other by an input $\sequence \in \getSequences{\actionSublass}$.
In general, adding an input to an input set would not only cover $\actionSublass$, but would also likely cover other blocks, that would then be covered by several inputs, thus introducing the possibility to remove some of them without affecting coverage.
To track of how a given block $\actionSublass$ is covered by inputs from a given input set $\TestSuite$, we introduce the concept of \emph{superposition} as $\superposition{\actionSublass}{\TestSuite} \eqdef \card{\getSequences{\actionSublass} \cap \TestSuite}$.
For instance, if $\superposition{\actionSublass}{\TestSuite} = 1$, then there is only one input in $\TestSuite$ covering $\actionSublass$.
In that case, this input is necessary to maintain the coverage of $\TestSuite$.
More generally, with the \emph{redundancy} metric, we quantify how much an input \TSE{2.5}{$\sequence$} is necessary to ensure the coverage of an input set \TSE{2.5}{$\TestSuite$ it belongs to}:
$\redundancy{\sequence}{\TestSuite} \eqdef \min \set{\superposition{\actionSublass}{\TestSuite}}{\actionSublass \in \getSubclasses{\sequence}}$ $- 1$.
The $-1$ is used to normalize the redundancy metric so that its range starts at $0$.
If $\redundancy{\sequence}{\TestSuite} = 0$, we say that $\sequence$ is \emph{necessary} in $\TestSuite$, otherwise we say that $\sequence$ is \emph{redundant}.
In the following, we denote $\Redundant{\TestSuite} \eqdef \set{\sequence \in \TestSuite}{\redundancy{\sequence}{\TestSuite} > 0}$ the set of the redundant inputs in $\TestSuite$.

To focus on the least costly input sets during the search (\Cref{sec:aim:genetic}), we quantify the gain obtained by removing redundant inputs.
If $\TestSuite$ contains a redundant input $\sequence$, then we call \emph{removal step} a transition from $\TestSuite$ to $\TestSuite \setminus \set{\sequence}{}$.
Otherwise, we say that $\TestSuite$ is already \emph{reduced}.
Unfortunately, given two redundant inputs $\sequence_1$ and $\sequence_2$, removing $\sequence_1$ may render $\sequence_2$ necessary.
Hence, when considering potential removal steps (e.g., removing either $\sequence_1$ or $\sequence_2$), one has to consider the order of these steps.
We represent a \emph{valid order} of removal steps by a \TSE{2.6}{sequence} of inputs $\dataList{\sequence_1, \dots, \sequence_n}$ to be removed from $\TestSuite$ such that, for each $0 \le i < n$, $\sequence_{i + 1}$ is redundant in $\TestSuite \setminus \set{\sequence_1, \dots, \sequence_i}{}$.
We denote $\ReducOrders{\TestSuite}$ the set of valid orders of removal steps in $\TestSuite$.
Removing redundant inputs $\sequence_1, \dots, \sequence_n$ leads to a reduction of cost $\costFunction{\sequence_1} + \dots + \costFunction{\sequence_n}$.
For each input set $\TestSuite$, we consider the maximal \emph{gain} from valid orders of removal steps:
$$
\begin{array}{r@{}l}
\gainExhaust{\TestSuite} \eqdef
{}& \max
\Big\{
\sum\limits_{1 \le i \le n}
\costFunction{\sequence_i}\\
{}& \Big|\ \dataList{\sequence_1, \dots, \sequence_n} \in \ReducOrders{\TestSuite} \Big\}
\end{array}
$$

To reduce the cost of computing this gain, we prove in the separate appendix (Theorem 1) that, to determine which orders of removal steps are valid, we can remove inputs in any arbitrary order, without having to resort to backtracking to previous inputs.
Moreover, in our approach, we need to compute the gain only in situations when the number of redundant inputs is small (\Cref{sec:aim:genetic}),
%Indeed, during the search (\Cref{sec:aim:genetic}), we do not consider input sets in general but input sets that are already reduced, to or from which we then add or remove only a few inputs.
%More precisely, the gain is only used to obtain the best removal steps while initializing the populations (\Cref{sec:aim:selectParents}) or mutating the offspring (\Cref{sec:aim:mutation}), and to compute the objective functions (\Cref{sec:aim:updtPops}).
thus exhaustively computing the gain is tractable.

%To define our objective functions, let us consider an input set $\TestSuite \subseteq \TestSuiteInit$ which is already reduced but does not cover a given input block $\actionSublass$.
%As described in \Cref{sec:aim:mao}, any input $\sequence_1 \in \getSequences{\actionSublass}$ could be added to $\TestSuite$ in order to cover $\actionSublass$.
%In that case, inputs in $\TestSuite$ that cover a block in common with $\sequence_1$ could become redundant, which would lead to a gain $\gainExhaust{\TestSuite \cup \set{\sequence_1}{}}$ after the corresponding removal steps.
Adding an input $\sequence_1 \in \getSequences{\actionSublass}$ to $\TestSuite$ would lead to a gain but would also result in additional cost, hence warranting we consider the benefit-cost balance $\gainExhaust{\TestSuite \cup \set{\sequence_1}{}} - \costFunction{\sequence_1}$ to evaluate how efficiently $\actionSublass$ is covered by $\sequence_1$.
More generally, we define the \emph{potential} of $\TestSuite$ to efficiently cover $\actionSublass$ as the maximum benefit-cost balance obtained by adding inputs $\sequence_1$ in covering $\actionSublass$.
But, as an input $\sequence_1$ may be necessary to cover $\actionSublass$ while leading to no or not enough removal steps, $\gainExhaust{\TestSuite \cup \set{\sequence_1}{}} - \costFunction{\sequence_1}$ may be negative.
To facilitate normalization, we need the potential to return a non-negative value, thus we shift all the benefit-cost balances for a given objective $\actionSublass$ by adding a dedicated term.
As the potential is a maximum, the worst case of $\gainExhaust{\TestSuite \cup \set{\sequence_2}{}} - \costFunction{\sequence_2}$ is when 
$\gainExhaust{\TestSuite \cup \set{\sequence_2}{}} = 0$ and $\costFunction{\sequence_2}$ is the minimal cost amongst the inputs able to cover $\actionSublass$.
Hence, we obtain the following definition for the potential of $\TestSuite$ in covering $\actionSublass$:
$$
\begin{array}{r@{}l}
\potential{\TestSuite}{\actionSublass} \eqdef
&{} \max \{\gainExhaust{\TestSuite \cup \set{\sequence_1}{}} - \costFunction{\sequence_1}\\
&{}\hspace{2.32cm}|\ \sequence_1 \in \getSequences{\actionSublass} \}\\
+
&{} \min \set{\costFunction{\sequence_2}}{\sequence_2 \in \getSequences{\actionSublass}}
\end{array}
$$

%This metric is more meaningful when the considered input set $\TestSuite$ is already reduced since, in that case, the gain would come only from reductions obtained after adding inputs able to cover the uncovered block.
%If an input set is not already reduced, then unrelated inputs contribute to the potential of the input set, hence blurring the information regarding the block to be covered.
%This is why during the search (\Cref{sec:aim:genetic}) we consider as candidates only reduced input sets and, consequently, we reduce each input set after an input is added.

%We consider again an input block $\actionSublass$ and two reduced input sets $\TestSuite_1$ and $\TestSuite_2$ , both of them not covering $\actionSublass$ but having the same cost.
%If $\potential{\TestSuite_1}{\actionSublass} > \potential{\TestSuite_2}{\actionSublass}$ then, due to its combination of inputs, $\TestSuite_1$ can be extended to cover $\actionSublass$ in a way that (after removal steps) is less costly than it would be for $\TestSuite_2$.
%Thus, we consider $\TestSuite_1$ to be a more desirable way to achieve the coverage of $\actionSublass$ than $\TestSuite_2$.
We thus use the potential to define the objective function associated with an objective $\actionSublass$.

To normalize our metrics, we rely on the normalization function $\normalization{x} \eqdef \tfrac{x}{x + 1}$, which is used to reduce a range of values from $[0, \infty)$ to $[0, 1)$ while preserving the ordering. When used during a search, it is less prone to precision errors and more likely to drive faster convergence towards an adequate solution than alternatives~\cite{Arc10}.
We use it to normalize the cost 
%$\normalization{\costFunction{\TestSuite}}$
between $0$ and $1$ and the smaller is the normalized cost, the better a solution is.
For coverage, since a high potential is more desirable, we use its complement $\tfrac{1}{x + 1} = 1 - \normalization{x}$ to reverse the order, so that the more potential an input set has, the lower its coverage objective function is.
We thus define the objective function corresponding to a block $\actionSublass_i$ as:
$$\objFun{\actionSublass_i}{\TestSuite} \eqdef
\left\{
\begin{array}{ll}
    0 & \text{if } \actionSublass_i \in \getSubclasses{\TestSuite} \\
    \frac{1}{\potential{\TestSuite}{\actionSublass_i} + 1} & \text{otherwise}
\end{array}
\right.$$

The lower this value, the better.
If $\actionSublass_i$ is covered, then $\objFun{\actionSublass_i}{\TestSuite} = 0$, otherwise $\objFun{\actionSublass_i}{\TestSuite} > 0$.
As expected, input sets that cover the objective are better than input sets that do not, and if two input sets do not cover the objective, then the normalized potential is used to break the tie.

%%%%%%%%%%%%%%%%
\subsection{Solutions to Our Problem}
\label{sec:aim:solutions}
\label{defi:pareto}
\label{defi:pbDef}

%To solve our problem, we first need to retrieve, for each input $\sequence \in \TestSuiteInit$ in the initial input set, 1) its cost $\costFunction{\sequence}$ without executing the considered MRs and 2) its coverage $\getSubclasses{\sequence}$, depending on the input characteristics that are relevant for vulnerability detection.
Each element in the decision space is an input set $\TestSuite \subseteq \TestSuiteInit$, which is associated with a \emph{fitness vector}:
$$\fitnessVector{\TestSuite} \eqdef \dataList{\normalization{\costFunction{\TestSuite}}, \objFun{\actionSublass_1}{\TestSuite}, \dots, \objFun{\actionSublass_\nbSubClasses}{\TestSuite}}$$
where $\getSubclasses{\TestSuiteInit} = \set{\actionSublass_1, \dots, \actionSublass_\nbSubClasses}{}$ denotes the $\nbSubClasses$ input blocks to be covered by input sets $\TestSuite \subseteq \TestSuiteInit$.

Hence, we can define the Pareto front formed by the non-dominated solutions in our decision space (\Cref{sec:aim:background:paretoFront}) and we formulate our problem definition as a many-objective optimization problem:
$$\minimize_{\TestSuite \subseteq \TestSuiteInit} \fitnessVector{\TestSuite}$$
where the $\minimize$ notation means that we want to find or at least approximate the non-dominated decision vectors having a fitness vector on the Pareto front~\cite{LLTY15}.
Because we want full input coverage (\Cref{sec:aim:goals}), the ultimate goal is a non-dominated solution $\TestSuiteFinal$ such that:$$\fitnessVector{\TestSuiteFinal} = \dataList{\normalization{\mathit{cost}_{\min}}, 0, \dots, 0}$$
where $\mathit{cost}_{\min}$ is the cost of the cheapest subset of $\TestSuiteInit$ with full input coverage.
 
%Moreover, even if at the end we are only interested in a final solution ensuring full input coverage at minimal cost, approximating other non-dominated solutions is useful during the search (\Cref{sec:aim:genetic}).
%For instance, non-dominated solutions with partial input coverage and a minimal cost for such coverage may be intermediary steps towards finding the optimal solution $\TestSuiteFinal$.
%Also, considering other points in the Pareto front instead of only focusing on approximating solutions with full input coverage is useful to maintain diversity during the search, preventing this search from being stuck at a local minimum.

\section{Overview of the approach}
\label{sec:aim:overview}

As stated in our problem definition (\Cref{sec:aim:framework}), we aim to reduce the cost of MST (\Cref{sec:aim:background:MST}) by minimizing an initial input set without removing inputs that are required to exercise distinct security vulnerabilities.
To do so, we need to tackle the following sub-problems (\Cref{defi:pbDef}):
\begin{enumerate}
    \item For each initial input, we need to determine its cost without executing the considered MRs.
    \item For each initial input, we need to determine its coverage. In the context of metamorphic testing for Web systems, we consider input blocks based on system outputs and input parameters.
    \item Amongst all potential input sets  $\TestSuite \subseteq \TestSuiteInit$, we search for a non-dominated solution $\TestSuiteFinal$ that preserves coverage while minimizing cost.
\end{enumerate}

\begin{figure*}[t]
\centering\includegraphics[width=0.75\linewidth]{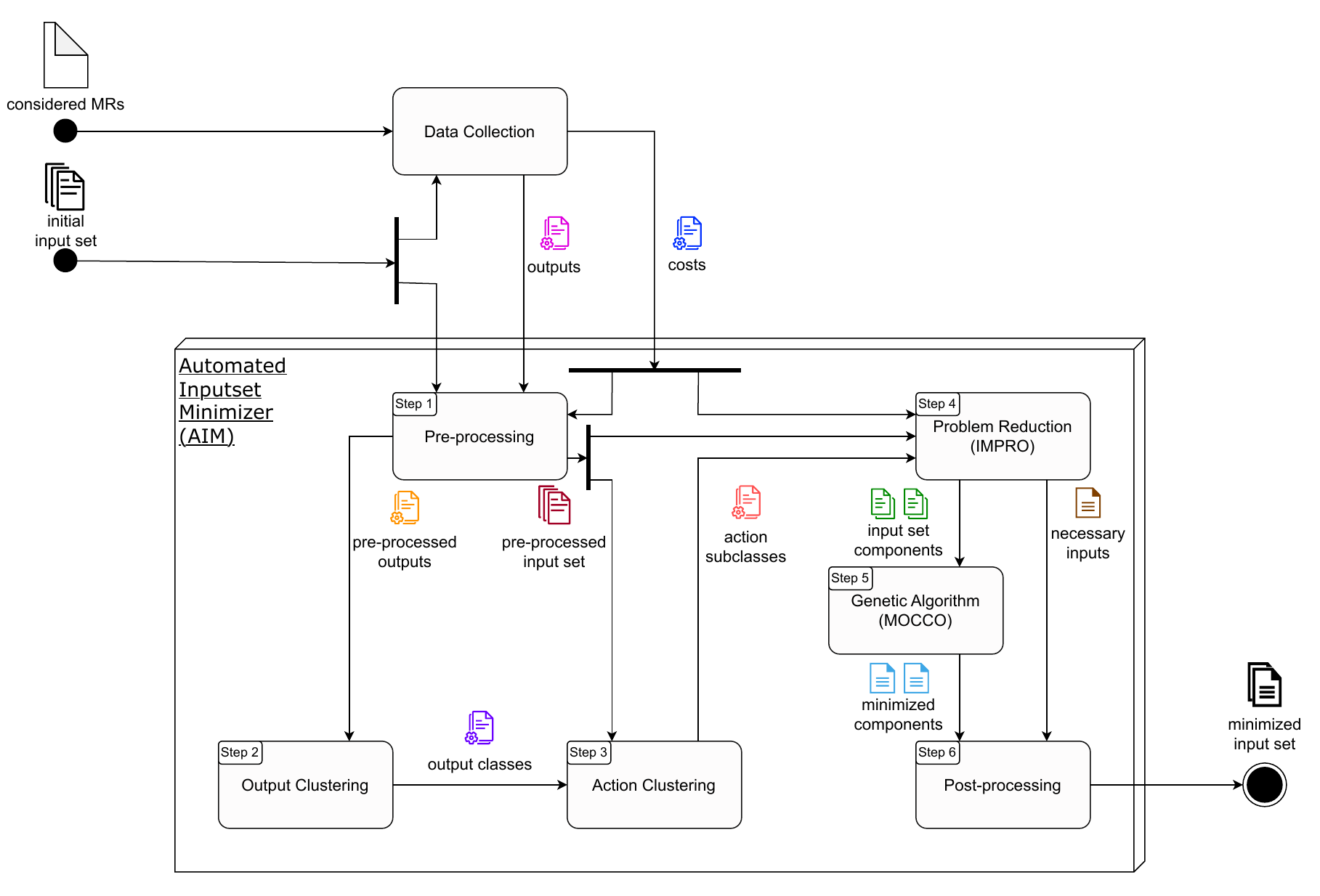}
\caption{Activity diagram of the \aimLong(\aim) approach.}
\label{fig:aim}
\end{figure*}

%The \emph{\aimLong (\aim)} approach works in six steps whose workflow is depicted in \Cref{fig:aim}.
%\aim takes an \emph{initial input set} and generates a \emph{minimized input set}, i.e., a subset of the initial input set.
%The inputs in both sets can be used as source inputs for MST.
%We optimized \aim to work with test inputs for Web systems; in such context, each input is a sequence of actions (e.g., URL requests or interactions with the widget of a Web page) leading to Web pages as output.
%Such action sequences can be automated through Web testing frameworks such as Selenium~\cite{selenium}.

The \emph{\aimLong (\aim)} approach relies on analyzing the output and cost corresponding to each input.
\aim obtains such information through a new feature added to the \mstWi toolset to execute each input on the system and retrieve the content of the corresponding Web pages.
Obtaining the outputs of the system is very inexpensive compared to executing the considered MRs.
Moreover, to address our first sub-problem, we also updated \mstWi to retrieve the cost of an input without executing the considered MRs.
We rely on a surrogate metric (\Cref{sec:aim:background:MST}),  linearly correlated with execution time, which is inexpensive to collect (\Cref{sec:aim:cost}).

In Step 1 (Pre-processing),
\aim pre-processes the initial input set and the output information, by extracting relevant textual content from each returned Web page (\Cref{sec:aim:outputClustering:preprocessing}).

To address the second sub-problem, \aim relies on a \emph{double-clustering} approach (\Cref{sec:aim:doubleCLustering}), which is implemented by Step 2 (Output Clustering) and Step 3 (Action Clustering).
For both steps, \aim relies on state-of-the-art clustering algorithms, which require to select hyper-parameter values  (\Cref{sec:aim:doubleClustering:hyperParameter}).
\emph{Output clustering} (\Cref{sec:aim:outputClustering}) is performed on the pre-processed outputs, each generated cluster corresponding to an \emph{output class}.
Then, for each output class identified by the Output Clustering Step, \emph{Action clustering} (\Cref{sec:aim:actionClustering}) first determines the actions whose output belongs to the considered output class, then partitions these actions based on action parameters such as URL, username, and password, obtaining \emph{action subclasses}.
On the completion of Step 3, \aim has mapped each input to a set of action subclasses, used for measuring input coverage as per our problem definition (\Cref{defi:coveringSeqs}).

To preserve diversity in our input set, and especially to retain inputs that are necessary to exercise vulnerabilities, we require the minimized input set gene\-rated by \aim to cover the same action subclasses as the initial input set.
That way, we increase the chances that the minimized input set contains at least one input able to exercise each vulnerability detectable with the initial input set.

Using cost and coverage information, \aim can address the last sub-problem.
Since the size of the search space exponentially grows with the number of initial inputs, the solution cannot be obtained by exhaustive search.
Actually, our problem is analogous to the knapsack pro\-blem~\cite{LHH07}, which is NP-hard, and is thus unlikely to be solved by deterministic algorithms.
Therefore, \aim relies on  meta-heuristic search to find a solution (Step 5) after reducing the search space (Step 4).
%to the maximum extent possible.

In Step 4, since the search space might be large, \aim first reduces the search space to the maximal extent possible (\Cref{sec:aim:search}) before resorting to meta-heuristic search.
Precisely, it relies on the \improLong (\impro) component for problem reduction, which determines the necessary inputs, removes inputs that cannot be part of the solution, and partition the remaining inputs into input set \emph{components} that can be independently minimized.

In Step 5, \aim applies a genetic algorithm (\Cref{sec:aim:genetic}) to minimize each component.
Because existing algorithms did not entirely fit our needs, we explain why we introduce \geneticAlgo (\geneticAlgoLong), a novel genetic algorithm
%based on two populations, one focusing in covering the objectives, and the other one on reducing cost.
%By restricting crossover so that one parent is selected in each population, we aim at
which converges towards a solution covering the objectives at a minimal cost, obtaining a \emph{minimized input set component}.

Finally, after the genetic search is completed for each component, in Step 6 (Post-processing), \aim generates the \emph{minimized input set} by combining the necessary inputs identified by \impro  and the inputs from the \geneticAlgo minimized components (\Cref{sec:aim:postprocessing}).

Note that, even though in our study we focus on Web systems, steps 4 (\impro), 5 (\geneticAlgo), and 6 (post-processing), which form the core of our solution, are generic and can be applied to any system.
Moreover, step 1 (pre-processing) and steps 2 and 3 (double-clustering) can be tailored to apply \aim to other domains (e.g., desktop applications, embedded systems).
And though we relied on \mstWi to collect our data, \aim does not depend on a particular data collector, and using or implementing another data collector would enable the use of our approach in other contexts.

\section{Step 1: Data Collection and Pre-Processing}
\label{sec:aim:system}

In Step 1, \aim determines the cost of each initial input (\Cref{sec:aim:cost}) and extract meaningful textual content from the Web pages obtained with the initial inputs (\Cref{sec:aim:outputClustering:preprocessing}).
%To address our first sub-problem (\Cref{sec:aim:overview}), we determine the cost of each initial input (\Cref{sec:aim:cost}).
%Then, since in this study we focus on Web systems, we detail how we extract meaningful textual content from the Web pages obtained with the initial inputs (\Cref{sec:aim:outputClustering:preprocessing}).

%%%%%%%%%%%%%%%%
\subsection{Input Cost}
\label{sec:aim:cost}

The cost of a source input is the number of actions executed by source and follow-up inputs for the considered MRs (\Cref{defi:cost}).
Note that counting the number of actions to be executed is inexpensive compared to executing them on the SUT, then checking for the verdict of the output relation.
For instance, counting the number of actions for eleven MRs with \jenkins' initial input set (\Cref{sec:aim:results}) took less than five minutes, while executing the MRs took days.

\subsection{Output Representation}
\label{sec:aim:outputClustering:preprocessing}

Since, in this study, we focus on Web systems, the outputs of the SUT are Web pages.
Fortunately, collecting these pages using a crawler and extracting their textual content is inexpensive compared to executing MRs.
Hence, we can use system outputs to determine relevant input blocks (\Cref{sec:aim:framework}).

We focus on textual content extracted from the Web pages returned by the Web system under test.
We remove from the textual content of each Web page all the data that is shared among many Web pages and thus cannot characterize a specific page, like system version, date, or (when present) the menu of the Web page.
Moreover, to focus on the meaning of the Web page, we consider the remaining textual content not as a string of characters but as a sequence of words.
Also, following standard practice in natural language processing, we apply a stemming algorithm
%to simplify words to their simplest form, thus considering
to consider distinct words with the same stem as equivalent, for instance the singular and plural forms of the same word.
Finally, we remove stopwords, numbers, and special characters, in order to focus on essential textual information.

\section{Steps 2 and 3: Double Clustering}
\label{sec:aim:doubleCLustering}

To reduce the cost of MST, we want to minimize an initial input set while preserving, for each vulnerability affecting the SUT, at least one input able to exercise it; of course, in practice, such vulnerabilities are not known in advance but should be discovered by MST.
Hence, we have to determine in which cases two inputs are distinct enough so that both should be kept in the minimized input set, and in which cases some inputs are redundant with the ones we already selected and thus can be removed.
To determine which inputs are similar and which significantly differ, we rely on clustering algorithms.
Precisely, we rely on the $\kmeans$, $\dbscan$, and $\hdbscan$ algorithms to cluster our data points.
Each of them has a set of hyper-parameters to be set and we first detail how these hyper-parameters are obtained using \emph{Silhouette analysis} (\Cref{sec:aim:doubleClustering:hyperParameter}).

Since, for practical reasons, we want to avoid making assumptions regarding the nature of the Web system under test (e.g., programming language or underlying middleware), we propose a black-box approach relying on input and output information to determine which inputs we have to keep or remove.
In the context of a Web system, each input is a sequence of actions, each action enables a user to access a Web page (using a POST or GET request method), and each output is a Web page.
%Furthermore, each action uses a request method, either a POST or GET method, to send an HTTP request to the server.
After gathering output and action information, we perform \emph{double-clustering} on our data points, i.e., two clustering steps performed in sequence:
\begin{enumerate}
    \item \emph{Output clustering} (\Cref{sec:aim:outputClustering}) uses the outputs of the Web system under test, i.e., textual data obtained by pre-processing content from Web pages (\Cref{sec:aim:outputClustering:preprocessing}).
    We define an output distance (\Cref{sec:aim:outputClustering:outputDistance}) to quantify similarity between these outputs, which is then used to run Silhouette analysis and clustering algorithms to partition outputs into \emph{output classes} (\Cref{sec:aim:outputClustering:classes}).
    \item \emph{Action clustering} (\Cref{sec:aim:actionClustering}) then determines input coverage. First, \aim collects in the same \emph{action set} actions leading to outputs in the same output class (\Cref{sec:aim:diversity:actionSets}).
    Then, \aim refines each action set by partitioning the actions it contains using action parameters.
    To do so, it first uses the request method (\Cref{sec:aim:requestPartition}) to split action sets into parts.
    Then, we define an action distance (\Cref{sec:aim:distanceActions}) based on the URL (\Cref{sec:aim:actionClustering:urlDistance}) and other parameters (\Cref{sec:aim:actionClustering:paramDistance}) of the considered actions.
    Finally, \aim relies on Silhouette analysis and cluste\-ring algorithms to partition each part of an action set into \emph{action subclasses} (\Cref{sec:aim:actionClustering:subclasses}), defining our input blocks (\Cref{defi:coveringSeqs}).
\end{enumerate}

Note that \emph{double-clustering} should not be confused with \emph{biclustering}~\cite{BPP08,PGAR15}, since the latter simultaneously clusters two distinct aspects (features and samples) of the data, while the former clusters only one aspect (actions, in our case, that can be seen as features) but in two consecutive steps (action outputs, then action parameters), the second refining the first one.

\subsection{Hyper-parameters Selection}
\label{sec:aim:doubleClustering:hyperParameter}

In this study, we rely on the common \kmeans~\cite{ASBZ17}, \dbscan~\cite{dbscan}, and \hdbscan~\cite{MHA17} clustering algorithms (\Cref{sec:aim:background:Clustering}) to determine output classes (\Cref{sec:aim:outputClustering}) and action subclasses (\Cref{sec:aim:actionClustering}).
%Hence, we introduce here a methodology shared between both the output and input clustering steps of our approach.
These clustering algorithms require a few hyper-parameters to be set.
One needs to select for \kmeans the number of clusters $k$, for \dbscan the distance threshold $\epsilon$ and the minimum number of neighbors $\minpts$, and for \hdbscan the minimum number $n$ of individuals required to form a cluster.

To select the best values for these hyper-parameters, we rely on \emph{Silhouette analysis}.
Though the Silhouette score is a common metric used to determine optimal values for hyper-parameters~\cite{AFPB23a,AFPB23b}, it is obtained from the average Silhouette score of the considered data points.
Thus, for instance, clusters with all data points having a medium Silhouette score cannot be distinguished from clusters where some data points have a very large Silhouette score while others have a very small one.
Hence, having a large Silhouette score does not guarantee that all the data points are well-matched to their cluster.
To quantify the variability in the distribution of Silhouette scores, we use Gini index, a common measure of statistical dispersion.
If the Gini index is close to $0$, then Silhouette scores are almost equal.
Conversely, if it is close to $1$, then the variability in Silhouette score across data points is large.

Hence, for our Silhouette analysis, we consider two objectives: (average) Silhouette score and the Gini index of the Silhouette scores.
The selection of hyper-parameters is therefore a multi-objective problem with two objectives.
We rely on the common \nsgaTwo evolutionary algorithm~\cite{DPAM02} to solve this problem and approximate the Pareto front regarding both Silhouette score and Gini index.
Then, we select the item in the Pareto front that has the highest Silhouette score.
%The results of the obtained hyper-parameter values for output clustering and action clustering are reported in \Cref{sec:aim:results:configurations}.

\subsection{Step 2: Output Clustering}
\label{sec:aim:outputClustering}
\label{defi:sequences}

\emph{Output clustering} consists in defining an output distance (\Cref{sec:aim:outputClustering:outputDistance}) to quantify dissimilarities between Web system outputs, and then to partition the outputs to obtain \emph{output classes} (\Cref{sec:aim:outputClustering:classes}).

A user communicates with a Web system using actions.
Hence, an input for a Web system is a sequence of actions (e.g., login, access to a Web page, logout).
%We denote by $\seqActions{\action_1, \dots, \action_n}$ an \emph{input} consisting in $n \ge 1$ actions $\action_1, \dots, \action_n$.
%The \emph{length} of an input $\sequence$ is its number of actions and is denoted by $\length{\sequence}$.
As the same action may occur several times in an input $\sequence$, a given occurrence of an action is identified by its position $i$ in the input and denoted $\getAction{\sequence}{i}$.
%If $1 \le i \le \length{\sequence}$, we denote by $\getAction{\sequence}{i}$ the action at position $i$ in input $\sequence$.
Outputs of a Web system are textual data obtained by pre-processing the content from Web pages (\Cref{sec:aim:outputClustering:preprocessing}).
The accessed Web page depends not only on the considered action, but also on the previous ones; for instance if the user has logged into the system.
Hence, we denote by $\getOutput{\sequence}{i}$ the output of the action at position $i$ in $\sequence$.

\subsubsection{Output distance}
\label{sec:aim:outputClustering:outputDistance}

In this study, we use system outputs (i.e., Web pages) to characterize system states.
%Hence, we first consider action (di)similarity by the Web pages they can access.
Hence, two actions that do not lead to the same output should be considered distinct because they bring the system into different states.
More generally, dissimilarity between outputs is quantified using an \emph{output distance}.
Since we deal with textual data, we consider both Levenshtein and bag distances.
Levenshtein distance is usually a good representation of the difference between two textual contents~\cite{HAB13,TSBB17}.
However, computing the minimal number of edits between two strings can be costly, since the complexity of the Levenshtein distance between two strings is $\complexity{\length{s_1}\times \length{s_2}}$, where $\length{.}$ is the length of the string~\cite{lev17}.
Thus, we consider the bag distance~\cite{Mer22} as an alternative to the Levenshtein distance, because its complexity is  only $\complexity{\length{s1}+\length{s_2}}$~\cite{BCP02}.
But it does not take into account the order of words and is thus less precise than Levenshtein distance.
%We compare in \Cref{sec:aim:results} the execution time required to compute both Levenshtein and bag distances.
%Then, we evaluate which distance leads to a minimized input set able to exercise the largest number possible of vulnerabilities.

\subsubsection{Output Classes}
\label{sec:aim:outputClustering:classes}

%Since we want to minimize an input set while preserving input diversity, for each output obtained by executing the initial input set, we want in the minimized input set at least one input that lead to this output. To do so, w
We partition  the textual content we obtained from Web pages (\Cref{sec:aim:outputClustering:preprocessing}) using the \kmeans, \dbscan, and \hdbscan clustering algorithms, setting the hyper-parameters using Silhouette analysis (\Cref{sec:aim:doubleClustering:hyperParameter}), and determining similarities between outputs using the chosen output distance.
We call \emph{output classes} the obtained clusters and we denote by $\getOutputClass{\sequence}{i}$ the unique output class $\getOutput{\sequence}{i}$ belongs to.
\label{defi:outputClassOfAction}
%Let $\sequence$ be an input and let $i \in 1 \le i \le \length{\sequence}$ be a position in this input.

%%%%%%%%
\subsection{Step 3: Action Clustering}
\label{sec:aim:actionClustering}

%The double clustering approach aims to identify the coverage entities that should be covered by a minimized input set, to ensure that inputs that may help discovering vulnerabilities are not discarded.
%In the context of Web systems, to ensure all possible vulnerabilities are discovered, it is likely necessary to access all the Web pages, or at least all the ones exercising a different feature. 
%Ensuring that all the diverse Web pages are exercised can be based on the output classes detected in the previous step (\Cref{sec:aim:outputClustering}).

Exercising all the Web pages is not sufficient to discover all the vulnerabilities; indeed, vulnerabilities might be detected through specific combinations of parameter values associated to an action 
(e.g., values belonging to a submitted form). 
%Precisely, actions on a Web system may have different \emph{parameters} 
%that can be used to distinguish them, 
Precisely, actions on a Web system can differ with respect to a number of \emph{parameters} 
that include  
%like 
the URL (allowing the action to perform a request to a Web server), the method of sending a request to the server (like GET or POST), 
%parameters like form inputs, username, or password.
URL parameters (e.g., \texttt{http://myDomain.com/myPage?urlParameter1=}
\texttt{value1\&urlParameter2=value2}), and entries in form inputs (i.e., textarea, textbox, options in select items, datalists). 

Based on the obtained output classes, \emph{action clustering} first determines \emph{action sets} (\Cref{sec:aim:diversity:actionSets}).
%such that actions leading to outputs in the same output class are in the same action set.
Then, action clustering refines each action set by partitioning the actions it contains using actions parameters.
First, we give priority to the method used to send a request to the server, so we split each action set using the request method (\Cref{sec:aim:requestPartition}).
Then, to quantify the dissimilarity between two actions, we define an action distance (\Cref{sec:aim:distanceActions}) based on URL (\Cref{sec:aim:actionClustering:urlDistance}) and other parameters (\Cref{sec:aim:actionClustering:paramDistance}).
That way, action clustering refines each action set into \emph{action subclasses} (\Cref{sec:aim:actionClustering:subclasses}).

\subsubsection{Action Sets}
\label{sec:aim:diversity:actionSets}

Based on the obtained output classes (\Cref{defi:outputClassOfAction}), \aim determines \emph{action sets} such that actions leading to outputs in the same output class $\outputClass$ are in the same action set:
\label{defi:actionClass}
$$
\begin{array}{r@{}r}
\getActionClass{\outputClass} \eqdef
& \{ \action\ |\ \exists \sequence, i: \getAction{\sequence}{i} = \action \\
& \land\ \getOutputClass{\sequence}{i} = \outputClass \}\\
\end{array}
$$

Note that, because an action can have different outputs depending on the considered input, it is possible for an action to belong to several actions sets, corresponding to several output classes.

\subsubsection{Request Partition} 
\label{sec:aim:requestPartition}

Each action uses either a POST or GET method to send an HTTP request to the server. 
Actions (such as login) that send the parameters to the server in the message body use the POST method, while actions that send the parameters through the URL use the GET method.
As this difference is meaningful for distinguishing different action types, we split each action set into two parts: the actions using a POST method and those using a GET method.

\subsubsection{Action Distance} 
\label{sec:aim:distanceActions}
\label{defi:actionDist}

After request partition (\Cref{sec:aim:requestPartition}), we consider one part of an action set at a time and we refine it using an action distance quantifying dissimilarity between actions based on remaining parameters  (e.g., URL or form entries).
%user name, password, etc.
In the context of a Web system, each Web content is identified by its URL, so we give more importance to this parameter.
We denote $\urlAct{\action_i}$ the URL of action $\action_i$.
For the sake of clarity, we call in the rest of the section \emph{residual parameters} the parameters of an action which are not its request method nor its URL and we denote $\paramsAct{\action_i}$ the residual parameters of action $\action_i$.
Since we give more importance to the URL, we represent the distance between two actions by a real value, where the integral part corresponds to the distance between their respective URLs and the decimal part to the distance between their respective residual parameters:
$$\begin{array}{r@{}l}
     \actionDist{\action_1}{\action_2} &{}\eqdef  \distURL{\urlAct{\action_1}}{\urlAct{\action_2}}\\
     &{} + \distParam{\paramsAct{\action_1}}{\paramsAct{\action_2}}
\end{array}$$
 where the URL distance $\distURL{.}{.}$ is defined in \Cref{sec:aim:actionClustering:urlDistance} and returns an integer, and the parameter distance $\distParam{.}{.}$ is defined in \Cref{sec:aim:actionClustering:paramDistance} and returns a real number between $0$ and $1$.

\subsubsection{URL distance}
\label{sec:aim:actionClustering:urlDistance}
\label{def:actionClustering:UrlDistance}

\label{def:actionClustering:Url}
\begin{figure}[t]
\centering
\begin{tikzpicture}[scale=1.1]
    % styles
    \tikzset{
        edge/.style={thick,color=black},
    }
    % words
    \node[black]
        (url1)
        at (0,0)
        {\texttt{http}};
    \node[black]
        (url2)
        at (0,-1)
        {\texttt{hostname}};
    \node[black]
        (url3)
        at (-1,-2)
        {\texttt{login}};
    \node[black]
        (url4)
        at (1,-2)
        {\texttt{job}};
    \node[black]
        (url5)
        at (1,-3)
        {\texttt{try1}};
    \node[black]
        (url6)
        at (1,-4)
        {\texttt{lastBuild}};
    % edges
	\draw[edge]
        (url1) -- (url2);
	\draw[edge]
        (url2) -- (url3);
	\draw[edge]
        (url2) -- (url4);
	\draw[edge]
        (url4) -- (url5);
	\draw[edge]
        (url5) -- (url6);
    % arrows
    \draw[edge,<->] (-2.5,-1) -- (-2.5,-2) node [midway, left] {$1$};
    \draw[edge,<->] (2.5,-1) -- (2.5,-4) node [midway, right] {$3$};
\end{tikzpicture}
\caption{The URL distance between \texttt{http://hostname/login} and \texttt{http://hostname/job/try1/lastBuild} is $1 + 3 = 4$.}
\label{fig:exampleURL}
\end{figure}
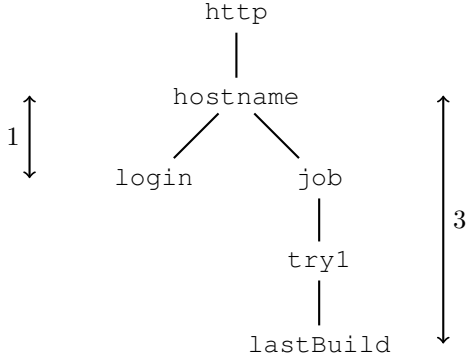

A URL is represented as a sequence of at least two words, separated by \aimChar{://} between the first and second word, then by \aimChar{/} between any other words.
The length of a URL $\dataUrl$ is its number of words, denoted $\length{\dataUrl}$.
Given two URLs, $\dataUrl_1$ and $\dataUrl_2$, their \emph{lowest common ancestor} is the longest prefix they have in common, denoted $\lca{\dataUrl_1}{\dataUrl_2}$.
We define the distance between two URLs as the total number of words separating them from their lowest common ancestor:
$$\begin{array}{r@{}l}
     \distURL{\dataUrl_1}{\dataUrl_2} \eqdef
     &{} \length{\dataUrl_1} + \length{\dataUrl_2} \\
     &{} - 2\times \length{\lca{\dataUrl_1}{\dataUrl_2}}
\end{array}$$
We provide an example in \Cref{fig:exampleURL}.

%\begin{ex}
%We consider two URLs: $\dataUrl_1$ = \texttt{http://hostname/login} and $\dataUrl_2$ = \texttt{http://hostname/} \texttt{job/try1/lastBuild}.
%Their LCA is $\texttt{http://hostname}$.
%There is one word (\texttt{login}) separating $\dataUrl_1$ from their LCA, hence $\length{\dataUrl_1} - \length{\lca{\dataUrl_1}{\dataUrl_2}} = 1$.
%Moreover, there are three words (\texttt{job}, \texttt{try1}, and \texttt{lastBuild}) separating $\dataUrl_2$ from their LCA, hence $\length{\dataUrl_2} - \length{\lca{\dataUrl_1}{\dataUrl_2}} = 3$.
%Therefore, the URL distance between $\dataUrl_1$ and $\dataUrl_2$ is $\distURL{\dataUrl_1}{\dataUrl_2} = 1 + 3 = 4$.
%\end{ex}

\subsubsection{Parameter Distance}
\label{sec:aim:actionClustering:paramDistance}
\label{def:actionClustering:paramDistVal}
\label{def:actionClustering:paramDist}

To quantify the dissimilarity between residual parameters, we first independently quantify the dissimilarity between pairs of parameters of the same type.
Since, in our context, we exercise vulnerabilities by only using string or numerical values, 
we ignore parameter values of other types such as byte arrays (e.g., an image uploaded to the SUT).
In other contexts, new parameter distance functions suited to other input types may be required. 
For strings, we use the Levenshtein distance~\cite{HAB13,TSBB17}, whereas for numerical values we consider the absolute value of their difference~\cite{BSRT19}:
$$
\begin{array}{l}
\distParamVal{v_1}{v_2}
\eqdef\\
\left\{
\begin{array}{ll}
     \editDistance{v_1}{v_2}&  \text{if } \typeOf{v_1} = \text{str} = \typeOf{v_2}\\
     \abs{v_1 - v_2}& \text{if } \typeOf{v_1} = \text{int} = \typeOf{v_2}\\
     \text{undefined}& \text{otherwise}
\end{array}
\right.
\end{array}
$$
%where $\typeOf{v}$ denotes the type of the value $v$.

Since we have parameters of different types, we normalize the parameter distance using the normalization function $\normalization{x} = \tfrac{x}{x + 1}$ (\Cref{sec:aim:problemDefinition}).
Then, we add these normalized distances together, and normalize the sum to obtain a result between $0$ and $1$.
We compute the parameter distance in case of \emph{matching parameters}, i.e., the number of parameters is the same and the corresponding parameters have the same type.
Otherwise, we assume the largest distance possible, which is $1$ due to the normalization.
This is the only case where the value $1$ is reached, as distance lies otherwise in [$0$ $1$[, as expected for a decimal part (\Cref{defi:actionDist}):
$$
\begin{array}{l}
\distParam{\params_1}{\params_2}
\eqdef\\
\left\{
\begin{array}{l}
     \normalization{\sum\limits_{0 \le i < \length{\params_1}}
         \normalization{\distParamVal{\listInd{\params_1}{i}}{\listInd{\params_2}{i}}}
     }\\
     \quad\text{if } \params_1 \text{ and } \params_2 \text{ have matching parameters}\\
     1 \text{ otherwise}
\end{array}
\right.
\end{array}
$$

where $\params_1 = \paramsAct{\action_1}$, $\params_2 = \paramsAct{\action_2}$, and $\listInd{\params}{i}$ is the $i$-th element of $\params$.

%\begin{ex}
%\label{ex:residuals}
For instance, we consider two actions $\action_1$ and $\action_2$ having matching parameters with the values in \Cref{tab:exampleResiduals} for page number, username, and password.
\begin{table}[t]
\centering
\caption{Values for the Example of Parameter Distance}
\label{tab:exampleResiduals}
\begin{tabular}{c|c c c|}
\cline{2-4}
& Page Number
& Username
& Password\\
\hline
\multicolumn{1}{|c|}{$\params_1$}
& $10$
& \aimStr{John}
& \aimStr{qwerty}\\
\multicolumn{1}{|c|}{$\params_2$}
& $42$
& \aimStr{Johnny}
& \aimStr{qwertyuiop}\\
\hline
\end{tabular}
\end{table}

%$$\begin{array}{r@{}l}
%     \params_1 &{}= \dataList{\text{pageNb: int} = 10, \text{username: str} = \aimStr{John}, \text{password: str} = \aimStr{qwerty}}\\
%     \params_2 &{}= \dataList{\text{pageNb: int} = 42, \text{username: str} = \aimStr{Johnny}, \text{password: str} = \aimStr{qwertyuiop}}
%\end{array}$$
The distance for the page number is $\distParamVal{10}{42} = 32$, normalized into
$\tfrac{32}{32 + 1} \approx 0.97$.
For the username, it is $\distParamVal{\aimStr{John}}{\aimStr{Johnny}} = 2$, normalized into
$\tfrac{2}{2 + 1} \approx 0.66$.
For the password, it is $\distParamVal{\aimStr{qwerty}}{\aimStr{qwertyuiop}} = 4$, normalized into
$\tfrac{4}{4 + 1} = 0.80$.
Thus, the parameter distance is
$\distParam{\params_1}{\params_2} \approx \tfrac{0.97 + 0.66 + 0.80}{0.97 + 0.66 + 0.80 + 1} \approx 0.71$.
%\end{ex}

\subsubsection{Action Subclasses}
\label{sec:aim:actionClustering:subclasses}
\label{defi:actionSubclass}
\label{defi:actionSubclassesFromSequence}

We partition both parts of each action set (\Cref{sec:aim:requestPartition}) using the \kmeans, \dbscan, or \hdbscan clustering algorithms, setting the hyper-parameters using our Silhouette analysis (\Cref{sec:aim:doubleClustering:hyperParameter}), and quantifying action dissimilarity using our action distance (\Cref{sec:aim:distanceActions}), obtaining clusters we call \emph{action subclasses}.
We denote by $\getActionSubclass{\action}{\actionClass}$ the unique action subclass $\actionSublass$ from the action set $\actionClass$ such that $\action \in \actionSublass$.
For the sake of simplicity, we denote $\getSubclassFromAct{\sequence}{i}$ the action subclass corresponding to the $i$-th action in input $\sequence$:
$$
\begin{array}{r@{}l}
\getSubclassFromAct{\sequence}{i} \eqdef
& \mathit{ActionSubclass}(\getAction{\sequence}{i},\\
& \getActionClass{\getOutputClass{\sequence}{i}}
%\getActionSubclass{\getAction{\sequence}{i}}{\getActionClass{\getOutputClass{\sequence}{i}}}
\end{array}$$

Finally, in our study, the objectives covered by an input %$\sequence$
are:
$$\getSubclasses{\sequence} \eqdef
\set{\getSubclassFromAct{\sequence}{i}}{1 \le i \le \length{\sequence}}$$

\section{Step 4: Problem Reduction}
\label{sec:aim:search}

The search space for our problem (\Cref{defi:pbDef}) consists of all the subsets of the initial input set, which leads to $2^m$ potential solutions, where $m$ is the number of initial inputs.

For this reason, \aim integrates a \emph{problem reduction} step, implemented by the \improLong (\impro) component, to minimize the search space before solving the search problem in the next step (\Cref{sec:aim:genetic}).
We apply the following techniques to reduce the size of the search space:
\begin{itemize}
\item \emph{Determining redundancy}:
Necessary inputs (\Cref{sec:aim:problemDefinition}) must be part of the solution, hence one can only investigate redundant inputs (\Cref{sec:aim:necessarySeqs}).
Moreover, one can restrict the search by removing the objectives already covered by necessary inputs.
Finally, if a redundant input does not cover any of the remaining objectives, it will not contribute to the final coverage, and hence can be removed.

\item \emph{Removing duplicates}:
Several inputs may have the same cost and coverage.
In this case, we consider them as \emph{duplicates} (\Cref{sec:aim:duplicates}).
Thus, we keep only one and we remove the others.

\item \emph{Removing locally-dominated inputs}:
For each input, if there exists other inputs that cover the same objectives at a same or lower cost, then the considered input is \emph{locally-dominated} by the other inputs (\Cref{sec:aim:localDominance}) and is removed.

\item \emph{Dividing the problem}:
We consider two inputs covering a common objective as being connected.
Using this relation, we partition the search space into connected \emph{components} that can be independently solved (\Cref{sec:aim:subproblem}), thus reducing the number of objectives and inputs to investigate at a time.
\end{itemize}

Before detailing these techniques, we first explain in which order there are applied (\Cref{sec:aim:reduction:order}).

%%%%%%%%
\subsection{Order of Reduction Techniques}
\label{sec:aim:reduction:order}

We want to perform first the least expensive reduction techniques, to sequentially reduce the cost of the following more expensive techniques.
Determining redundancy requires $\bigO{m \times c}$ steps, removing duplicates requires $\bigO{m^2}$ steps,  and removing locally-dominated inputs requires $\bigO{m \times 2^n}$ steps,
where $m$ is the number of inputs, $c$ is the maximal number of objectives covered by an input, and $n$ is the maximal number of neighbors for an input (i.e., the number of other inputs that cover an objective shared with the considered input).
In our study, we assume $c < m$.
Hence, we first determine redundancy, then  remove duplicates, and remove locally-dominated inputs.
Dividing the problem requires exploring neighbors and comparing non-visited inputs with visited ones, so it is potentially the most costly of these reduction techniques; hence, it is performed at the end.

After determining redundancy, the removal of already covered objectives may lead to new inputs being duplicates or locally-dominated.
Moreover, the removal of duplicates or locally-dominated inputs may lead to changes in redundancy, making some previously redundant inputs necessary.
Hence, these reduction techniques should be iteratively applied, until a stable output is reached.
Such output can be detected by checking if inputs were removed during an iteration.
%Indeed, when no input is removed then, during the next iteration, no new input becomes necessary, so the objectives to be covered do not change, and thus there is no opportunity to remove more inputs.

Therefore, the order is as follows.
We first initialize variables (\Cref{sec:aim:probRed:init}).
Then we repeat, until no input is removed, the following steps: determine redundancy (\Cref{sec:aim:necessarySeqs}), remove duplicates (\Cref{sec:aim:duplicates}), and remove locally-dominated inputs (\Cref{sec:aim:localDominance}).
Finally, we divide the problem into sub-problems (\Cref{sec:aim:subproblem}).

%%%%%%%%
\subsection{Initializing Variables}
\label{sec:aim:probRed:init}
\label{defi:probRed:init}

During problem reduction, we consider three variables: $\NecessaryInputs$, the set of the inputs that has to be part of the final solution, $\SearchSpace$, the remaining inputs to be investigated, and $\Objectives$, the objectives that remain to be covered by subsets of $\SearchSpace$.
$\NecessaryInputs$ is initially empty.
$\SearchSpace$ is initialized as the initial input set.
$\Objectives$ is initialized as the coverage of the initial input set (\Cref{defi:actionSubclassesFromSequence,sec:aim:goals}).

%%%%%%%%
\subsection{Determining Redundancy}
\label{sec:aim:necessarySeqs}
\label{defi:necessaryInitSeq}
\label{defi:searchSpace}

\begin{algorithm}[t]
\caption{Redundancy determination technique.}
\label{algo:redundancy}
\begin{algorithmic}[1]
\Procedure{redundancy}{$\NecessaryInputs, \SearchSpace, \Objectives$}
\State $\RedundantInputs \gets \Redundant{\SearchSpace}$ \label{line:redundancy:redundant}
\State $ \NecessaryInputsNew \gets \SearchSpace \setminus \RedundantInputs$ \label{line:redundancy:new}
\State $ \NecessaryInputs \gets \NecessaryInputs \cup \NecessaryInputsNew$ \label{line:redundancy:necessary}
\State $ \Objectives \gets \Objectives \setminus \getSubclasses{\NecessaryInputsNew}$ \label{line:redundancy:objectives}
\State $ \SearchSpace \gets \{\sequence \in \RedundantInputs |\ \getSubclasses{\sequence} \cap \Objectives \neq \varnothing \}$\label{line:redundancy:searchSpace}
\State \textbf{return} $\NecessaryInputs, \SearchSpace, \Objectives$
\EndProcedure
\end{algorithmic}
\end{algorithm}

~\TSE{1.4.2 }{This technique} is presented in \Cref{algo:redundancy}.
Each time it is repeated, the redundancy of the remaining inputs is computed (Line~\ref{line:redundancy:redundant}).
Among them, inputs which are necessary (\Cref{sec:aim:problemDefinition}) in $\SearchSpace$ (Line~\ref{line:redundancy:new}) for the objectives in $\Objectives$ have to be included in the final solution (Line~\ref{line:redundancy:necessary}), otherwise some objectives will not be covered.
%$$\NecessaryInputs \leftarrow \NecessaryInputs \cup \NecessaryInputsNew$$
%where $\NecessaryInputsNew = \SearchSpace \setminus \Redundant{\SearchSpace}$.
Then, the objectives already covered by the necessary inputs are removed (Line~\ref{line:redundancy:objectives}). 
%$$\Objectives \leftarrow \ObjectivesNew$$
%where $\ObjectivesNew = \getSubclasses{\SearchSpace} \setminus \getSubclasses{\NecessaryInputsNew}$.
Hence, in the following, we only consider, for each remaining input $\sequence \in \SearchSpace$, their coverage regarding the remaining objectives, i.e., $\getSubclasses{\sequence} \cap \Objectives$, instead of $\getSubclasses{\sequence}$.

Finally, some redundant inputs may cover only objectives that are already covered by necessary inputs.
In that case, they cannot be part of the final solution because they would contribute to the cost but not to the coverage of the objectives.
Hence, we restrict without loss the search space for our problem by considering only redundant inputs that can cover the remaining objectives (Line~\ref{line:redundancy:searchSpace}).
%$$
%\begin{array}{r@{}l}
%\SearchSpace \leftarrow
%&{}\{\sequence \in \Redundant{\SearchSpace}\\
%&{}\quad  |\ \getSubclasses{\sequence} \cap \ObjectivesNew \neq \varnothing \}
%\end{array}
%$$

%%%%%%%%
\subsection{Removing Duplicates}
\label{sec:aim:duplicates}
\label{defi:equivSeqs}

In the many-objective problem described in \Cref{sec:aim:problemDefinition}, inputs are characterized by their coverage (\Cref{defi:actionSubclassesFromSequence}) and their cost (\Cref{defi:cost}).
%Hence, two inputs are considered duplicates if they belong to the same cluster and they have the same coverage and cost.
Hence, two inputs with the same coverage and cost are considered duplicates.
In that case, \impro selects one and remove the other.
%Hence, two inputs are considered equivalent if they have the same coverage and cost.
%For each equivalence class containing at least two inputs, we say these inputs are \emph{duplicates}.
%\impro selects one representative per equivalence class in $\SearchSpace$ and removes the others.
%In practice, \impro keeps the duplicate with the smallest index.

%%%%%%%%
\subsection{Removing Locally-dominated Inputs}
\label{sec:aim:localDominance}
\label{defi:localDominance}
\label{defi:overlapRel}
\label{defi:probRed:updAfterLocalDom}

For a given input $\sequence \in \SearchSpace$, if the same coverage can be achieved by one or several other input(s) for at most the same cost, then $\sequence$ is not required for the solution.
Formally, we say that the input $\sequence \in \SearchSpace$ is \emph{locally dominated} by the subset $S \subseteq \SearchSpace$, denoted $\sequence \localDom S$, if $\sequence \not\in S$, $\getSubclasses{\sequence} \subseteq \getSubclasses{S}$, and $\costFunction{\sequence} \ge \costFunction{S}$.
In order to simplify the problem, inputs that are locally dominated should be removed from the remaining inputs $\SearchSpace$.

Removing a redundant input $\sequence$ (\Cref{defi:redundantSeq}) can only affect the redundancy of the inputs in $\TestSuite$ that cover objectives in $\getSubclasses{\sequence}$.
Hence, we consider two inputs as being connected if they cover at least one common objective.
Formally, we say that two inputs $\sequence_1$ and $\sequence_2$ \emph{overlap}, denoted by $\sequence_1 \overlap \sequence_2$, if $\getSubclasses{\sequence_1} \cap \getSubclasses{\sequence_2} \neq \varnothing$.
The name of the \emph{local} dominance relation comes from the fact proved in the separate appendix (Proposition 3) that, to determine if an input is locally-dominated, one has only to check amongst its neighbors for the overlapping relation instead of amongst all the remaining inputs, thus making this step tractable.
%Formally, if $\sequence_1 \localDom S$, then $\sequence_1 \localDom S \cap \set{\sequence_2 \in \SearchSpace}{\sequence_1 \overlap \sequence_2}$.

One concern is that removing a locally-dominated input could alter the local dominance of other inputs.
%Hence, like reduction steps (\Cref{sec:aim:reduction}), \impro would have to take into account the order for removing locally-dominated inputs.
Fortunately, this is not the case for local dominance.
We prove in the separate appendix (Theorem 2) that, for every locally-dominated input $\sequence \in \SearchSpace$, there always exists a subset $S \subseteq \SearchSpace$ of not locally-dominated inputs such that $\sequence \localDom S$.
Hence, the locally dominated inputs can be removed in any order without reducing coverage or preventing cost reduction, both being ensured by non locally-dominated inputs.
Therefore, \impro keeps in the search space only the remaining inputs that are not locally-dominated:
$$\begin{array}{r@{}l}
     \SearchSpace &{} \leftarrow \set{\sequence \in \SearchSpace}{\forall S \subseteq \SearchSpace: \sequence \not\localDom S} \\
\end{array}$$

%%%%%%%%
\subsection{Dividing the Problem}
\label{sec:aim:subproblem}
\label{defi:overlapGraph}
\label{defi:redundancySubgraph}
\label{defi:subprob}

After removing as many inputs as possible, we leverage the overlapping relation (\Cref{defi:overlapRel}) to partition the remaining inputs into connected components, in a divide-and-conquer approach.
We denote $\OverlapGraph{\TestSuite}$ the \emph{overlapping graph} of the input set $\TestSuite$, i.e., the undirected graph such that vertices are inputs in $\TestSuite$ and edges correspond to the overlapping relation $\overlap$,
%Formally, for a given input set $\TestSuite$, we denote $\OverlapGraph{\TestSuite} \eqdef \dataList{\OverlapVertices{\TestSuite}, \OverlapEdges{\TestSuite}}$ its \emph{overlapping graph}, i.e., the undirected graph such that vertices $\OverlapVertices{\TestSuite} \eqdef \TestSuite$ are inputs in $\TestSuite$ and edges $\OverlapEdges{\TestSuite} \eqdef \set{\set{\sequence_1, \sequence_2}{}}{\sequence_1, \sequence_2 \in \OverlapVertices{\TestSuite} \land \sequence_1 \overlap \sequence_2}$ correspond to the overlapping relation,
and $\RedundantComponents{\TestSuite}$
%\eqdef \Components{\OverlapGraph{\TestSuite}}$
the set of the \emph{connected components} of $\OverlapGraph{\TestSuite}$.

\begin{figure*}[t]
\centering
\begin{minipage}[b]{0.5\textwidth}
    \centering
    \input{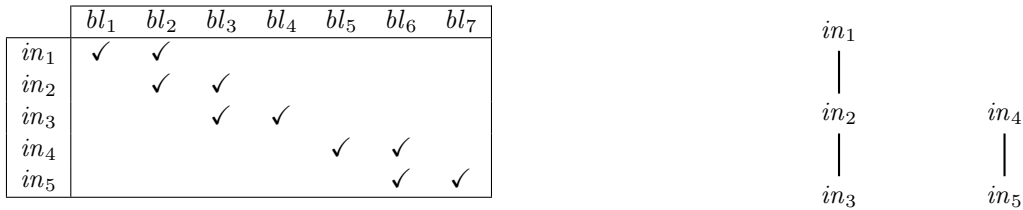}
    %\caption{Coverage Matrix}
    \label{fig:overlappingGraph:left}
\end{minipage}
\begin{minipage}[b]{0.5\textwidth}
    \centering
    \begin{tikzpicture}[scale=1.1]
    % styles
    \tikzset{
        component/.style={very thick},
		edge/.style={thick,color=black},
    }
    % sequences
    \node[black]
        (seq1)
        at (-1,1)
        {$\sequence_1$};
    \node[black]
        (seq2)
        at (-1,0)
        {$\sequence_2$};
    \node[black]
        (seq3)
        at (-1,-1)
        {$\sequence_3$};
%    \node[black]
%        (seq4)
%        at (0,-1)
%        {$\sequence_4$};
    \node[black]
        (seq5)
        at (1,-1)
        {$\sequence_5$};
    \node[black]
        (seq6)
        at (1,0)
        {$\sequence_4$};
    % sequences
	\draw[edge]
        (seq1) -- (seq2);
	\draw[edge]
        (seq2) -- (seq3);
%	\draw[edge]
%        (seq3) -- (seq4);
%	\draw[edge]
%        (seq4) -- (seq5);
	\draw[edge]
        (seq5) -- (seq6);
\end{tikzpicture}
    %\caption{Overlapping Graph}
    \label{fig:overlappingGraph:right}
\end{minipage}
\caption{Inputs covering one objective in common (left) are connected in the corresponding overlapping graph (right).}
\label{fig:overlappingGraph}
\end{figure*}

%\begin{ex}
%\label{ex:overlappingGraph}
For instance, in \Cref{fig:overlappingGraph}, we represent on the left the input blocks covered by each input in the search space and the corresponding overlapping graph on the right.
The connected components of the graph are $\set{\sequence_1, \sequence_2, \sequence_3}{}$ and $\set{\sequence_4, \sequence_5}{}$.
%\end{ex}

Such connected components are important because we prove in the separate appendix (Proposition 8) that inputs in a connected component can be removed without altering the redundancy of inputs in other connected components.
%Formally, if $\RedComp_1, \dots, \RedComp_c \in \RedundantComponents{\TestSuite}$ denote $c$ connected components of $\TestSuite$ and, in each component $\RedComp_i$, there exists a valid order of removal steps $\dataList{\sequence^i_1, \dots, \sequence^i_{n_i}}$, then the concatenation $\dataList{\sequence^1_1, \dots, \sequence^1_{n_1}} + \dots + \dataList{\sequence^c_1, \dots, \sequence^c_{n_c}}$ is a valid order of removal steps in $\TestSuite$.
Similarly, we prove in the separate appendix (Theorem 3) that the gain, i.e., the maximal cost reduction from removing redundant inputs (\Cref{sec:aim:gain}), can be independently computed on each connected component, i.e., for each input set $\TestSuite$,
$\gainExhaust{\TestSuite} = \sum_{\RedComp \in \RedundantComponents{\TestSuite}} \gainExhaust{\RedComp}$.

Hence, instead of searching a solution on $\SearchSpace$ to solve our initial problem (\Cref{defi:pbDef}), we use a divide-and-conquer strategy to split the problem into more manageable sub-problems that can be independently solved on each connected component $\RedComp \in \RedundantComponents{\SearchSpace}$.

We denote $\Objectives(\RedComp) \eqdef \Objectives \cap \getSubclasses{\RedComp}$ the remaining objectives to be covered by inputs in $\RedComp$ and we formulate the sub-problem on the connected component similarly to the initial problem:
$$\minimize_{\TestSuite \subseteq \RedComp} \fitnessVectorComponent{\TestSuite} \eqdef \dataList{\normalization{\costFunction{\TestSuite}}, \objFun{\actionSublass_1}{\TestSuite}, \dots, \objFun{\actionSublass_\nbSubClasses}{\TestSuite}}$$
where $\Objectives(\RedComp) = \set{\actionSublass_1, \dots, \actionSublass_\nbSubClasses}{}$ and the $\minimize$ notation is detailed in \Cref{sec:aim:background:paretoFront}.
We denote $\TestSuite_{\RedComp}$ a non-dominated solution with full coverage, such that $\fitnessVectorComponent{\TestSuite_{\RedComp}} = \dataList{\normalization{\mathit{cost}_{\min}}, 0, \dots, 0}$.

\section{Step 5: Genetic Search}
\label{sec:aim:genetic}

\TSE{2.2}{Our initial goal of obtaining a subset $\TestSuiteFinal \subseteq \TestSuiteInit$ with total input coverage at minimal cost (\Cref{sec:aim:goals}) can be expressed
as a weighted set cover problem.
Given a universe $\mathcal{U}$, a set $S = \set{S_1, S_2, \dots}{}$ of subsets $S_i \subseteq \mathcal{U}$, and a weight function mapping each subset $S_i$ to a positive real number, the weighted set cover problem consists in finding a subset $T \subseteq S$ such that subsets $S_i$ in $T$ cover $\mathcal{U}$ at a minimal cumulative cost.
To express our initial goal as a weighted set cover problem, we consider as universe the input blocks $\mathcal{U} = \getSubclasses{\TestSuiteInit}$ to cover, as subsets of the universe the input blocks $\getSubclasses{\sequence}$ covered by each input $\sequence \in \TestSuiteInit$, and as weight of a subset $\getSubclasses{\sequence}$ the cost $\costFunction{\sequence}$ of the corresponding input.
A solution to this set cover problem is a set of subsets $T = \set{\getSubclasses{\sequence_1}, \dots, \getSubclasses{\sequence_n}}{}$, providing a minimized input set $\set{\sequence_1, \dots, \sequence_n}{}$, given that each subset $\getSubclasses{\sequence}$ can be uniquely mapped to its initial input $\sequence$.
This is the case for each connected component $\RedComp$, since \impro removed duplicates (\Cref{sec:aim:duplicates}) and locally-dominated inputs (\Cref{sec:aim:localDominance}).
While the set cover problem is NP-hard~\cite{KV18}, it admits a very efficient~\cite{Sla96} polynomial time approximation using a greedy algorithm, as well as various techniques to find approximate solutions to an ILP formulation of the problem~\cite{Vaz01}.
Unfortunately, the set cover problem considers only the total coverage and the cumulative cost of solutions, not input blocks as individual objectives (\Cref{sec:aim:mao}) as in the formulation of our many-objective problem (\Cref{defi:pbDef}) or sub-problem (\Cref{sec:aim:subproblem}).
Hence, an algorithm solving the set cover problem may not take into account relevant information about how each input block is covered. Indeed, some blocks may be more difficult to cover and should thus receive priority.
Further, an input selected by the greedy algorithm, because it covers many blocks at small cost, may be less optimal than several inputs covering  more blocks at slightly larger cost.
For instance, if
$\getSubclasses{\sequence_1} = \set{\actionSublass_1, \actionSublass_2}{}$ with $\costFunction{\sequence_1} = 2$, 
$\getSubclasses{\sequence_2} = \set{\actionSublass_1, \actionSublass_3}{}$ with $\costFunction{\sequence_2} = 3$, and
$\getSubclasses{\sequence_3} = \set{\actionSublass_2, \actionSublass_4}{}$ with $\costFunction{\sequence_3} = 3$, 
%$\getSubclasses{\sequence_1} = \set{\actionSublass_2, \actionSublass_3, \actionSublass_4}{}$ with $\costFunction{\sequence_1} = 3$, 
%$\getSubclasses{\sequence_2} = \set{\actionSublass_1, \actionSublass_2, \actionSublass_5}{}$ with $\costFunction{\sequence_2} = 5$, 
%$\getSubclasses{\sequence_3} = \set{\actionSublass_1, \actionSublass_2, \actionSublass_3}{}$ with $\costFunction{\sequence_3} = 4$, and 
%$\getSubclasses{\sequence_4} = \set{\actionSublass_4, \actionSublass_5}{}$ with $\costFunction{\sequence_4} = 3$, 
the greedy algorithm selects $\sequence_1$ because it is locally optimal, then $\sequence_2$ and $\sequence_3$ to cover respectively $\actionSublass_3$ and $\actionSublass_4$, at a total cost of $8$. However, selecting only $\sequence_2$ and $\sequence_3$ leads to full coverage but at a lower cost of $6$.
In this example, the greedy algorithm selects $\sequence_1$ covering blocks $\actionSublass_1$ and $\actionSublass_2$, which are easier to cover because each block can be covered by two inputs, while blocks $\actionSublass_3$ and $\actionSublass_4$ can only be covered by one input.
%then $\sequence_2$ to complete the coverage at a total cost of $8$, while selecting $\sequence_3$ and $\sequence_4$ also leads to full coverage but at a cost of $7$.
Therefore, it is unlikely for a polynomial time approximation of the set cover problem to find, in general, the best solution to our many-objective problem.}

Alternatively, obtaining an optimal solution $\TestSuite_{\RedComp}$ to our sub-problem on a connected component $\RedComp$ is similar to solving the knapsack problem, which is also NP-hard~\cite{LHH07}.
To be more precise, our problem is equivalent to the 0-1 knapsack problem, which consists in selecting a subset of items to maximize a total value, while satisfying a weight capacity.
Since we consider a many-objective problem (\Cref{sec:aim:mao}), we must address the multidimensional variant of the 
0-1 knapsack problem, where each item has many \quotes{weights}, one per considered objective.
In our case, we minimize the total cost instead of maximizing total value and ensure the coverage of each action objective instead of making sure that each weight capacity is not exceeded.
Furthermore, the 0-1 multidimensional knapsack problem is harder than the initial knapsack problem as it does not admit a fully polynomial time approximation scheme~\cite{KS81}, hence the need for a meta-heuristic.

For our approach to scale, we adopt a genetic algorithm because it is known to find good approximations in reasonable execution time~\cite{PKT15} and has been widely used in software testing.
An input set $\TestSuite \subseteq \RedComp$ can thus be seen as a chromosome, where each gene corresponds to an input $\sequence \in \RedComp$, the gene value being $1$ if $\sequence \in \TestSuite$ and $0$ otherwise.
Though several many-objective algorithms have been successfully applied within the software engineering community, like \nsgaThree~\cite{DJ14, JD14} and \mosa~\cite{PKT15} (\Cref{sec:aim:background:MaO}), these algorithms do no entirely fit our needs (\Cref{sec:aim:genetic:motivation}).
Hence, we propose \geneticAlgo (\geneticAlgoLong), a novel genetic algorithm based on two populations and summarized in \Cref{algo:mocco}.
We first explain how these populations are initialized (\Cref{sec:aim:initPops}).
Then, for each generation, \geneticAlgo performs the following standard steps:
selection of the parents (\Cref{sec:aim:selectParents}),
crossover of the parents to produce an offspring (\Cref{sec:aim:crossover}),
mutation of the offspring (\Cref{sec:aim:mutation}),
and update of the populations (\Cref{sec:aim:updtPops}) to obtain the next generation.
The process continues until a termination criterion is met (\Cref{sec:aim:termination}).
Then, we detail how \geneticAlgo determines the solution $\TestSuite_{\RedComp}$ to each sub-problem.

\begin{algorithm}[t]
\caption{\geneticAlgo overview.}
\label{algo:mocco}
\begin{algorithmic}[1]
\Procedure{MOCCO}{$\RedComp, \sizePop, \nbGens, \timeBudget$}
\State $\timeStart \gets \getTime$ \label{line:mocco:getTime}
\State $\RoofersVar \gets \initRoofers{\RedComp, \sizePop}$ \label{line:mocco:roofers}
\State $\MisersVar \gets \varnothing$ 
\State $n \gets 1$
\State $\mathit{stillTime} \gets \mathit{True}$
\While{$n \le \nbGens \land \mathit{stillTime}$}
    \State $n \gets n + 1$
    \State $\TestSuite_1, \TestSuite_2 \gets \mathit{selectParents}(\RoofersVar, \MisersVar)$
    \State $\TestSuite_3, \TestSuite_4 \gets \mathit{crossover}(\TestSuite_1, \TestSuite_2)$
    \For{$\TestSuite \in \set{\TestSuite_3, \TestSuite_4}{}$}
        \State $\TestSuite \gets \mutation{\TestSuite}$
        \State $\TestSuite \gets \reduction{\TestSuite}$
        \If{$\getTime - \timeStart > \timeBudget$}
            \State $\mathit{stillTime} \gets \mathit{False}$
            \State \textbf{break}
        \EndIf
        \If{$\TestSuite \in \RoofersVar \cup \MisersVar$}
            \State \textbf{continue}
        \EndIf
        \If{$\getSubclasses{\TestSuite}  = \Objectives(\RedComp)$}
            \State $\RoofersVar \gets \mathit{updRoofers}(\RoofersVar, \TestSuite)$
        \Else
            \State $\MisersVar \gets \mathit{updMisers}(\MisersVar, \TestSuite)$
        \EndIf
    \EndFor
\EndWhile
\State $\TestSuite_{\RedComp} \gets \mathit{selectSolution}(\RoofersVar)$
\State \textbf{return} $\TestSuite_{\RedComp}$
\EndProcedure
\end{algorithmic}
\end{algorithm}

%%%%%%%%
\subsection{Motivation for a Novel Genetic Algorithm}
\label{sec:aim:genetic:motivation}

While our problem is similar to the multidimensional 0-1 knapsack problem, it is not exactly equivalent, since standard solutions to the multidimensional knapsack problem have to ensure that, for each \quotes{weight type}, the total weight of the items in the knapsack is below weight capacity, while we want to ensure that, for each objective, at least one input in the minimized input set covers it.
Hence, standard solutions based on genetic search are not applicable in our case, and we focus on genetic algorithms able to solve many-objective problems in the context of test case generation or minimization.
We have explained earlier (\Cref{sec:aim:background:MaO}) the challenges raised by many-objective problems and how the \nsgaThree~\cite{DJ14, JD14} and \mosa~\cite{PKT15} genetic algorithms tackle such challenges.

\nsgaThree has the advantage of letting users choose the parts of the Pareto front they are interested in, by providing reference points.
Otherwise, it relies on a systematic approach to place points on a normalized hyperplane.
While this approach is useful in general, %this is not what we want to achieve.
%Indeed,
we are interested only in solutions that are close to a utopia point $\dataList{0, 0, \dots, 0}$ (\Cref{sec:aim:problemDefinition}) covering every objective at no cost.
Hence, we do not care about diversity over the Pareto front, and we want to explore a very specific region of the search space.
Moreover, apart from the starting points, the main use of the reference points in \nsgaThree is to determine, after the first non\-domination fronts are obtained from \nsgaTwo~\cite{DPAM02}, the individuals to be selected from the last considered front, so that the population reaches a predefined number of individuals.
This is not a problem we face because we know there is only one point (or several points, but at the same coordinates) in the Pareto front that would satisfy our constraint of full coverage at minimal cost.
Hence, we do not use the Pareto front as a set of solutions, even if we intend to use individuals in the Pareto front as intermediate steps to reach relevant solutions.
%while avoiding getting stuck in a local optimum and providing pressure for reducing the cost of the candidates.

Regarding \mosa~\cite{PKT15}, trade-offs obtained from approximating the Pareto front are only used for maintaining diversity during the search, which is similar to what we intend to do.
But, as opposed to the use case tackled by \mosa, in our case determining inputs covering a given objective is straightforward.
Indeed,
%each objective is covered by at least one input from the initial input set, and
for each objective, we can easily determine inputs that are able to cover it (\Cref{sec:aim:mao}). 
Hence, individuals ensuring the coverage of the objectives are easy to obtain, while the hard part of our problem is to determine a combination of inputs able to cover all the objectives at a minimal cost.
Hence, even if \mosa may find a reasonable solution, because it focuses on inputs individually covering an objective and not on their collective coverage and cost, it is unlikely to find the best solution.

%Because common many-objective algorithms for software engineering problems do not fit our needs,
Hence, we propose a novel genetic algorithm, named \geneticAlgo. 
We take inspiration from \mosa~\cite{PKT15} by considering two populations:
1) a population of solutions (like MOSA’s archive), called the \emph{roofers} because they cover all the objectives (\Cref{defi:actionSubclass}), and  
2) a population of individuals on the Pareto front (\Cref{defi:pareto}), called the \emph{misers} because they minimize the cost, while not covering all objectives.
Like \nsgaThree and \mosa, \geneticAlgo has to tackle challenges raised by many-objective problems (\Cref{sec:aim:background:MaO}).

\label{defi:exposure}
To address challenge 1, we take inspiration from the whole suite approach~\cite{FA13} which counts covered branches as a single objective, by defining the \emph{exposure} as the sum of the coverage objective functions (\Cref{sec:aim:problemDefinition}):
$$\exposure{\TestSuite} \eqdef \sum_{\actionSublass_i \in \Objectives(\RedComp)} \objFun{\actionSublass_i}{\TestSuite}$$
Since $\objFun{\actionSublass_i}{.}$ is zero when the objective $\actionSublass_i$ is covered, the larger the exposure, the smaller the input set coverage.
As described in \Cref{sec:aim:mao}, we do not use the exposure as objective because we want to distinguish between input blocks. 
But we use it as a weight when randomly selecting a parent amongst the misers (\Cref{sec:aim:selectParents}), so that the further away a miser is from complete coverage, the less likely it is to be selected. 
%In other words, instead of using a binary preference criterion as in \mosa~\cite{PKT15}, we use a preference bias to get a more nuanced result. 
That way, we aim to benefit from the large number of dimensions to avoid getting stuck in a local optimum and to have a better convergence rate~\cite{LLTY15}, while still focusing the search on the region of interest.

Since we want to deeply explore this particular region, we do not need to preserve diversity over the whole Pareto front.
Therefore, we do not use diversity operators, avoiding challenge 2.

Finally, we address challenge 3 by 1) restricting the recombination operations and 2) tailoring them to our problem, as follows:
\begin{enumerate}
\item A crossover between roofers can only happen during the first generations, when no miser is present in the population.    After the first miser is generated, crossover (\Cref{sec:aim:crossover}) is allowed only between a roofer and a miser. 
Hence, the roofer parent provides full coverage while the miser parent provides almost full coverage at low cost. 
Moreover, because of how the objective functions are computed (\Cref{sec:aim:problemDefinition}), the not-yet-covered objectives are likely to be covered in an efficient way. 
That way, we hope to increase our chances of obtaining offspring with both large coverage and low cost.
\item Not only our recombination strategy is designed to be computationally efficient (by minimizing the number of introduced redundancies), but we exploit our knowledge of input coverage to determine a meaningful crossover between parents, with inputs from one parent for one half of the objectives and inputs from the other parent for the other half.
\end{enumerate}

%%%%%%%%
\subsection{Population Initialization}
\label{sec:aim:initPops}

During the search, because we need diversity to explore the search space, we consider a population (with size $\sizePop \ge 2$) of the least costly individuals generated so far that satisfy full coverage.
We call \emph{roofers} such individuals, by analogy with covering a roof, and we denote $\Roofers{n}$ the roofer population at generation $n$.

But focusing only on the roofers would prevent us to exploit the least expensive solutions obtained in the Pareto front while trying to minimize the connected component (\Cref{defi:subprob}).
Instead, inputs that do not cover all the objectives, and will thus not be retained for the final solution, but are efficient at minimizing cost, are thus useful as intermediary steps towards finding an efficient solution.
Hence, we maintain a second population, formed by individuals that are non-dominated so far and minimize cost while not covering all the objectives.
We call \emph{misers} such individuals, because they focus on cost reduction more than objective coverage, and we denote $\Misers{n}$ the miser population at generation $n$.

The reason for maintaining two distinct populations is to restrict the crossover strategy (\Cref{sec:aim:crossover}) so that (in most cases) one parent is a roofer and one parent is a miser.
Since misers prioritize cost over coverage, a crossover with a miser tends to reduce cost.
Because roofers prioritize coverage over cost, a crossover with a roofer tends to increase coverage.
Hence, with such a strategy, we intend to converge towards a solution minimizing cost and maximizing coverage.

For both populations, we want to ensure that the individuals are reduced (\Cref{sec:aim:reduction}), i.e., they contain no redundant inputs.
Hence, during the initialization and updates of these populations, we ensure that removal steps are performed.
Because, as detailed in the following, the number of redundant inputs obtained for each generation is small, the optimal order of removal steps can be exhaustively computed.
We denote by $\reduction{\TestSuite}$ the input set $\TestSuite$ after these removal steps.
This limits the exploration space, %by avoiding parts of the search space that are not relevant.
%Indeed,
since non-reduced input sets are likely to have a large cost and hence to be far away for the utopia point of full coverage at no cost we intend to focus on.
%Considering such non-reduced input sets would consist in exploring the search space by over-covering the objectives, while with the misers we explore it by under-covering it, thus focusing the search on the region of interest.

\begin{algorithm}[t]
\caption{Roofer population initialization.}
\label{algo:rofferInit}
\begin{algorithmic}[1]
\Procedure{initRoofers}{$\RedComp, \sizePop$}
\State $\RoofersVar \gets \varnothing$
\While{$\card{\RoofersVar} < \sizePop$}
    \State $\TestSuite \gets \varnothing$
    \While{$\getSubclasses{\TestSuite} \neq \Objectives(\RedComp)$}
        \State $\actionSublass \gets \select{\Objectives(\RedComp) \setminus \getSubclasses{\TestSuite}}{\distrib{\textit{unif}}}$
        \State $\sequence \gets \select{\getSequences{\actionSublass}}{\distrib{\textit{init}}}$
        \State $\TestSuite \gets \TestSuite \cup \set{\sequence}{}$
        \State $\TestSuite \gets \reduction{\TestSuite}$%\Comment{Only inputs that overlap with $\sequence$ (\Cref{defi:overlapRel})}
    \EndWhile
    \If{$\TestSuite \not\in \RoofersVar$}
        \State $\RoofersVar \gets \RoofersVar \cup \set{\TestSuite}{}$
    \EndIf
\EndWhile
\State \textbf{return} $\RoofersVar$
\EndProcedure
\end{algorithmic}
\end{algorithm}

The miser population is initially empty, i.e., $\Misers{0} \eqdef \varnothing$, as misers are generated during the search through mutations (\Cref{sec:aim:updtPops}).
We detail in \Cref{algo:rofferInit} how the roofer population $\Roofers{0}$ is initialized, where $\select{X}{\distrib{}}$ randomly select one element in $X$ using distribution $\distrib{}$, $\distrib{\textit{unif}}$ denotes the uniform distribution, and $\distrib{\textit{init}}$ denotes the following distribution:
$$\proba{\textit{init}}{\sequence_1} \eqdef \frac{\frac{1}{1 + \occurrence{\sequence_1}}}{\sum\limits_{\sequence_2 \in \getSequences{\actionSublass}}\frac{1}{1 + \occurrence{\sequence_2}}}$$
where $\occurrence{\sequence}$ denotes the number of times the input $\sequence$ was selected in the roofer population so far.
This distribution ensures that inputs that were not selected so far are more likely to be selected, so that the initial roofer population can be more diverse.
Note that computing $\reduction{\TestSuite}$ is tractable since adding a new input can only affect the redundancy of inputs that overlap with it (\Cref{defi:overlapRel}).
\subsection{Parents Selection}
\label{sec:aim:selectParents}

For each generation $n$, parents are selected as follows.
If $\Misers{n} \neq \varnothing$, then one parent is selected from the miser population and one from the roofer population.
Otherwise, two distinct parents are selected from the roofer population.
A parent $\TestSuite_1 \in \Misers{n}$ is randomly selected from the miser population using the following distribution:
$$\proba{\textit{misers}}{\TestSuite_1} \eqdef \frac{\frac{1}{\exposure{\TestSuite_1}}}{\sum\limits_{\TestSuite_2 \in \Misers{n}} \frac{1}{\exposure{\TestSuite_2}}}$$
where the exposure is defined in \Cref{defi:exposure}.
The purpose of this distribution is to ensure that input sets with large coverage or at least large potential (\Cref{sec:aim:problemDefinition}) are more likely to be selected.
%Note that this distribution is well-defined because, according to \Cref{theo:misers}, no miser $\TestSuite \in \Misers{n}$ covers all the objectives, so the objective value for at least one objective $\actionSublass_i \in \Objectives(\RedComp)$ is positive, $\objFun{\actionSublass_i}{\TestSuite} > 0$, hence so does the exposure.
A parent $\TestSuite_1 \in \Roofers{n}$ is randomly selected from the roofer population using the following distribution:
$$\proba{\textit{roofers}}{\TestSuite_1} \eqdef \frac{\frac{1}{\costFunction{\TestSuite_1}}}{\sum\limits_{\TestSuite_2 \in \Roofers{n}}\frac{1}{\costFunction{\TestSuite_2}}}$$
The purpose of this distribution is to ensure that less costly input sets are more likely to be selected.
% Note that this distribution is well-defined because the cost of every input of actions $\sequence$ is positive: $\costFunction{\sequence} > 0$ (\pref{theo:costPositive}).

%%%%%%%%
\subsection{Parents Crossover}
\label{sec:aim:crossover}
\label{defi:crossover}

After selecting two distinct parents $\TestSuite_1$ and $\TestSuite_2$, we detail how they are used to generate the offspring $\TestSuite_3$ and $\TestSuite_4$.
Our crossover strategy exploits the fact that, for each objective $\actionSublass$ to cover, it is easy to infer inputs in $\getSequences{\actionSublass}$ able to cover $\actionSublass$ (\Cref{sec:aim:mao}).
For each crossover, we randomly split the objectives in two halves $\halfObj_1$ and $\halfObj_2$ such that $\halfObj_1 \cup \halfObj_2 = \Objectives(\RedComp)$ and $\halfObj_1 \cap \halfObj_2 = \varnothing$.
We consider here a balanced split, to prevent cases where one parent massively contributes to offspring coverage.

Then, we use this split to define the crossover: inputs in the connected component $\RedComp$ are split between $\halfSeqs_1 \eqdef \getSequences{\halfObj_1}$, the ones covering the first half of the objectives, and $\halfSeqs_2 \eqdef \getSequences{\halfObj_2}$, the ones covering the second half.
Note that some inputs may cover objectives both in $\halfObj_1$ and $\halfObj_2$, so we call the \emph{edge} of the split the intersection $\halfSeqs_1 \cap \halfSeqs_2$.
Because we assume both parents are reduced, this means that redundant inputs can only happen at the edge of the split.
%This is another reason for our crossover strategy: by using a crossover based on the objective coverage, we ensure that only a few redundant inputs are likely to occur after recombination, so that we can reduce them with only a few removal steps.
The genetic material of both parents $\TestSuite_1$ and $\TestSuite_2$ is then split in two parts: inputs in $\halfSeqs_1$ and inputs in $\halfSeqs_2$, as follows:
$$
\begin{array}{r@{}l}
     \TestSuite_1 = {}& (\TestSuite_1 \cap \halfSeqs_1) \cup (\TestSuite_1 \cap \halfSeqs_2) \\
     \TestSuite_2 = {}& (\TestSuite_2 \cap \halfSeqs_1) \cup (\TestSuite_2 \cap \halfSeqs_2)
\end{array}
$$

Then, these parts are swapped to generated the offspring, as follows:
$$
\begin{array}{r@{}l}
     \TestSuite_3 \eqdef {}& (\TestSuite_1 \cap \halfSeqs_1) \cup (\TestSuite_2 \cap \halfSeqs_2) \\
     \TestSuite_4 \eqdef {}& (\TestSuite_2 \cap \halfSeqs_1) \cup (\TestSuite_1 \cap \halfSeqs_2)
\end{array}
$$

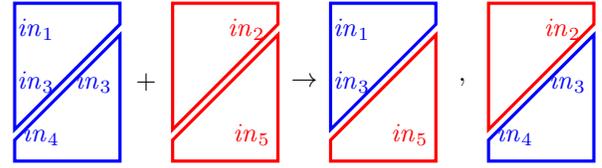
\begin{figure}[t]
\centering
\begin{tikzpicture}[scale=0.7]
    % styles
    \tikzset{
        chromosome/.style={very thick},
    }
    % other nodes
    \node
        (arrow)
        at (0, 0)
        {$\rightarrow$};
    \node
        (plus)
        at (-3, 0)
        {$+$};
    \node
        (comma)
        at (3, 0)
        {$,$};
    % parent a
    \draw[chromosome,blue]
        (-5.5, 1.5) --
        (-3.5, 1.5) --
        (-3.5, 1.1) --
        (-5.5, -0.9) --
        cycle;
    \node[blue]
        (a1)
        at (-5.1,1)
        {$\sequence_1$};
    \node[blue]
        (a3a)
        at (-5.1,0)
        {$\sequence_3$};
    \draw[chromosome,blue]
        (-3.5, 0.9) --
        (-5.5, -1.1) --
        (-5.5, -1.5) --
        (-3.5, -1.5) --
        cycle;
    \node[blue]
        (a3b)
        at (-4,0)
        {$\sequence_3$};
    \node[blue]
        (a4)
        at (-5,-1)
        {$\sequence_4$};
    % parent b
    \draw[chromosome,red]
        (-2.5, 1.5) --
        (-0.5, 1.5) --
        (-0.5, 1.1) --
        (-2.5, -0.9) --
        cycle;
    \node[red]
        (b2)
        at (-1.1,1)
        {$\sequence_2$};
    \draw[chromosome,red]
        (-0.5, 0.9) --
        (-2.5, -1.1) --
        (-2.5, -1.5) --
        (-0.5, -1.5) --
        cycle;
    \node[red]
        (b5)
        at (-1,-1)
        {$\sequence_5$};
    % offspring c
    \draw[chromosome,blue]
        (0.5, -0.9) --
        (2.5, 1.1) --
        (2.5, 1.5) --
        (0.5, 1.5) --
        cycle;
    \node[blue]
        (c1)
        at (0.9,1)
        {$\sequence_1$};
    \node[blue]
        (c3a)
        at (0.9,0)
        {$\sequence_3$};
    \draw[chromosome,red]
        (2.5, -1.5) --
        (0.5, -1.5) --
        (0.5, -1.1) --
        (2.5, 0.9) --
        cycle;
    \node[red]
        (c5)
        at (2,-1)
        {$\sequence_5$};
    % offspring d
    \draw[chromosome,red]
        (3.5, -0.9) --
        (5.5, 1.1) --
        (5.5, 1.5) --
        (3.5, 1.5) --
        cycle;
    \node[red]
        (d2)
        at (4.9,1)
        {$\sequence_2$};
    \draw[chromosome,blue]
        (5.5, -1.5) --
        (3.5, -1.5) --
        (3.5, -1.1) --
        (5.5, 0.9) --
        cycle;
    \node[blue]
        (d3b)
        at (5,0)
        {$\sequence_3$};
    \node[blue]
        (d4)
        at (4,-1)
        {$\sequence_4$};
    % genes b
\end{tikzpicture}
\caption{Crossover Example}
\label{fig:crossover}
\end{figure}

%\begin{ex}
%\label{ex:crossover}
For illustration purpose we consider in \Cref{fig:crossover} a small connected component $\RedComp = \set{\sequence_1, \sequence_2, \sequence_3, \sequence_4, \sequence_5}{}$ and the following split for the inputs: $\getSequences{\halfObj_1} = \set{\sequence_1, \sequence_2, \sequence_3}{}$ and $\getSequences{\halfObj_2} = \set{\sequence_3, \sequence_4, \sequence_5}{}$.
The edge of the split is $\getSequences{\halfObj_1} \cap \getSequences{\halfObj_2} = \set{\sequence_3}{}$.
The offspring of parents $\TestSuite_1 = \set{\sequence_1, \sequence_3, \sequence_4}{}$ and $\TestSuite_2 = \set{\sequence_2, \sequence_5}{}$ is $\TestSuite_3 = \set{\sequence_1, \sequence_3, \sequence_5}{}$ and $\TestSuite_4 = \set{\sequence_2, \sequence_3, \sequence_4}{}$.
%\end{ex}

%Note that, according to \Cref{defi:crossover}, if both parents are roofers then they cover all the objectives and so do their offspring.
%Hence, a miser can only appear during the mutation step in \Cref{sec:aim:mutation}.

%Nevertheless,
%The swapping of genetic material may lead to some inputs in the offspring being redundant.
%But, at this stage, \geneticAlgo does not remove them and it waits for the mutation step to occur, to benefit from the cumulative effect of the randomness from both crossover and mutation steps, as detailed at the end of \Cref{sec:aim:mutation}.

%%%%%%%%
\subsection{Offspring Mutation}
\label{sec:aim:mutation}
\label{defi:mutation}

Each gene of an offspring $\TestSuite$ corresponds to an input $\sequence \in \RedComp$,
%in the considered connected component,
the gene value being $1$ if $\sequence \in \TestSuite$ and $0$ otherwise.
A mutation happens when this gene value is changed, hence mutation randomly adds or removes one input from an offspring.
The crossover (at the edge of the split) and mutation (when an input is added) steps may result in inputs being redundant in the offspring.
Since redundancies in the connected component were already reduced by \impro (\Cref{defi:subprob}) and changes in redundancy could happen only amongst neighbors (for the overlapping relation) of the changed inputs (\Cref{defi:overlapRel}), we only expect a few redundant inputs.
Therefore, removal steps can be exhaustively computed to replace each offspring $\TestSuite$ by its reduced counterpart $\reduction{\TestSuite}$ (\Cref{sec:aim:initPops}).
%We waited for the mutation step to occur before removing redundant inputs because, to improve the exploration of the search space, we want the randomness from the crossover (\Cref{sec:aim:crossover}) to interact with the randomness of the mutation.
%%Indeed, performing removal steps before mutation could prevent the search from reaching interesting candidates.
%For example, the input removed through mutation may differ from the one that would have been selected by removal steps.
%Further, when the mutation adds an input, it randomly generates more redundancies, potentially leading to a less costly individual after removal steps.
%In short, premature removal steps could prevent the search from exploring relevant combinations of inputs.

%%%%%%%%
\subsection{Population Update}
\label{sec:aim:updtPops}

%We presented how, starting from the roofer and miser populations at generation $n$, $\Roofers{n}$ and $\Misers{n}$, \geneticAlgo generates, mutates, and reduces the offspring.
~\TSE{1.4.3}{We detail how} the offspring is used to obtain the populations $\Roofers{n + 1}$ and $\Misers{n + 1}$ at generation $n + 1$.

First, we discard any offspring that is a duplicate of an individual already present in either $\Roofers{n}$ or $\Misers{n}$.
Indeed, the duplication of individuals would only result in altering their weight for being selected as parents (thus, the intended procedure), reducing roofer diversity, and increasing the number of miser comparisons (as detailed below).
%and thus the purpose of the selection procedure (\Cref{sec:aim:selectParents}).
%Moreover, the roofer population has a fixed size.
%Hence, for duplicates to be part of the population, other roofers would have to be removed, leading to a reduction of diversity in the population.
%Finally, as detailed below, each miser is characterized by its fitness vector in order to be compared with other misers.
%Hence, including duplicates would only make these comparisons more costly without bringing any benefit regarding diversity on the Pareto front.

If a remaining offspring $\TestSuite_1$ covers all the objectives in $\Objectives(\RedComp)$, then it is a candidate for the roofer population.
Otherwise, it is a candidate for the miser population.

For each candidate $\TestSuite_1$ for the roofer population, \geneticAlgo computes its cost.
If $\costFunction{\TestSuite_1} \le \max$ $\set{\costFunction{\TestSuite}}{\TestSuite \in \Roofers{n}}$, then it selects the most costly roofer $\TestSuite_2$ (or, in case of a tie, one of the most costly roofers), removes $\TestSuite_2$ from the population, and adds $\TestSuite_1$.
Otherwise, $\TestSuite_1$ is rejected.
Note that we chose $\le$ instead of $<$ for the above cost criterion because, in case of a tie, we prefer to evolve the population instead of maintaining the status quo, to increase the odds of exploring new regions of the search space.
%After completing this procedure for each candidate to the roofer population, we obtain the population $\Roofers{n + 1}$ for the next generation.
%We prove in the technical report~\cite{NMPB24} (Theorem 4) that roofers satisfy desirable properties, including that the cheapest roofer of the last generation has not only full coverage but is the cheapest roofer observed during the search to date.

%We now detail how the miser population is updated.
For each candidate $\TestSuite_1$ to the miser population, we compute its fitness vector $\fitnessVectorComponent{\TestSuite_1}$ (\Cref{defi:subprob}).
Then, for each $\TestSuite_2 \in \Misers{n}$ we compare $\fitnessVectorComponent{\TestSuite_1}$ and $\fitnessVectorComponent{\TestSuite_2}$.
If $\TestSuite_2 \dominates \TestSuite_1$ in the sense of Pareto-dominance (\Cref{sec:aim:background:paretoFront}), then we stop the process and $\TestSuite_1$ is rejected.
If $\TestSuite_1 \dominates \TestSuite_2$, then $\TestSuite_2$ is removed from $\Misers{n}$.
That way, we ensure that the miser population contains only non-dominated individuals.
After completing the comparisons, if the process was not stopped, then $\TestSuite_1$ itself is non-dominated, so it is added to the miser population.
In that case, $\fitnessVectorComponent{\TestSuite_1}$ is stored for future comparisons.
%After completing this procedure for each candidate to the miser population, we obtain the population $\Misers{n + 1}$ for the next generation.
%We prove in the technical report~\cite{NMPB24} (Theorem 5) that misers satisfy desirable properties during the search, including that no miser encountered during the search Pareto-dominates misers in the last generation.

Properties satisfied by roofers and misers are detailed in the separate appendix (Theorems 4 and 5).

\subsection{Termination}
\label{sec:aim:termination}

\geneticAlgo repeats the process until it reaches a fixed number %$\nbGens$
of generations or exhausts a given time budget.
Then, amongst the least costly roofers (several may have the same cost), it randomly selects one individual $\TestSuite_{\RedComp}$ %(the solution may not be unique as several may have the same cost)
as solution to our sub-problem (\Cref{defi:subprob}).
$\TestSuite_{\RedComp}$ covers all the objectives and, amongst the input sets covering those objectives, $\TestSuite_{\RedComp}$ has the smallest cost encountered during the search.
%Some misers may have a lower cost, but they do not cover all the objectives, and hence cannot be considered a solution to our sub-problem.

\section{Step 6: Data Post-Processing}
\label{sec:aim:postprocessing}
\label{defi:finalSolution}

The set $\NecessaryInputs$ was initially empty (\Cref{defi:probRed:init}) and then accumulated necessary inputs each time redundancy was determined (\Cref{defi:necessaryInitSeq}).
After removing inputs and reducing the objectives to be covered accordingly (\Cref{sec:aim:search}), \impro obtained a set $\SearchSpace$ of remaining inputs and objectives.
Then, \impro divided the remaining problem into sub-problems (\Cref{defi:subprob}), one for each connected component $\RedComp$.
Finally, for each connected component $\RedComp$, the corresponding sub-problem was solved using \geneticAlgo (\Cref{sec:aim:genetic}), obtaining the corresponding minimized component $\TestSuite_{\RedComp}$.
At the end of the search, \aim merges inputs from each minimized component $\TestSuite_{\RedComp}$ with the necessary inputs $\NecessaryInputs$ to obtain a \emph{minimized input set} $\TestSuiteFinal$ as solution to our initial problem (\Cref{defi:pbDef}):
$$\TestSuiteFinal \eqdef \NecessaryInputs \cup \bigcup_{\RedComp \in \RedundantComponents{\SearchSpace}} \TestSuite_{\RedComp}$$

\section{Empirical Evaluation}
\label{sec:aim:results}

In this \namecref{sec:aim:results}, we report our results on the assessment of our approach with two Web systems. 
We investigate the following Research Questions (RQs):
\begin{description}
 \item[RQ1]
    \textbf{What is the vulnerability detection effectiveness of \aim, compared to alternatives?}
        This research question aims to determine if and to what extent \aim reduces the effectiveness of MST by comparing the vulnerabilities detected between the initial and the minimized input sets. Also, we further compare the vulnerability detection rate of \aim with simpler alternative approaches.
    \item[RQ2] 
    % \textbf{What is the magnitude of input set minimization?} 
    \textbf{What is the input set minimization effectiveness of \aim, compared to alternatives?}
        This research question aims to analyze the magnitude of minimization in terms of the number of inputs, cost (\Cref{sec:aim:goals,sec:aim:cost}), and execution time for the considered MRs, both for \aim and alternative approaches.  
  
 \TSE{2.2}{   \item[RQ3] 
    \textbf{What is the input set minimization effectiveness of \geneticAlgo, compared to alternatives?}
    This research question aims to determine the particular contribution of the novel \geneticAlgo genetic algorithm in minimizing cost while preserving vulnerability detection, by comparing it to alternative approaches for different time budgets.}
\end{description}

\subsection{Experiment Design}
\label{sec:aim:aim:methodology}

\subsubsection{Subjects of the Study}
\label{sec:aim:aim:benchmarks}

To assess our approach with MRs and input sets that successfully detect real-world vulnerabilities, we rely on the same input sets and settings as \mstWi~\cite{BPGB23}.

The targeted Web systems under test are \jenkins~\cite{Ecl} and \joomla~\cite{Joomla}.
\jenkins is a leading open source automation server while Joomla is a content management system (CMS) that relies on the MySQL RDBMS and the Apache HTTP server.
We chose these Web systems because of their plug-in architecture and Web interface with advanced features (such as Javascript-based login and AJAX interfaces), which makes \jenkins and \joomla good representatives of modern Web systems.

Further, these systems present differences in their output interface and input types that, since inputs and outputs are key drivers for our approach,  contribute to improve the generalizability of our results. Concerning outputs, \joomla is a CMS where 
Web pages tend to contain a large amount of static text that differ in every page, while \jenkins provides mainly structured content that may continuously change (e.g., seconds from the last execution of a Jenkins task). The input interfaces of \jenkins are mainly  short forms and buttons whereas the inputs interfaces of \joomla often include long text areas and several selection interfaces (e.g., for tags annotation).

The selected versions of \jenkins and \joomla---2.121.1 and 3.8.7, respectively---are affected by known vulnerabilities that can be triggered from the Web interface; we describe them in \Cref{sec:aim:results:vulnerabilities}.

The input set provided in the \mstWi's replication package has been collected by running \crawljax with, respectively, four users for Jenkins and six users for Joomla having different roles, e.g., admin.
For each role, \crawljax has been executed for a maximum of 300 minutes, to prevent the crawler from running indefinitely, thereby avoiding excessive resource consumption.
Further, to exercise features not reached by \crawljax, a few additional Selenium~\cite{selenium}-based test scripts (four for \jenkins and one for \joomla) have been added to the input set.
In total, we have 160 initial inputs for \jenkins and 148 for \joomla, which are all associated to a unique identifier.

\TSE{2.2}{Since \geneticAlgo assumes  redundancy in the input set to be already reduced by \impro (e.g., to ensure the reduction step after mutation is tractable), we do not consider the initial input sets for RQ3 but their reduced versions.
Thus, we run the double-clustering (\Cref{sec:aim:doubleCLustering}) and problem reduction (\Cref{sec:aim:search}) steps to obtain a set of necessary inputs and several connected components.
Then, the input set for RQ3 is the union of these necessary inputs and connected components, called a \emph{reduced input set}.
We repeat this process several times (to reduce the impact of randomness),
% , as detailed in \Cref{sec:aim:results:performance}), 
thus obtaining several reduced input sets.}

\TSE{2.2}{Finally, since we know necessary inputs should be part of the solution and minimization can be performed independently on each connected component (\Cref{sec:aim:search}), we can exhaustively search the best order of removal steps for each connected component, in order to determine the \emph{optimal solution} for each reduced input set.
Thus, we can compare the solutions obtained by different 
% many-objective 
search algorithms for different time budgets to this optimal solution.
Note that this approach is tractable for the considered connected components only because they are small, but it is intractable for larger input sets, e.g., the initial or reduced input sets.}

\subsubsection{Security Vulnerabilities} 
\label{sec:aim:results:vulnerabilities}

The replication package for \mstWi~\cite{Replicability} includes 76 metamorphic relations (MRs). These MRs can identify nine vulnerabilities in \jenkins and three vulnerabilities in \joomla using the initial input set, as detailed in \Cref{table:subjectVulnrabilities,table:vulnerabilitiesJoomla}, respectively.

\begin{table*}[t]
\center
\scriptsize
\caption{\jenkins Vulnerabilities.}
\label{table:subjectVulnrabilities}
\begin{tabular}{|p{1.25cm}|p{5.5cm}|p{2.25cm}|p{7cm}|}
\hline
CVE &
  Description &
  Vulnerability Type &
  Input Identifiers \\
\hline
CVE-2018-1000406 \cite{MITa} &
  In the file name parameter of a Job configuration, users with Job / Configure permissions can specify a relative path escaping the base directory. Such path can be used to upload a file on the \jenkins host, resulting in an arbitrary file write vulnerability. &
  CWE\_22  &
  160 \\
\hline
CVE-2018-1000409 \cite{MITb} &
   A session fixation vulnerability prevents \jenkins from invalidating the existing session and creating a new one when a user signed up for a new user account. &
  CWE\_384 &
  112, 113, 114 \\
\hline
CVE-2018-1999003 \cite{MITc}  &
  \jenkins does not perform a permission check for URLs handling cancellation of queued builds, allowing users with Overall / Read permission to cancel queued builds. &
  CWE\_280, CWE\_863  &
  116, 157 \\
\hline
CVE-2018-1999004 \cite{MITd} &
  \jenkins does not perform a permission check for the URL that initiates agent launches, 
  allowing users with Overall / Read permission to initiate agent launches. &
  CWE\_863, CWE\_285&
  2, 116 \\
\hline
CVE-2018-1999006 \cite{MITe} &
  A exposure of sensitive information vulnerability allows attackers to determine the date and time when a plugin was last extracted. &
  CWE\_200, CWE\_668&
  33, 55, 57, 61, 62, 63, 64, 75, 107, 108, 110, 135, 136, 156, 160 \\
\hline
CVE-2018-1999046 \cite{MITf} &
  Users with Overall / Read permission are able to access the URL serving agent logs on the UI due to a lack of permission checks. &
  CWE\_200 &
  2, 116 \\
\hline
CVE-2020-2162 \cite{MITg} &
  Jenkins does not set Content-Security-Policy headers for files uploaded as file parameters to a build, resulting in a stored XSS vulnerability. & 
CWE\_79
&
  1, 18, 19, 23, 26, 75, 156, 158 \\
\hline
Password aging problem in \jenkins  &  Jenkins does not integrate any mechanism for managing password aging; consequently, users aren't incentivized to update passwords periodically.
  &
  CWE\_262&
  1, 2, 3, 4, 5, 6, 7, 8, 9, 10, 11, 12, 13, 14, 15, 16, 18, 19, 20, 22, 23, 24, 25, 26, 27, 28, 30, 32, 110, 33, 34, 35, 38, 39, 41, 42, 43, 44, 45, 46, 47, 58, 61, 62, 64, 65, 66, 69, 70, 71, 73, 74, 75, 104, 108, 116, 117, 118, 119, 120, 121, 122, 123, 124, 125, 126, 127, 128, 129, 130, 131, 133, 134, 143, 145, 146, 159, 160 \\
\hline
Weak password in \jenkins  &   Jenkins does not require users to have strong passwords, which makes it easier for attackers to compromise user accounts.
  &
  CWE\_521&
  112, 113, 114\\
\hline
\end{tabular}%
\end{table*}
\begin{table*}[t]
\centering
\scriptsize
\caption{\joomla Vulnerabilities.}
\label{table:vulnerabilitiesJoomla}
\begin{tabular}{|p{1.25cm}|p{5.5cm}|p{2.25cm}|p{7cm}|}
\hline
CVE &
  Description &
  Vulnerability Type &
  Input Identifiers \\
\hline
CVE-2018-11327 \cite{MITi} &
  Inadequate checks allow users to see the names of tags that were either unpublished or published with restricted view permission. &
  CWE\_200  &
  37 with 22, 23, 24, 25, 50 \\
\hline
CVE-2018-17857 \cite{MITh} &
  Inadequate checks on the tag search fields can lead to an access level violation. &
  CWE\_863  &
  1 with 22, 23, 24, 25 \\
\hline
Password aging problem in Joomla  &
  Joomla does not integrate any mechanism for managing password aging; consequently, users aren't incentivized to update passwords periodically.  &
  CWE\_262&
  2, 3, 5, 6, 7, 8, 11, 12, 15, 17, 20, 22, 23, 24, 25, 26, 27, 28, 29, 30, 66, 110, 144, 146 \\
\hline
\end{tabular}%
\end{table*}

For both tables, the first column contains, when available, the CVE identifiers of the considered vulnerabilities.
The password aging problem (for both \jenkins and \joomla) and weak password (for \jenkins) are vulnerabilities that were identified during the \mstWi study~\cite{BPGB23} and therefore do not have CVE identifiers.
The second column provides a short description of the vulnerabilities.
The third column reports the CWE ID for each vulnerability. We present two CWE IDs (e.g., CWE 863 and 280) in cases where the CVE report denotes a general vulnerability type (e.g., CWE 863 for incorrect authorization~\cite{MITc}), though a more precise identification (e.g., CWE 280 concerning improper handling of privileges that may result in incorrect authorization) could be applied. Since the 12 considered vulnerabilities are associated to nine different CWE IDs and each vulnerability has a unique CWE ID, we can conclude that the selected subjects cover a diverse set of vulnerability types, thus further improving the generalizability of our results.

The last column in \Cref{table:subjectVulnrabilities,table:vulnerabilitiesJoomla} lists identifiers for inputs which were able to trigger the vulnerability using one of the corresponding MRs.
For instance, one can detect vulnerability CVE-2018-1999046 in \jenkins by running the MR written for CWE\_200 with inputs 2 or 116.
For the first two \joomla vulnerabilities, two inputs need to be present at the same time in the input set in order to trigger the vulnerability because, as opposed to most MRs, the corresponding MRs requires two source inputs to generate follow-up inputs.
For instance, to detect vulnerability CVE-2018-17857 in \joomla, one needs input 1 and at least one input amongst inputs 22, 23, 24, or 25.

\subsubsection{\aim configurations}
\label{sec:aim:results:configurations}

\aim can be configured in different ways to obtain a minimized input set from an initial input set.
Such a \emph{configuration} consists in a choice of distance function and algorithm for output clustering (\Cref{sec:aim:outputClustering}), and a choice of algorithm for action clustering (\Cref{sec:aim:actionClustering}).

For the sake of conciseness, in  \Cref{tab:vulnsCoverage,tab:jenkins:duels:sizes,tab:joomla:duels:sizes,tab:jenkins:duels:costs,tab:joomla:duels:costs,tab:jenkins:duels:times,tab:joomla:duels:times}, we denote each configuration by three letters, where \configLev{} and \configBag{} respectively denote the Levenshtein and Bag distances, and \configKmeans{}, \configDbscan{}, and \configHdbscan{} respectively denote the \kmeans, \dbscan, and \hdbscan clustering algorithms.
For instance, \BagDbscanHdbscan denotes that Bag distance and \dbscan were used for output clustering, and then \hdbscan for action clustering.
These notations are summarized in \Cref{tab:configs}.

\aim performs Silhouette analysis (\Cref{sec:aim:doubleClustering:hyperParameter}) to determine the hyper-parameters required for these clustering algorithms.
We considered the same ranges of values for the hyper-parameters in both output clustering (\Cref{sec:aim:outputClustering}) and action clustering steps (\Cref{sec:aim:actionClustering}).
For \kmeans, we select the range $\interval{1}{70}$ for the number of clusters $k$.
In the case of \dbscan, the range for the distance threshold $\epsilon$ is $\interval{2}{10}$ for \jenkins and $\interval{3}{15}$ for \joomla.
The range is larger for \joomla because  \joomla has a larger number of Web pages than \jenkins.
Finally, the range for the minimum number of neighbors $\minpts$ is $\interval{1}{5}$ for both systems.
For \hdbscan, the range for the minimum number $n$ of individuals required to form a cluster is $\interval{2}{8}$ for both systems.

We also determine the hyper-parameters for the genetic search (\Cref{sec:aim:genetic}).
Related work on whole test suite generation successfully relied on a population of 80 individuals~\cite{FA13}.
Since we reduced the problem (\Cref{sec:aim:search}) before applying \geneticAlgo independently to each connected component, which includes fewer inputs than the whole test suite generation~\cite{FA13}, we experimented with a lower population size of 20 individuals.
\TSE{3.1.2}{Additionally, for RQ1 and RQ2, we set the number of generations for the genetic algorithm to 100, similar to the value considered in previous work~\cite{FA13}.}
% \TSE{3.1.2}{For RQ1 and RQ2, we set the termination criterion for \geneticAlgo to the number of generations.} 
\TSE{2.2}{However, for RQ3, to compare the minimized input sets obtained by search algorithms with different time budgets, we use a maximum time budget of 600 seconds as the termination criterion for  \geneticAlgo, consistent with \mosa's study~\cite{PKT15}, while recording the intermediate results over time.}

% However, for RQ3, we use a maximum time budget of 600 seconds, consistent with \mosa's study~\cite{PKT15}, as the termination criterion for  \geneticAlgo.

\begin{table}[t]
\centering
\scriptsize
\caption{\aim configurations and baselines for RQ1 and RQ2.}
\label{tab:configs}
\begin{tabular}{|c|c|c|c|c|}
\cline{3-5}
    \multicolumn{2}{c|}{} & \multicolumn{2}{c|}{Output clustering}
    & \multicolumn{1}{c|}{Action clustering}\\
\cline{3-5}
    \multicolumn{2}{c|}{} & Distance & Algorithm & Algorithm \\
\hline
\multirow{18}{*}{\aim configurations}
% Levenshtein and Bag
% \kmeans, \dbscan, and \hdbscan
    & \LevKmeansKmeans & Levenshtein & \kmeans & \kmeans \\
    & \LevKmeansDbscan & Levenshtein & \kmeans & \dbscan \\
    & \LevKmeansHdbscan & Levenshtein & \kmeans & \hdbscan \\
    & \LevDbscanKmeans & Levenshtein & \dbscan & \kmeans \\
    & \LevDbscanDbscan & Levenshtein & \dbscan & \dbscan \\
    & \LevDbscanHdbscan & Levenshtein & \dbscan & \hdbscan \\
    & \LevHdbscanKmeans & Levenshtein & \hdbscan & \kmeans \\
    & \LevHdbscanDbscan & Levenshtein & \hdbscan & \dbscan \\
    & \LevHdbscanHdbscan & Levenshtein & \hdbscan & \hdbscan \\
    & \BagKmeansKmeans & Bag & \kmeans & \kmeans \\
    & \BagKmeansDbscan & Bag & \kmeans & \dbscan \\
    & \BagKmeansHdbscan & Bag & \kmeans & \hdbscan \\
    & \BagDbscanKmeans & Bag & \dbscan & \kmeans \\
    & \BagDbscanDbscan & Bag & \dbscan & \dbscan \\
    & \BagDbscanHdbscan & Bag & \dbscan & \hdbscan \\
    & \BagHdbscanKmeans & Bag & \hdbscan & \kmeans \\
    & \BagHdbscanDbscan & Bag & \hdbscan & \dbscan \\
    & \BagHdbscanHdbscan & Bag & \hdbscan & \hdbscan \\
\hline
\multirow{4}{*}{Baselines}
    & \Rt & \multicolumn{2}{c|}{$\times$} & Random \\
    & \ArtKmeans & \multicolumn{2}{c|}{$\times$} & \kmeans \\
    & \ArtDbscan & \multicolumn{2}{c|}{$\times$} & \dbscan \\
    & \ArtHdbscan & \multicolumn{2}{c|}{$\times$} & \hdbscan \\
\hline
\end{tabular}
\end{table}

\subsubsection{Baselines}
\label{sec:aim:results:baselines}

For RQ1 and RQ2, we identify the following baselines against which to compare \aim configurations.

A 2016 survey reported that 57\% of metamorphic testing (MT) work used Random Testing (RT) to generate source inputs~\cite{SFSRC16} and, in 2021, 84\% of the publications related to MT adopted traditional or improved RT methods to generate source inputs~\cite{HWHY21}.
In the context of test suite minimization, random search is a straightforward baseline against which to compare \aim that is commonly used~\cite{WAG15,ZAY19}.
This baseline consists in randomly selecting a given number of inputs from the initial input set.
This number is determined based on \aim runs.
Each \aim run is performed using 18 different configurations (\Cref{sec:aim:results:configurations}), 
each leading to a different minimized input set.
So, for a fair comparison, we configure random search to select $n$ inputs from the initial input, where $n$ is the size of the largest input set produced by the 18 \aim configurations.
We repeat this process, for each \aim run, to obtain the same number of input sets for random search as for \aim.

\TSE{2.4}{Moreover, Adaptive Random Testing (ART) was proposed to enhance the performance of RT.
It is based on the intuition that inputs close to each other are more likely to have similar failure behaviors than inputs further away from each other.
Thus, ART generates inputs widely spread across the input domain, in order to find failure with fewer inputs than RT~\cite{BCK+16}.
ART is also commonly used  in the context of test suite minimization~\cite{HAB13,CMVB19}.
It is similar to our action clustering step (\Cref{sec:aim:actionClustering}), since it is based on partitioning the input space and generating new inputs in blocks that are not already covered~\cite{HWHY21}.}

\TSE{2.4}{
To perform ART, we need to group inputs based on the similarity between their actions. So, we use \aim to perform action clustering directly on the initial input set instead of output classes. Then, for each cluster, we randomly select one input that covers it. Finally, we group the selected inputs to obtain an input set for the ART baseline. This algorithm will stop when we have selected one input from each cluster, and thus, it is not limited by a time budget.}
Again, we repeat this process for each \aim run.
Since we considered the \kmeans, \dbscan, and \hdbscan clustering algorithms, there are three variants of this baseline.

In
\Cref{tab:vulnsCoverage,tab:jenkins:duels:sizes,tab:joomla:duels:sizes,tab:jenkins:duels:costs,tab:joomla:duels:costs,tab:jenkins:duels:times,tab:joomla:duels:times}, \Rt denotes the random search baseline while \ArtKmeans, \ArtDbscan, and \ArtHdbscan denote the ART baselines, using respectively the \kmeans, \dbscan, and \hdbscan clustering algorithms.
These notations are summarized in \Cref{tab:configs}.

\TSE{2.2}{RQ3 aims at determining the contribution of the \geneticAlgo genetic algorithm to input set minimization and comparing it with alternative approaches.
We consider random search, a greedy algorithm for the set cover problem (\Cref{sec:aim:genetic}), as well as \mosa  and \nsgaThree (\Cref{sec:aim:background:MaO}), which are well-known many-objective search algorithms~\cite{GUPTA,KIRAN, NAZ, ZOHA}.}

\TSE{2.2}{We use random search to provide insights on how difficult the problem is and to help assess if search algorithms are necessary to solve it.
Random search starts with an empty input set and, for each iteration, randomly selects one input from the reduced input set (\Cref{sec:aim:aim:benchmarks}).
This process continues until all the objectives are covered.}
%Then, the obtained input set is compared with the ones obtained from the other algorithms.}

\TSE{2.2}{The greedy algorithm for the set cover problem~\cite{Vaz01} is one of the best-possible polynomial time approximation algorithm for this problem, with a tight bound on the cost of the solution for the unweighted variant of the problem~\cite{Sla96}, matching theoretical bounds~\cite{DS14}, and that can be efficiently adapted to the weighted variant~\cite{Vaz01}.
Since the weighted set cover problem is close to our problem (\Cref{sec:aim:genetic}), we adapt this greedy algorithm to obtain a minimized input set from the reduced input set.
The minimized input set is initially empty.
For each iteration, the input with the best cost effectiveness is selected, where the cost effectiveness of input $\sequence$ is computed as $\frac{\card{\getSubclasses{\sequence} \cap \mathit{Uncovered}}}{\costFunction{\sequence}}$. Recall that $\getSubclasses{\sequence}$ and $\costFunction{\sequence}$ are defined in \Cref{sec:aim:goals}, and $\mathit{Uncovered}$ denotes the set of the objectives not yet covered by selected inputs.
This process continues until all the objectives are covered.}

\TSE{2.2}{The initial population for both \mosa and \nsgaThree is the reduced input set.
For the other parameters, such as population size, mutation rate, and (for \nsgaThree) the number of reference points (\Cref{sec:aim:background:Optimization}), we use the default values recommended in the original studies~\cite{PKT15, DJ14,JD14}.
More precisely, the population size for \mosa is set to 50, while it is automatically computed for \nsgaThree based on the number of reference points.
We set the crossover rate to $1$ for both \mosa and \nsgaThree, the same as \geneticAlgo.
Finally, we use the same objective functions and mutation operator for both \mosa and \nsgaThree,  as in \geneticAlgo.
Finally, for \geneticAlgo and all baselines, we use the same termination criterion as in \mosa's study~\cite{PKT15}, allocating a maximal time budget of 600 seconds for all search algorithms, while recording intermediate results to determine the minimized input sets for smaller time budgets.}

\subsubsection{Evaluation Metrics}
\label{sec:aim:results:performance}

To reduce the impact of randomness in our experiment, each configuration and baseline was run 50 times on each system, obtaining one minimized input set for each run.
Moreover, for the sake of performance analysis, we also recorded the  execution time required by \aim to generate minimized input sets.
The purpose of the metrics we use for RQ1 and RQ2 is to determine the \quotes{best} configuration to run \aim on a given system.
But one cannot know, before experimenting with the target system, which configuration would be the \quotes{best} for this system.
Based on the results from \jenkins and \joomla (see \Cref{sec:aim:aim:benchmarks}), we determine the overall \quotes{best} configuration, to be recommended as default for a new system.

\TSE{3.1.1, 3.2}{For RQ1, we consider the vulnerabilities described in \Cref{table:subjectVulnrabilities} for \jenkins and in \Cref{table:vulnerabilitiesJoomla} for \joomla.
We manually investigated the results of the initial input sets to identify the inputs capable of detecting vulnerabilities in the systems under test, so that we can map inputs to vulnerabilities. For each system, we consider that a vulnerability is detected by a minimized input set if it contains at least one input or pair of inputs able to trigger this vulnerability.}
For the first two \joomla vulnerabilities requiring pairs of inputs, the vulnerability is detected if both inputs are present in the input set.
Hence, for each configuration or baseline, our metric is the \emph{vulnerability detection rate (VDR)}, i.e., the total number of vulnerabilities detected by the minimized input sets obtained for the 50 runs, divided by the total number of vulnerabilities detected by the corresponding initial input sets.
If VDR is $100\%$, then we say the configuration or baseline leads to full vulnerability coverage.
The overall \quotes{best} configuration for \jenkins and \joomla, regarding vulnerability detection, should have a large VDR for both systems, ideally 100\%.
For each system, we reject configurations and baselines which do not lead to full vulnerability coverage, and then compare the remaining ones to answer RQ2.

RQ2 aims at evaluating the effectiveness of \aim in minimizing the initial input set.
Our goal is to identify the \aim configuration generating minimized input sets leading to the minimal execution time for the 76 considered MRs, across the two case studies, and reporting on the execution time saved, compared to executing \mstWi on the full input set.
But, to have a fair comparison between MRs execution time obtained respectively with the initial and minimized input sets, we have to take into account the \aim execution time required to minimize the initial input set.
Thus, the input set minimization effectiveness is quantified as the sum of \aim execution time to obtain the minimized input set plus MRs execution time with the minimized input set, divided by that of the initial input set.
\TSE{3.3.1}{However, since MR execution time is usually large, we cannot collect the time required to execute our 76 MRs on all the input sets generated by all \aim configurations. \TSE{3.3.1}{ We estimate it would take thousands of hours for the 1800 runs, resulting from 18 configurations $\times$ 50 repetitions $\times$ 2 case study subjects.}
For this reason, we rely on three additional metrics, that can be inferred without executing MRs, to identify the \quotes{best} configuration.
Then, we report on the input set minimization effectiveness obtained by such configuration.
Further, to keep the experiment within feasible computation resources, amongst the 50 minimized input sets of the best configuration, we select one with a median cost (\Cref{sec:aim:goals,sec:aim:cost}), to be representative of the 50 runs.}

To determine the \quotes{best} \aim configuration, we 
consider the size of the generated input set (i.e., the number of inputs in it), its cost (\Cref{sec:aim:goals,sec:aim:cost}), and the time required by the configuration to generate results. 
Input set size is a direct measure of effectiveness, while cost is an indirect measure, specific to our approach, that is linearly correlated with MR execution time (\Cref{sec:aim:background:MST}).
For these three metrics (size, cost, \aim execution time), we compare, for each system, the \aim configurations and baselines leading to full vulnerability coverage.
More precisely, for each metric, we denote by $\configRandomVar_i$ the value of the metric obtained for the $i^\text{th}$ approach (\aim configurations or baseline); the 50 runs of approach $i$ leading to a sample containing 50 data points.

To compare two samples for $\configRandomVar_1$ and $\configRandomVar_2$, we perform a Mann-Whitney-Wilcoxon test, which is recommended to assess differences in stochastic order for software engineering data analysis~\cite{AB14}.
This is a non-parametric test of the null hypothesis that $\proba{}{\configRandomVar_1 > \configRandomVar_2} = \proba{}{\configRandomVar_1 < \configRandomVar_2}$, i.e., $\configRandomVar_1$ and $\configRandomVar_2$ are stochastically equal~\cite{VD00}.
Hence, from $\configRandomVar_1$ and $\configRandomVar_2$ samples, we obtain the p-value $\pValue$ indicating how likely is the observation of these samples, assuming that $\configRandomVar_1$ and $\configRandomVar_2$ are stochastically equal.
If $\pValue \le 0.05$, we consider it is unlikely that $\configRandomVar_1$ and $\configRandomVar_2$ are stochastically equal.

To assess practical significance, we also consider a metric for effect size.
An equivalent reformulation of the null hypothesis is $\proba{}{\configRandomVar_1 > \configRandomVar_2} + 0.5\times\proba{}{\configRandomVar_1 = \configRandomVar_2} = 0.5$, which can be estimated by counting in the samples the number of times a value for $\configRandomVar_1$ is larger than a value for $\configRandomVar_2$ (ties counting for 0.5), then by dividing by the number of comparisons.
That way, we obtain the Vargha and Delaney's $\vdaLong$ metric~\cite{VD00} which, for the sake of conciseness, we simply denote $\vda$ in  \Cref{tab:jenkins:duels:sizes,tab:joomla:duels:sizes,tab:jenkins:duels:costs,tab:joomla:duels:costs,tab:jenkins:duels:times,tab:joomla:duels:times}.
$\vda$ is considered to be a robust metric for representing effect size in the context of non-parametric methods~\cite{KMB+17}.
$\vda$ ranges from $0$ to $1$, where $\vda = 0$ indicates that $\proba{}{\configRandomVar_1 < \configRandomVar_2} = 1$, $\vda = 0.5$ indicates that $\proba{}{\configRandomVar_1 > \configRandomVar_2} = \proba{}{\configRandomVar_1 < \configRandomVar_2}$, and $\vda = 1$ indicates that $\proba{}{\configRandomVar_1 > \configRandomVar_2} = 1$.

\TSE{2.2}{RQ3 aims at determining the contribution of \geneticAlgo to input set minimization and comparing it with alternative approaches.
We use the results of the double-clustering and problem reductions steps to obtain reduced input sets (\Cref{sec:aim:aim:benchmarks}) from the 50 runs of the \quotes{best} configuration.
For each reduced input set, each considered algorithm (\geneticAlgo or a baseline) is executed to obtain the corresponding minimized input set.
Then, the minimized input sets are checked to determine if the algorithms lead to full vulnerability coverage for this run, and their cost (\Cref{sec:aim:goals,sec:aim:cost}) is recorded.
Again, we denote by $\configRandomVar_i$ the cost for the $i^\text{th}$ approach. The 50 runs of approach $i$ yield a sample containing 50 data points.
As opposed to RQ2, where any run from a configuration/baseline can be compared to any run of another configuration/baseline, we want to compare the cost of the $n^\text{th}$ minimized input set for an algorithm to the cost of the $n^\text{th}$ minimized input set of another algorithm, since they are both obtained from the same $n^\text{th}$ reduced input set.
Hence, to compare two samples for $\configRandomVar_1$ and $\configRandomVar_2$, we perform a Wilcoxon signed-rank test, which is a non-parametric paired test~\cite{AB14}, with a level of significance of $0.05$.}

%Since we want to determine the smaller cost, we perform this test for the null hypothesis that $\configRandomVar_1 \ge \configRandomVar_2$.
%If $\pValue \le 0.05$, we reject the null hypothesis in favor of the alternative hypothesis $\configRandomVar_1 < \configRandomVar_2$.}

\TSE{2.2}{To assess practical significance, we also consider a metric for effect size.
Metrics for this test are often defined in terms of the positive-rank sum $\tPlus$ and the negative-rank sum $\tMinus$~\cite{Dem06,Ker14}.
Similarly to the Vargha and Delaney's $\vdaLong$ metric~\cite{VD00} we used for the Mann-Whitney-Wilcoxon test to answer RQ2, we use for RQ3 the effect size $\effectSize \eqdef \frac{\tPlus}{\tPlus + \tMinus}$, so that $\effectSize$ ranges from $0$ to $1$, where $\effectSize = 0$ indicates that $\configRandomVar_1 < \configRandomVar_2$ in every case and $\effectSize = 1$ indicates that $\configRandomVar_1 > \configRandomVar_2$ in every case.}

\subsection{Empirical Results}
\label{sec:aim:results:results}

We first describe the system configurations used to obtain our results (\Cref{sec:aim:results:systemConfig}).
To answer RQ1, we report the VDR associated with the obtained minimized input sets (\Cref{sec:aim:results:detectedVulnerabilities}).
Then, we describe the 
%magnitude
effectiveness of the input set reduction of the whole \aim approach to answer RQ2 (\Cref{sec:aim:results:magnitudeOfReduction}) and of the \geneticAlgo component to answer RQ3  (\Cref{sec:aim:results:geneticAlgos}).

\subsubsection{System Configurations}
\label{sec:aim:results:systemConfig}

We performed all the experiments on a system with the following configurations: a virtual machine installed on professional desktop PCs (Dell G7 7500, RAM 16Gb, Intel(R) Core(TM) i9-10885H CPU @ 2.40GHz) and terminal access to a shared remote server with Intel(R) Xeon(R) Gold 6234 CPU (3.30GHz) and 8 CPU cores.

\subsubsection{RQ1 - Detected Vulnerabilities}
\label{sec:aim:results:detectedVulnerabilities}

Results are presented in \Cref{tab:vulnsCoverage}.
Configurations and baselines that lead to full vulnerability coverage for both systems are in green, in yellow if they lead to full vulnerability coverage for one system, and in red if they never lead to full vulnerability coverage.
\TSE{3.2}{ As shown in \Cref{table:subjectVulnrabilities} and \Cref{table:vulnerabilitiesJoomla}, there are 9 vulnerabilities in \jenkins and 3 vulnerabilities in \joomla. Each \aim configuration is executed 50 times to reduce the effect of randomness in our experiments. We consider an \aim configuration to achieve full vulnerability coverage on \jenkins if it achieves $9*50 = 450$ vulnerability detections across all runs. Similarly, full vulnerability coverage on \joomla across all runs is reached if the configuration achieves $3*50 = 150$ vulnerability detections.}
\TSE{3.1.3}{
The execution time of the \aim configurations ranges from 15 to 24 minutes on \jenkins and from 25 to 47 minutes on \joomla.
We conclude that such variation across configurations is not significant compared to the time required to execute MRs.}

 % The execution time of the \aim configurations ranges from 15 to 24 minutes on \jenkins and from 25 to 47 minutes on \joomla, which is considerably smaller than the time required to execute MRs. Moreover, there is not much variation across configurations for a given input set.

\begin{table}[t]
\centering
\scriptsize
\caption{Coverage of the \jenkins and \joomla vulnerabilities after 50 runs of each configuration and baseline.}
\label{tab:vulnsCoverage}
\begin{tabular}{|c|c|c|c|c|}
\hline
    Vulnerability
    & \multicolumn{4}{c|}{System Under Test}\\
\cline{2-5}
    Coverage
    & \multicolumn{2}{c|}{ \jenkins}
    & \multicolumn{2}{c|}{ \joomla}\\
\hline
    Configurations & Nb of detected & \multirow{2}{*}{VDR} & Nb of detected & \multirow{2}{*}{VDR} \\
    or baselines & vulnerabilities & & vulnerabilities & \\
\hline
\cellcolor{yellow!50} \LevKmeansKmeans & 
\cellcolor{green!50} 450 & \cellcolor{green!50} 100.0\% & \cellcolor{red!50} 146 & \cellcolor{red!50} 97.3\% \\
\cellcolor{red!50} \LevKmeansDbscan & \cellcolor{red!50} 371 & \cellcolor{red!50} 82.4\% & \cellcolor{red!50} 50 & \cellcolor{red!50} 33.3\% \\
\cellcolor{yellow!50} \LevKmeansHdbscan & \cellcolor{red!50} 379 & \cellcolor{red!50} 84.2\% & \cellcolor{green!50} 150 & \cellcolor{green!50} 100.0\% \\
\cellcolor{green!50} \LevDbscanKmeans & \cellcolor{green!50} 450 & \cellcolor{green!50} 100.0\% & \cellcolor{green!50} 150 & \cellcolor{green!50} 100.0\% \\
\cellcolor{red!50} \LevDbscanDbscan & \cellcolor{red!50} 400 & \cellcolor{red!50} 88.9\% & \cellcolor{red!50} 50 & \cellcolor{red!50} 33.3\% \\
\cellcolor{red!50} \LevDbscanHdbscan & \cellcolor{red!50} 400 & \cellcolor{red!50} 88.9\% & \cellcolor{red!50} 50 & \cellcolor{red!50} 33.3\% \\
\cellcolor{yellow!50} \LevHdbscanKmeans & \cellcolor{green!50} 450 & \cellcolor{green!50} 100.0\% & \cellcolor{red!50} 100 & \cellcolor{red!50} 66.7\% \\
\cellcolor{red!50} \LevHdbscanDbscan & \cellcolor{red!50} 403 & \cellcolor{red!50} 89.6\% & \cellcolor{red!50} 100 & \cellcolor{red!50} 66.7\% \\
\cellcolor{red!50} \LevHdbscanHdbscan & \cellcolor{red!50} 447 & \cellcolor{red!50} 99.3\% & \cellcolor{red!50} 100 & \cellcolor{red!50} 66.7\% \\
\hline
\cellcolor{yellow!50} \BagKmeansKmeans & \cellcolor{green!50} 450 & \cellcolor{green!50} 100.0\% & \cellcolor{red!50} 133 & \cellcolor{red!50} 88.7\% \\
\cellcolor{red!50} \BagKmeansDbscan & \cellcolor{red!50} 403 & \cellcolor{red!50} 89.6\% & \cellcolor{red!50} 50 & \cellcolor{red!50} 33.3\% \\
\cellcolor{yellow!50} \BagKmeansHdbscan & \cellcolor{red!50} 410 & \cellcolor{red!50} 91.1\% & \cellcolor{green!50} 150 & \cellcolor{green!50} 100.0\% \\
\cellcolor{green!50} \BagDbscanKmeans & \cellcolor{green!50} 450 & \cellcolor{green!50} 100.0\% & \cellcolor{green!50} 150 & \cellcolor{green!50} 100.0\% \\
\cellcolor{red!50} \BagDbscanDbscan & \cellcolor{red!50} 338 & \cellcolor{red!50} 75.1\% & \cellcolor{red!50} 50 & \cellcolor{red!50} 33.3\% \\
\cellcolor{yellow!50} \BagDbscanHdbscan & \cellcolor{green!50} 450 & \cellcolor{green!50} 100.0\% & \cellcolor{red!50} 50 & \cellcolor{red!50} 33.3\% \\
\cellcolor{yellow!50} \BagHdbscanKmeans & \cellcolor{green!50} 450 & \cellcolor{green!50} 100.0\% & \cellcolor{red!50} 100 & \cellcolor{red!50} 66.7\% \\
\cellcolor{red!50} \BagHdbscanDbscan & \cellcolor{red!50} 404 & \cellcolor{red!50} 89.8\% & \cellcolor{red!50} 100 & \cellcolor{red!50} 66.7\% \\
\cellcolor{red!50} \BagHdbscanHdbscan & \cellcolor{red!50} 428 & \cellcolor{red!50} 95.1\% & \cellcolor{red!50} 100 & \cellcolor{red!50} 66.7\% \\
\hline
\cellcolor{red!50} \Rt & \cellcolor{red!50} 339 & \cellcolor{red!50} 75.3\% & \cellcolor{red!50} 74 & \cellcolor{red!50} 49.3\% \\
\cellcolor{red!50} \ArtKmeans & \cellcolor{red!50} 447 & \cellcolor{red!50} 99.3\% & \cellcolor{red!50} 125 & \cellcolor{red!50} 83.3\% \\
\cellcolor{red!50} \ArtDbscan & \cellcolor{red!50} 77 & \cellcolor{red!50} 17.1\% & \cellcolor{red!50} 22 & \cellcolor{red!50} 14.7\% \\
\cellcolor{red!50} \ArtHdbscan & \cellcolor{red!50} 350 & \cellcolor{red!50} 77.8\% & \cellcolor{red!50} 68 & \cellcolor{red!50} 45.3\% \\
\hline
\end{tabular}
\end{table}

First, note that \textbf{the choice of distance function for output clustering does not have a significant impact on vulnerability coverage}.
Indeed, apart from \LevDbscanHdbscan and \BagDbscanHdbscan, the results using the Levenshtein or Bag distances are fairly similar (e.g., both LKK and BKK discover 450 vulnerabilities in Jenkins) and seem to only depend on the choice of clustering algorithms.
This indicates that the order of words in a Web page is not a relevant distinction when performing clustering for vulnerability coverage.
Considering now \LevDbscanHdbscan and \BagDbscanHdbscan, taking into account the order of words can even be detrimental, since they perform equally poorly for \joomla but they differ for \jenkins, where only \BagDbscanHdbscan leads to full vulnerability coverage.

Second, \textbf{the choice of clustering algorithm for action clustering seems to be the main factor determining vulnerability coverage}.
Configurations using \dbscan as algorithm for the action cluste\-ring step never lead to full vulnerability coverage for any system.
This indicates that this clustering algorithm poorly fits the data in the input space.
This is confirmed by the results obtained for the \ArtDbscan baseline, which only uses \dbscan on the input space and performs the worst (amongst baselines and \aim configurations) regarding vulnerability coverage.
After investigation, the minimized input sets acquired for \ArtDbscan are much smaller compared to those obtained for the other baseline methods. 
These results cannot be explained by the hyper-parameter as we employed a large range of values (\Cref{sec:aim:results:configurations}).
We conjecture that \dbscan merges together many action clusters even when the URLs involved in these actions are distinct.

On the other hand, \textbf{configurations using \kmeans for the action clustering step always lead to full vulnerability coverage for \jenkins and lead to the largest vulnerability coverage for \joomla}.
This is confirmed by the results obtained for the \ArtKmeans baseline, which only uses \kmeans on the input space and performs the best (amongst baselines) regarding vulnerability coverage.
Indeed, even if this configuration does not lead to full vulnerability coverage, it is very close.
In fact, even if it tends to perform worse than \aim configurations that use \kmeans for action clustering, it tends to perform better than \aim configurations that do not use \kmeans for action clustering.
The success of \kmeans in achieving better vulnerability coverage on these datasets can be attributed to its ability to handle well-separated clusters.
In our case, these clusters are well-separated because of the distinct URLs occurring in the datasets.

Finally, \textbf{no baseline reached full vulnerability coverage}.
On top of the already mentioned \ArtKmeans and \ArtDbscan baselines, \ArtHdbscan performed similarly to random testing (\Rt), indicating that the effect of the \hdbscan algorithm for action clustering is neutral.
The only \aim configuration that performed worse than random testing is \BagDbscanDbscan, combining \dbscan (as mentioned before, the worst clustering algorithm regarding vulnerability detection) for both output and action clustering with Bag distance.
Only \LevDbscanKmeans and \BagDbscanKmeans lead to full vulnerability coverage for both \jenkins and \joomla, and hence are our candidate \quotes{best} configurations in terms of VDR.
The combination of \dbscan and \kmeans was very effective on our dataset since \dbscan was able to identify dense regions of outputs and \kmeans allowed for further refinement, forming well-defined action clusters based on URLs.

\subsubsection{RQ2 - Input Set Reduction Effectiveness}
\label{sec:aim:results:magnitudeOfReduction}

To answer RQ2 on the effectiveness of minimization, we compare the input set reduction of baselines and configurations for both \jenkins and \joomla.
Amongst them, only the \LevKmeansKmeans, \LevDbscanKmeans, \LevHdbscanKmeans, \BagKmeansKmeans, \BagDbscanKmeans, \BagDbscanHdbscan, and \BagHdbscanKmeans configurations lead to full vulnerability coverage for \jenkins.
Their input set sizes are compared in \Cref{tab:jenkins:duels:sizes}, their costs in \Cref{tab:jenkins:duels:costs}, and their \aim execution time in \Cref{tab:jenkins:duels:times}.
Similarly, only the \LevKmeansHdbscan, \LevDbscanKmeans, \BagKmeansHdbscan, and \BagDbscanKmeans configurations lead to full vulnerability coverage for \joomla.
Their input set sizes are compared in \Cref{tab:joomla:duels:sizes}, their costs in \Cref{tab:joomla:duels:costs}, and their \aim execution time in \Cref{tab:joomla:duels:times}.
Configurations with full vulnerability coverage for both \jenkins and \joomla (i.e., \LevDbscanKmeans and \BagDbscanKmeans) are in bold.

\begin{table*}[t]
\centering
\scriptsize
\caption{Comparison of \jenkins input set sizes for configurations with full vulnerability coverage.}
\label{tab:jenkins:duels:sizes}
\begin{tabular}{cc|c|c|c|c|c|c|c|}
\hline
\multicolumn{2}{|c|}{sizes} & \LevKmeansKmeans & \textbf{\LevDbscanKmeans} & \LevHdbscanKmeans & \BagKmeansKmeans & \textbf{\BagDbscanKmeans} & \BagDbscanHdbscan & \BagHdbscanKmeans \\
\hline
\multicolumn{1}{|c|}{\multirow{2}{*}{\LevKmeansKmeans}} & $\pValue$
	 & 
	 & \cellcolor{green!91}\num{5.4e-15}
	 & \num{5.1e-01}
	 & \num{4.2e-01}
	 & \num{8.8e-01}
	 & \cellcolor{red!100}\num{3.1e-20}
	 & \num{7.2e-01}\\
\multicolumn{1}{|c|}{} & $\vda$
	 & 
	 & \cellcolor{green!91}0.05
	 & 0.54
	 & 0.55
	 & 0.49
	 & \cellcolor{red!100}1.0
	 & 0.48\\
\hline
\multicolumn{1}{|c|}{\multirow{2}{*}{\textbf{\LevDbscanKmeans}}} & $\pValue$
	 & \cellcolor{red!91}\num{5.4e-15}
	 & 
	 & \cellcolor{red!99}\num{1.8e-17}
	 & \cellcolor{red!93}\num{1.4e-15}
	 & \cellcolor{red!89}\num{1.8e-14}
	 & \cellcolor{red!100}\num{3.2e-20}
	 & \cellcolor{red!88}\num{3.8e-14}\\
\multicolumn{1}{|c|}{} & $\vda$
	 & \cellcolor{red!91}0.95
	 & 
	 & \cellcolor{red!99}0.99
	 & \cellcolor{red!93}0.96
	 & \cellcolor{red!89}0.94
	 & \cellcolor{red!100}1.0
	 & \cellcolor{red!88}0.94\\
\hline
\multicolumn{1}{|c|}{\multirow{2}{*}{\LevHdbscanKmeans}} & $\pValue$
	 & \num{5.1e-01}
	 & \cellcolor{green!99}\num{1.8e-17}
	 & 
	 & \num{9.8e-01}
	 & \num{3.4e-01}
	 & \cellcolor{red!100}\num{3.0e-20}
	 & \num{1.6e-01}\\
\multicolumn{1}{|c|}{} & $\vda$
	 & 0.46
	 & \cellcolor{green!99}0.01
	 & 
	 & 0.5
	 & 0.44
	 & \cellcolor{red!100}1.0
	 & 0.42\\
\hline
\multicolumn{1}{|c|}{\multirow{2}{*}{\BagKmeansKmeans}} & $\pValue$
	 & \num{4.2e-01}
	 & \cellcolor{green!93}\num{1.4e-15}
	 & \num{9.8e-01}
	 & 
	 & \num{4.1e-01}
	 & \cellcolor{red!100}\num{3.2e-20}
	 & \num{2.1e-01}\\
\multicolumn{1}{|c|}{} & $\vda$
	 & 0.45
	 & \cellcolor{green!93}0.04
	 & 0.5
	 & 
	 & 0.45
	 & \cellcolor{red!100}1.0
	 & 0.43\\
\hline
\multicolumn{1}{|c|}{\multirow{2}{*}{\textbf{\BagDbscanKmeans}}} & $\pValue$
	 & \num{8.8e-01}
	 & \cellcolor{green!89}\num{1.8e-14}
	 & \num{3.4e-01}
	 & \num{4.1e-01}
	 & 
	 & \cellcolor{red!100}\num{3.2e-20}
	 & \num{6.9e-01}\\
\multicolumn{1}{|c|}{} & $\vda$
	 & 0.51
	 & \cellcolor{green!89}0.06
	 & 0.56
	 & 0.55
	 & 
	 & \cellcolor{red!100}1.0
	 & 0.48\\
\hline
\multicolumn{1}{|c|}{\multirow{2}{*}{\BagDbscanHdbscan}} & $\pValue$
	 & \cellcolor{green!100}\num{3.1e-20}
	 & \cellcolor{green!100}\num{3.2e-20}
	 & \cellcolor{green!100}\num{3.0e-20}
	 & \cellcolor{green!100}\num{3.2e-20}
	 & \cellcolor{green!100}\num{3.2e-20}
	 & 
	 & \cellcolor{green!100}\num{3.2e-20}\\
\multicolumn{1}{|c|}{} & $\vda$
	 & \cellcolor{green!100}0.0
	 & \cellcolor{green!100}0.0
	 & \cellcolor{green!100}0.0
	 & \cellcolor{green!100}0.0
	 & \cellcolor{green!100}0.0
	 & 
	 & \cellcolor{green!100}0.0\\
\hline
\multicolumn{1}{|c|}{\multirow{2}{*}{\BagHdbscanKmeans}} & $\pValue$
	 & \num{7.2e-01}
	 & \cellcolor{green!88}\num{3.8e-14}
	 & \num{1.6e-01}
	 & \num{2.1e-01}
	 & \num{6.9e-01}
	 & \cellcolor{red!100}\num{3.2e-20}
	 & \\
\multicolumn{1}{|c|}{} & $\vda$
	 & 0.52
	 & \cellcolor{green!88}0.06
	 & 0.58
	 & 0.57
	 & 0.52
	 & \cellcolor{red!100}1.0
	 & \\
\hline
\end{tabular}
\end{table*}

\begin{table*}[t]
\centering
\scriptsize
\caption{Comparison of \jenkins input set costs for configurations with full vulnerability coverage.}
\label{tab:jenkins:duels:costs}
\begin{tabular}{cc|c|c|c|c|c|c|c|}
\hline
\multicolumn{2}{|c|}{costs} & \LevKmeansKmeans & \textbf{\LevDbscanKmeans} & \LevHdbscanKmeans & \BagKmeansKmeans & \textbf{\BagDbscanKmeans} & \BagDbscanHdbscan & \BagHdbscanKmeans \\
\hline
\multicolumn{1}{|c|}{\multirow{2}{*}{\LevKmeansKmeans}} & $\pValue$
	 & 
	 & \cellcolor{green!95}\num{2.5e-16}
	 & \cellcolor{red!47}\num{5.9e-05}
	 & \cellcolor{red!28}\num{1.5e-02}
	 & \cellcolor{red!33}\num{4.4e-03}
	 & \cellcolor{red!100}\num{4.1e-18}
	 & \cellcolor{red!32}\num{5.4e-03}\\
\multicolumn{1}{|c|}{} & $\vda$
	 & 
	 & \cellcolor{green!95}0.02
	 & \cellcolor{red!47}0.73
	 & \cellcolor{red!28}0.64
	 & \cellcolor{red!33}0.67
	 & \cellcolor{red!100}1.0
	 & \cellcolor{red!32}0.66\\
\hline
\multicolumn{1}{|c|}{\multirow{2}{*}{\textbf{\LevDbscanKmeans}}} & $\pValue$
	 & \cellcolor{red!95}\num{2.5e-16}
	 & 
	 & \cellcolor{red!100}\num{7.0e-18}
	 & \cellcolor{red!100}\num{9.5e-18}
	 & \cellcolor{red!100}\num{7.0e-18}
	 & \cellcolor{red!100}\num{4.1e-18}
	 & \cellcolor{red!100}\num{7.0e-18}\\
\multicolumn{1}{|c|}{} & $\vda$
	 & \cellcolor{red!95}0.98
	 & 
	 & \cellcolor{red!100}1.0
	 & \cellcolor{red!100}1.0
	 & \cellcolor{red!100}1.0
	 & \cellcolor{red!100}1.0
	 & \cellcolor{red!100}1.0\\
\hline
\multicolumn{1}{|c|}{\multirow{2}{*}{\LevHdbscanKmeans}} & $\pValue$
	 & \cellcolor{green!47}\num{5.9e-05}
	 & \cellcolor{green!100}\num{7.0e-18}
	 & 
	 & \num{1.0e-01}
	 & \num{2.4e-01}
	 & \cellcolor{red!100}\num{4.1e-18}
	 & \num{1.4e-01}\\
\multicolumn{1}{|c|}{} & $\vda$
	 & \cellcolor{green!47}0.27
	 & \cellcolor{green!100}0.0
	 & 
	 & 0.41
	 & 0.43
	 & \cellcolor{red!100}1.0
	 & 0.41\\
\hline
\multicolumn{1}{|c|}{\multirow{2}{*}{\BagKmeansKmeans}} & $\pValue$
	 & \cellcolor{green!28}\num{1.5e-02}
	 & \cellcolor{green!100}\num{9.5e-18}
	 & \num{1.0e-01}
	 & 
	 & \num{5.8e-01}
	 & \cellcolor{red!100}\num{4.1e-18}
	 & \num{6.1e-01}\\
\multicolumn{1}{|c|}{} & $\vda$
	 & \cellcolor{green!28}0.36
	 & \cellcolor{green!100}0.0
	 & 0.59
	 & 
	 & 0.53
	 & \cellcolor{red!100}1.0
	 & 0.53\\
\hline
\multicolumn{1}{|c|}{\multirow{2}{*}{\textbf{\BagDbscanKmeans}}} & $\pValue$
	 & \cellcolor{green!33}\num{4.4e-03}
	 & \cellcolor{green!100}\num{7.0e-18}
	 & \num{2.4e-01}
	 & \num{5.8e-01}
	 & 
	 & \cellcolor{red!100}\num{4.1e-18}
	 & \num{8.2e-01}\\
\multicolumn{1}{|c|}{} & $\vda$
	 & \cellcolor{green!33}0.33
	 & \cellcolor{green!100}0.0
	 & 0.57
	 & 0.47
	 & 
	 & \cellcolor{red!100}1.0
	 & 0.49\\
\hline
\multicolumn{1}{|c|}{\multirow{2}{*}{\BagDbscanHdbscan}} & $\pValue$
	 & \cellcolor{green!100}\num{4.1e-18}
	 & \cellcolor{green!100}\num{4.1e-18}
	 & \cellcolor{green!100}\num{4.1e-18}
	 & \cellcolor{green!100}\num{4.1e-18}
	 & \cellcolor{green!100}\num{4.1e-18}
	 & 
	 & \cellcolor{green!100}\num{4.1e-18}\\
\multicolumn{1}{|c|}{} & $\vda$
	 & \cellcolor{green!100}0.0
	 & \cellcolor{green!100}0.0
	 & \cellcolor{green!100}0.0
	 & \cellcolor{green!100}0.0
	 & \cellcolor{green!100}0.0
	 & 
	 & \cellcolor{green!100}0.0\\
\hline
\multicolumn{1}{|c|}{\multirow{2}{*}{\BagHdbscanKmeans}} & $\pValue$
	 & \cellcolor{green!32}\num{5.4e-03}
	 & \cellcolor{green!100}\num{7.0e-18}
	 & \num{1.4e-01}
	 & \num{6.1e-01}
	 & \num{8.2e-01}
	 & \cellcolor{red!100}\num{4.1e-18}
	 & \\
\multicolumn{1}{|c|}{} & $\vda$
	 & \cellcolor{green!32}0.34
	 & \cellcolor{green!100}0.0
	 & 0.59
	 & 0.47
	 & 0.51
	 & \cellcolor{red!100}1.0
	 & \\
\hline
\end{tabular}
\end{table*}

\begin{table*}[t]
\centering
\scriptsize
\caption{Comparison of \jenkins \aim execution times for configurations with full vulnerability coverage.}
\label{tab:jenkins:duels:times}
\begin{tabular}{cc|c|c|c|c|c|c|c|}
\hline
\multicolumn{2}{|c|}{times} & \LevKmeansKmeans & \textbf{\LevDbscanKmeans} & \LevHdbscanKmeans & \BagKmeansKmeans & \textbf{\BagDbscanKmeans} & \BagDbscanHdbscan & \BagHdbscanKmeans \\
\hline
\multicolumn{1}{|c|}{\multirow{2}{*}{\LevKmeansKmeans}} & $\pValue$
	 & 
	 & \cellcolor{red!55}\num{1.5e-06}
	 & \cellcolor{red!79}\num{1.0e-11}
	 & \cellcolor{red!27}\num{2.0e-02}
	 & \cellcolor{red!47}\num{3.8e-05}
	 & \cellcolor{red!100}\num{3.1e-18}
	 & \cellcolor{red!70}\num{1.3e-09}\\
\multicolumn{1}{|c|}{} & $\vda$
	 & 
	 & \cellcolor{red!55}0.78
	 & \cellcolor{red!79}0.89
	 & \cellcolor{red!27}0.63
	 & \cellcolor{red!47}0.74
	 & \cellcolor{red!100}1.0
	 & \cellcolor{red!70}0.85\\
\hline
\multicolumn{1}{|c|}{\multirow{2}{*}{\textbf{\LevDbscanKmeans}}} & $\pValue$
	 & \cellcolor{green!55}\num{1.5e-06}
	 & 
	 & \cellcolor{red!50}\num{1.2e-05}
	 & \cellcolor{green!33}\num{3.9e-03}
	 & \num{5.4e-01}
	 & \cellcolor{red!100}\num{2.6e-18}
	 & \cellcolor{red!37}\num{1.1e-03}\\
\multicolumn{1}{|c|}{} & $\vda$
	 & \cellcolor{green!55}0.22
	 & 
	 & \cellcolor{red!50}0.75
	 & \cellcolor{green!33}0.33
	 & 0.54
	 & \cellcolor{red!100}1.0
	 & \cellcolor{red!37}0.69\\
\hline
\multicolumn{1}{|c|}{\multirow{2}{*}{\LevHdbscanKmeans}} & $\pValue$
	 & \cellcolor{green!79}\num{1.0e-11}
	 & \cellcolor{green!50}\num{1.2e-05}
	 & 
	 & \cellcolor{green!69}\num{1.7e-09}
	 & \cellcolor{green!37}\num{1.4e-03}
	 & \cellcolor{red!96}\num{3.2e-17}
	 & \num{8.7e-01}\\
\multicolumn{1}{|c|}{} & $\vda$
	 & \cellcolor{green!79}0.11
	 & \cellcolor{green!50}0.25
	 & 
	 & \cellcolor{green!69}0.15
	 & \cellcolor{green!37}0.32
	 & \cellcolor{red!96}0.98
	 & 0.49\\
\hline
\multicolumn{1}{|c|}{\multirow{2}{*}{\BagKmeansKmeans}} & $\pValue$
	 & \cellcolor{green!27}\num{2.0e-02}
	 & \cellcolor{red!33}\num{3.9e-03}
	 & \cellcolor{red!69}\num{1.7e-09}
	 & 
	 & \cellcolor{red!30}\num{8.0e-03}
	 & \cellcolor{red!100}\num{3.0e-18}
	 & \cellcolor{red!56}\num{6.7e-07}\\
\multicolumn{1}{|c|}{} & $\vda$
	 & \cellcolor{green!27}0.37
	 & \cellcolor{red!33}0.67
	 & \cellcolor{red!69}0.85
	 & 
	 & \cellcolor{red!30}0.65
	 & \cellcolor{red!100}1.0
	 & \cellcolor{red!56}0.79\\
\hline
\multicolumn{1}{|c|}{\multirow{2}{*}{\textbf{\BagDbscanKmeans}}} & $\pValue$
	 & \cellcolor{green!47}\num{3.8e-05}
	 & \num{5.4e-01}
	 & \cellcolor{red!37}\num{1.4e-03}
	 & \cellcolor{green!30}\num{8.0e-03}
	 & 
	 & \cellcolor{red!97}\num{1.5e-17}
	 & \cellcolor{red!28}\num{1.1e-02}\\
\multicolumn{1}{|c|}{} & $\vda$
	 & \cellcolor{green!47}0.26
	 & 0.46
	 & \cellcolor{red!37}0.68
	 & \cellcolor{green!30}0.35
	 & 
	 & \cellcolor{red!97}0.99
	 & \cellcolor{red!28}0.65\\
\hline
\multicolumn{1}{|c|}{\multirow{2}{*}{\BagDbscanHdbscan}} & $\pValue$
	 & \cellcolor{green!100}\num{3.1e-18}
	 & \cellcolor{green!100}\num{2.6e-18}
	 & \cellcolor{green!96}\num{3.2e-17}
	 & \cellcolor{green!100}\num{3.0e-18}
	 & \cellcolor{green!97}\num{1.5e-17}
	 & 
	 & \cellcolor{green!92}\num{6.3e-16}\\
\multicolumn{1}{|c|}{} & $\vda$
	 & \cellcolor{green!100}0.0
	 & \cellcolor{green!100}0.0
	 & \cellcolor{green!96}0.02
	 & \cellcolor{green!100}0.0
	 & \cellcolor{green!97}0.01
	 & 
	 & \cellcolor{green!92}0.04\\
\hline
\multicolumn{1}{|c|}{\multirow{2}{*}{\BagHdbscanKmeans}} & $\pValue$
	 & \cellcolor{green!70}\num{1.3e-09}
	 & \cellcolor{green!37}\num{1.1e-03}
	 & \num{8.7e-01}
	 & \cellcolor{green!56}\num{6.7e-07}
	 & \cellcolor{green!28}\num{1.1e-02}
	 & \cellcolor{red!92}\num{6.3e-16}
	 & \\
\multicolumn{1}{|c|}{} & $\vda$
	 & \cellcolor{green!70}0.15
	 & \cellcolor{green!37}0.31
	 & 0.51
	 & \cellcolor{green!56}0.21
	 & \cellcolor{green!28}0.35
	 & \cellcolor{red!92}0.96
	 & \\
\hline
\end{tabular}
\end{table*}

\begin{table}[t]
\centering
\scriptsize
\caption{Comparison of \joomla input set sizes for configurations with full vulnerability coverage.}
\label{tab:joomla:duels:sizes}
\begin{tabular}{cc|c|c|c|c|}
\hline
\multicolumn{2}{|c|}{sizes} & \LevKmeansHdbscan & \textbf{\LevDbscanKmeans} & \BagKmeansHdbscan & \textbf{\BagDbscanKmeans} \\
\hline
\multicolumn{1}{|c|}{\multirow{2}{*}{\LevKmeansHdbscan}} & $\pValue$
	 & 
	 & \cellcolor{green!88}\num{4.8e-15}
	 & \num{1.3e-01}
	 & \cellcolor{green!88}\num{2.9e-16}\\
\multicolumn{1}{|c|}{} & $\vda$
	 & 
	 & \cellcolor{green!88}0.06
	 & 0.42
	 & \cellcolor{green!88}0.06\\
\hline
\multicolumn{1}{|c|}{\multirow{2}{*}{\textbf{\LevDbscanKmeans}}} & $\pValue$
	 & \cellcolor{red!88}\num{4.8e-15}
	 & 
	 & \cellcolor{red!88}\num{1.1e-14}
	 & \cellcolor{red!88}\num{4.5e-16}\\
\multicolumn{1}{|c|}{} & $\vda$
	 & \cellcolor{red!88}0.94
	 & 
	 & \cellcolor{red!88}0.94
	 & \cellcolor{red!88}0.94\\
\hline
\multicolumn{1}{|c|}{\multirow{2}{*}{\BagKmeansHdbscan}} & $\pValue$
	 & \num{1.3e-01}
	 & \cellcolor{green!88}\num{1.1e-14}
	 & 
	 & \cellcolor{green!88}\num{7.4e-16}\\
\multicolumn{1}{|c|}{} & $\vda$
	 & 0.58
	 & \cellcolor{green!88}0.06
	 & 
	 & \cellcolor{green!88}0.06\\
\hline
\multicolumn{1}{|c|}{\multirow{2}{*}{\textbf{\BagDbscanKmeans}}} & $\pValue$
	 & \cellcolor{red!88}\num{2.9e-16}
	 & \cellcolor{green!88}\num{4.5e-16}
	 & \cellcolor{red!88}\num{7.4e-16}
	 & \\
\multicolumn{1}{|c|}{} & $\vda$
	 & \cellcolor{red!88}0.94
	 & \cellcolor{green!88}0.06
	 & \cellcolor{red!88}0.94
	 & \\
\hline
\end{tabular}
\end{table}

\begin{table}[t]
\centering
\scriptsize
\caption{Comparison of \joomla input set costs for configurations with full vulnerability coverage.}
\label{tab:joomla:duels:costs}
\begin{tabular}{cc|c|c|c|c|}
\hline
\multicolumn{2}{|c|}{costs} & \LevKmeansHdbscan & \textbf{\LevDbscanKmeans} & \BagKmeansHdbscan & \textbf{\BagDbscanKmeans} \\
\hline
\multicolumn{1}{|c|}{\multirow{2}{*}{\LevKmeansHdbscan}} & $\pValue$
	 & 
	 & \cellcolor{green!88}\num{1.3e-14}
	 & \num{6.9e-01}
	 & \cellcolor{green!90}\num{3.5e-15}\\
\multicolumn{1}{|c|}{} & $\vda$
	 & 
	 & \cellcolor{green!88}0.06
	 & 0.48
	 & \cellcolor{green!90}0.05\\
\hline
\multicolumn{1}{|c|}{\multirow{2}{*}{\textbf{\LevDbscanKmeans}}} & $\pValue$
	 & \cellcolor{red!88}\num{1.3e-14}
	 & 
	 & \cellcolor{red!88}\num{1.3e-14}
	 & \cellcolor{red!88}\num{7.0e-15}\\
\multicolumn{1}{|c|}{} & $\vda$
	 & \cellcolor{red!88}0.94
	 & 
	 & \cellcolor{red!88}0.94
	 & \cellcolor{red!88}0.94\\
\hline
\multicolumn{1}{|c|}{\multirow{2}{*}{\BagKmeansHdbscan}} & $\pValue$
	 & \num{6.9e-01}
	 & \cellcolor{green!88}\num{1.3e-14}
	 & 
	 & \cellcolor{green!90}\num{3.6e-15}\\
\multicolumn{1}{|c|}{} & $\vda$
	 & 0.52
	 & \cellcolor{green!88}0.06
	 & 
	 & \cellcolor{green!90}0.05\\
\hline
\multicolumn{1}{|c|}{\multirow{2}{*}{\textbf{\BagDbscanKmeans}}} & $\pValue$
	 & \cellcolor{red!90}\num{3.5e-15}
	 & \cellcolor{green!88}\num{7.0e-15}
	 & \cellcolor{red!90}\num{3.6e-15}
	 & \\
\multicolumn{1}{|c|}{} & $\vda$
	 & \cellcolor{red!90}0.95
	 & \cellcolor{green!88}0.06
	 & \cellcolor{red!90}0.95
	 & \\
\hline
\end{tabular}
\end{table}

\begin{table}[t]
\centering
\scriptsize
\caption{Comparison of \joomla \aim execution times for configurations with full vulnerability coverage.}
\label{tab:joomla:duels:times}
\begin{tabular}{cc|c|c|c|c|}
\hline
\multicolumn{2}{|c|}{times} & \LevKmeansHdbscan & \textbf{\LevDbscanKmeans} & \BagKmeansHdbscan & \textbf{\BagDbscanKmeans} \\
\hline
\multicolumn{1}{|c|}{\multirow{2}{*}{\LevKmeansHdbscan}} & $\pValue$
	 & 
	 & \cellcolor{green!100}\num{2.5e-18}
	 & \cellcolor{green!33}\num{3.6e-03}
	 & \cellcolor{green!100}\num{1.2e-18}\\
\multicolumn{1}{|c|}{} & $\vda$
	 & 
	 & \cellcolor{green!100}0.0
	 & \cellcolor{green!33}0.33
	 & \cellcolor{green!100}0.0\\
\hline
\multicolumn{1}{|c|}{\multirow{2}{*}{\textbf{\LevDbscanKmeans}}} & $\pValue$
	 & \cellcolor{red!100}\num{2.5e-18}
	 & 
	 & \cellcolor{red!100}\num{2.8e-18}
	 & \cellcolor{red!99}\num{1.5e-18}\\
\multicolumn{1}{|c|}{} & $\vda$
	 & \cellcolor{red!100}1.0
	 & 
	 & \cellcolor{red!100}1.0
	 & \cellcolor{red!99}1.0\\
\hline
\multicolumn{1}{|c|}{\multirow{2}{*}{\BagKmeansHdbscan}} & $\pValue$
	 & \cellcolor{red!33}\num{3.6e-03}
	 & \cellcolor{green!100}\num{2.8e-18}
	 & 
	 & \cellcolor{green!100}\num{1.3e-18}\\
\multicolumn{1}{|c|}{} & $\vda$
	 & \cellcolor{red!33}0.67
	 & \cellcolor{green!100}0.0
	 & 
	 & \cellcolor{green!100}0.0\\
\hline
\multicolumn{1}{|c|}{\multirow{2}{*}{\textbf{\BagDbscanKmeans}}} & $\pValue$
	 & \cellcolor{red!100}\num{1.2e-18}
	 & \cellcolor{green!99}\num{1.5e-18}
	 & \cellcolor{red!100}\num{1.3e-18}
	 & \\
\multicolumn{1}{|c|}{} & $\vda$
	 & \cellcolor{red!100}1.0
	 & \cellcolor{green!99}0.0
	 & \cellcolor{red!100}1.0
	 & \\
\hline
\end{tabular}
\end{table}

In these six tables, configurations in each row are compared with configurations in each column.
$\pValue$ denotes the statistical significance and $\vda$ the effect size (\Cref{sec:aim:results:performance}).
When $\pValue > 0.05$, we consider the metric values obtained from the two configurations not to be significantly different, and hence the cell is left white.
Otherwise, the cell is colored, either in green or red.
Since we consider input set size and cost and \aim execution time, the smaller the values the better.
Thus, green (resp. red) indicates that the configuration in the row is better (resp. worse) than the configuration in the column.
The intensity of the color is proportional to the effect size.
More precisely, the intensity is $\abs{\cliffsDelta}$, where $\cliffsDelta = 2\times\vda - 1$ is Cliff's delta~\cite{KMB+17}.
$\abs{\cliffsDelta}$ is a number between $0$ and $1$, where $0$ indicates the smallest intensity (the lightest color) and $1$ indicates the largest intensity (the darkest color).

For \jenkins, 
among the candidate best configurations (i.e., \LevDbscanKmeans and \BagDbscanKmeans), \textbf{\BagDbscanKmeans performed significantly better than \LevDbscanKmeans for input set size and cost}, and even if the difference is smaller for \aim execution time, the effect size is also in favor of \BagDbscanKmeans. As for the other configurations,
\Cref{tab:jenkins:duels:sizes} on input set sizes and \Cref{tab:jenkins:duels:costs} on input set costs consistently indicate that \BagDbscanHdbscan is the best configuration while \LevDbscanKmeans is the worst configuration.
The other configurations seem equivalent in terms of size.
Regarding cost, \LevKmeansKmeans tends to be the second to last configuration, the other configurations being equivalent.
Regarding \aim execution time in \Cref{tab:jenkins:duels:times}, the results are more nuanced, \BagDbscanHdbscan is again the best configuration, but this time \LevKmeansKmeans is the worst configuration instead of \LevDbscanKmeans.
\BagDbscanHdbscan is the only configuration that reached full vulnerability coverage for \jenkins without using the \kmeans clustering algorithm and it performs significantly better than the other configurations, especially the ones involving \kmeans for both output and action clustering steps.
This indicates, without surprise, that the \kmeans algorithm takes more resources to be executed.
\BagDbscanHdbscan did not lead to full vulnerability coverage for \joomla, so we do not consider it as a candidate for \quotes{best} configuration.

For \joomla, 
\textbf{\BagDbscanKmeans performed significantly better than \LevDbscanKmeans} for the considered metrics.
As for the other configurations, \Cref{tab:joomla:duels:sizes} for input set sizes and \Cref{tab:joomla:duels:costs} for input set costs provide identical results, indicating that \LevKmeansHdbscan and \BagKmeansHdbscan dominate the others while being equivalent.
Moreover, \BagDbscanKmeans dominates \LevDbscanKmeans, which is the worst configuration.
The results are almost identical for \aim execution time in \Cref{tab:joomla:duels:times}, with the small difference that \LevKmeansHdbscan performs slightly better than \BagKmeansHdbscan.
However, \LevKmeansHdbscan and \BagKmeansHdbscan did not lead to full vulnerability coverage for \jenkins, as opposed to \BagDbscanKmeans and \LevDbscanKmeans.

Since we obtained similar results for both \jenkins and \joomla, \textbf{we consider \BagDbscanKmeans to be the \quotes{best} \aim configuration}.
This is not surprising since Bag distance is less costly to compute than Levenshtein distance (\Cref{sec:aim:outputClustering}) and we already observed that the order of words in a Web page does not appear to be a relevant distinction for vulnerability coverage (\Cref{sec:aim:results:detectedVulnerabilities}).

As mentioned in \Cref{sec:aim:results:detectedVulnerabilities}, no baseline leads to full vulnerability coverage.
\ArtDbscan fared poorly and \ArtHdbscan performed similarly to random testing \Rt, but \ArtKmeans was much better, with 99.3\% VDR for \jenkins and 83.3\% for \joomla.
But even if \ArtKmeans had reached full vulnerability coverage for both systems, it would be at a disadvantage compared to \aim configurations.
Indeed, over 50 \jenkins runs, the average input set size for \ArtKmeans was 94.92 inputs, while it ranges from 38 inputs (40\% of \ArtKmeans) for \BagDbscanHdbscan to 74.8 inputs (79\%) for \LevDbscanKmeans.
%, and 63.56 (67\%) for our best configuration \BagDbscanKmeans.
The average input set cost for \ArtKmeans was \num{193698.94} actions, while it ranges from \num{70500.76} actions (36\%) for \BagDbscanHdbscan to \num{152373.54} actions (79\%) for \LevDbscanKmeans.
%, and \num{117171.86} (60\%) for our best configuration \BagDbscanKmeans.
Over 50 \joomla runs, the average input set size for \ArtKmeans was 70 inputs, while it ranges from 36.02 inputs (51\%) for \LevKmeansHdbscan to 41.46 inputs (59\%) for \LevDbscanKmeans.
%, and 40.16 (57\%) for our best configuration \BagDbscanKmeans.
The average input set cost for \ArtKmeans was \num{2312784.58} actions, while it ranges from \num{580705.24} actions (25\%) for \LevKmeansHdbscan to \num{872352.72} actions (38\%) for \LevDbscanKmeans.
%, and \num{862591.18} (37) for our best configuration \BagDbscanKmeans.
In short, \textbf{all \aim configurations with full vulnerability coverage outperformed the best baseline \ArtKmeans}, which highlights the relevance of our approach in reducing the cost of testing.

\begin{table}[t]
\centering
\footnotesize
\caption{Comparison of MRs execution time before and after input set minimization.
%The total execution time after minimization is MRs execution time after minimization plus \aim execution time.
The percentage of reduction is one minus the ratio between total execution time after minimization and MRs execution time before minimization.}
\label{tab:finalResults}
 \normalem

\begin{tabular}{|l|r|r|}
\hline
\textbf{Execution time (minutes)} & \textbf{Jenkins} & \textbf{Joomla} \\ \hline
MRs with initial input set                                                              & \num{38307}            & \num{20703}           \\ \hline
MRs with minimized input set                                                            & 6119          & 3675            \\
+ \aim execution time                                                 & 22           & 22            \\ 
= Total execution time & 6141             & 3697            \\ \hline
Percentage of Reduction & \textbf{84\%} & \textbf{82\%} \\ \hline
\end{tabular}%

\end{table}

\TSE{3.3.1}{Finally, in \Cref{tab:finalResults}, we present the results of executing the MRs using both the initial input set and the minimized input set derived from the best configuration (BDK).
In total, by applying AIM, we reduced the execution time of all 76 MRs from \num{38307} minutes to \num{6119} minutes for \jenkins and from \num{20703} minutes to \num{3675} minutes for \joomla, using the minimized input set with median cost.
Moreover, executing \aim to obtain this minimized input set required 22 minutes for both systems.
Hence, we have a total execution time of \num{6141} minutes for \jenkins and \num{3697} minutes for \joomla.}
As a result, the ratio of the total execution time for the minimized input sets divided by the execution time for the initial input sets is $16.03\%$ for \jenkins and $17.85\%$ for \joomla.
In other words, \textbf{\aim reduced the execution time by about $\textbf{84}\%$ for \jenkins and more than $\textbf{82}\%$ for \joomla}.
This large reduction in execution time demonstrates the effectiveness of our approach in reducing the cost of metamorphic security testing.
%\TSE{3.3.1}{Furthermore, we estimate that executing all 76 MRs with the minimized input sets obtained from all 18 different \aim configurations, for 50 runs and without parallelization, would take at least \num{5500000} minutes on \jenkins and at least \num{3300000} minutes on \joomla.}

\subsubsection{RQ3 - Comparison of Search Algorithms}
\label{sec:aim:results:geneticAlgos}

\begin{figure*}[tb]
\begin{center}
\includegraphics[width=1.0\textwidth]{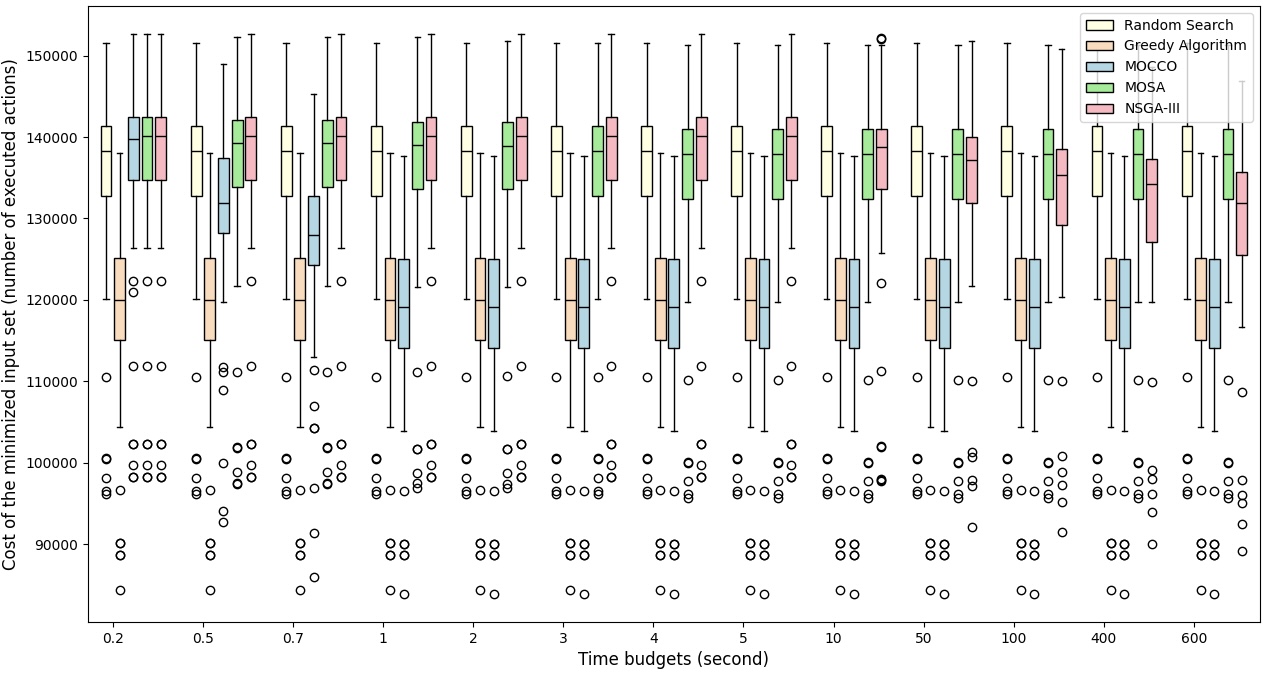}
\caption{\jenkins: Cost of the minimized input sets using Random Search, Greedy Algorithm, \geneticAlgo, \mosa, and \nsgaThree under different time budgets for 50 runs of BDK.} 
\label{fig:BoxplotJenkins}
\end{center}
\end{figure*}

\begin{figure*}[tb]
\begin{center}
\includegraphics[width=1.0\textwidth]{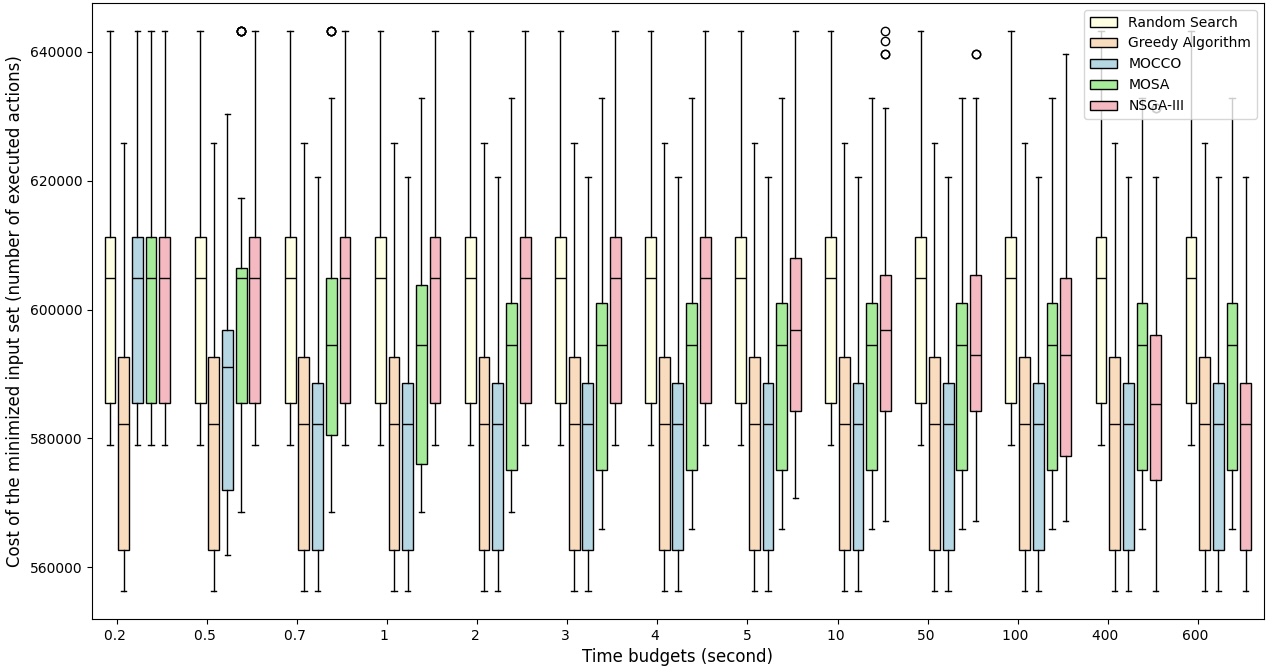}
\caption{\joomla: Cost of the minimized input sets using Random Search, Greedy Algorithm, \geneticAlgo, \mosa, and \nsgaThree under different time budgets for 50 runs of BDK.} 
\label{fig:BoxplotJoomla}
\end{center}
\end{figure*}

\TSE{2.2, 2.3}{To answer RQ3, we consider the 50 reduced input sets obtained from the best \aim configuration, namely BDK, and we compare the cost of the minimized input sets obtained by \geneticAlgo and baselines.
The cost of the 50 minimized input sets obtained for each genetic algorithm is represented using box plots in \Cref{fig:BoxplotJenkins,fig:BoxplotJoomla} for \jenkins and \joomla, respectively.
The results are presented for different time budgets, ranging from 0.2 to 600 seconds, by which time greedy and \geneticAlgoLong have converged. 
For both \jenkins and \joomla, \textbf{random search, greedy algorithm, and \mosa quickly converge. Random search and \mosa converge toward sub-optimal solutions}, which is expected since random search is unlikely to determine the best order of removal steps by chance, and \mosa minimizes the cost for each individual objective instead of considering the collective coverage of the selected inputs.
\textbf{The greedy algorithm finds a good approximation for all runs on both systems}.
It finds the optimal solution for most runs (41 out of 50 runs) on \joomla but for only a few runs (6 out of 50 runs) on \jenkins, likely because \jenkins reduced input sets are larger than those of \joomla.
%matching theoretical bounds~\cite{DS14,Vaz01}, thus suggesting that the larger the reduced input set, the further away the greedy algorithm is from the optimal solution.
\textbf{\geneticAlgo finds the optimal solution} (\Cref{sec:aim:aim:benchmarks}) \textbf{for all 50 runs in 1 second for \jenkins and 0.7 seconds for \joomla}, which is expected since \geneticAlgo is designed to solve this many-objective problem (\Cref{sec:aim:genetic:motivation}).
\nsgaThree slowly converges for both systems. \textbf{\nsgaThree does not find the optimal solution within the 600-second time budget for \jenkins, but does so in 600 seconds for \joomla}.
This is also expected since \nsgaThree explores the entire Pareto front while \geneticAlgo focuses on the region of interest, i.e, around the utopia point of full coverage at no cost.
\nsgaThree is the only baseline that finds the optimal solution for all runs and within the time budget, but only for one system, \textbf{while \geneticAlgo consistently finds the optimal solution in nearly three orders of magnitude faster}.}

\TSE{2.2}{Furthermore, we conducted statistical tests reported in \Cref{tab:jenkins:algos600} for \jenkins and \Cref{tab:joomla:algos400,tab:joomla:algos600} for \joomla.
% \NB{test}
Since \nsgaThree achieves the same result as \geneticAlgo within 600 seconds on \joomla, we also report their results at 400 seconds for a more comprehensive comparison.
In these three tables, algorithms in each row are compared with algorithms in each column.
$\pValue$ denotes the statistical significance and $\effectSize$ the effect size (\Cref{sec:aim:results:performance}).
%Since the test for the alternative hypothesis $\configRandomVar_1 < \configRandomVar_2$ is asymmetric, p-values for $\configRandomVar_1 < \configRandomVar_2$ and $\configRandomVar_2 < \configRandomVar_1$ cannot be small at the same time.
%Hence, as opposed to RQ2, each cell is colored based only on the effect size.
As for RQ2, when $\pValue > 0.05$, we consider the costs obtained from the two algorithms not to be significantly different, and hence the cell is left white.
Otherwise, the cell is colored, either in green or red.
Since we consider input set cost, the smaller the values the better.
Thus, green (resp. red) indicates that the algorithm in the row is better (resp. worse) than the algorithm in the column.
Finally, as for RQ2, the intensity of the color is computed with $\abs{2\times\effectSize - 1}$.}

\begin{table}[t]
\centering
\setlength\tabcolsep{3.5pt}
\footnotesize
\caption{Comparison of genetic algorithms for \jenkins (600 s).}
\label{tab:jenkins:algos600}
\begin{tabular}{cc|c|c|c|c|c|}
\hline
\multicolumn{2}{|c|}{costs} & \algoRandom & \algoGreedy & \geneticAlgo & \mosa & \nsgaThree \\
\hline
\multicolumn{1}{|c|}{\multirow{2}{*}{\algoRandom}} & $\pValue$
	 & 
	 & \cellcolor{red!100}\num{1.8e-15}
	 & \cellcolor{red!100}\num{1.8e-15}
	 & \cellcolor{red!100}\num{1.8e-15}
	 & \cellcolor{red!100}\num{1.8e-15}\\
\multicolumn{1}{|c|}{} & $\effectSize$
	 & 
	 & \cellcolor{red!100}1.0
	 & \cellcolor{red!100}1.0
	 & \cellcolor{red!100}1.0
	 & \cellcolor{red!100}1.0\\
\hline
\multicolumn{1}{|c|}{\multirow{2}{*}{\algoGreedy}} & $\pValue$
	 & \cellcolor{green!100}\num{1.8e-15}
	 & 
	 & \cellcolor{red!98}\num{1.1e-09}
	 & \cellcolor{green!100}\num{1.8e-15}
	 & \cellcolor{green!100}\num{1.8e-15}\\
\multicolumn{1}{|c|}{} & $\effectSize$
	 & \cellcolor{green!100}0.0
	 & 
	 & \cellcolor{red!98}0.99
	 & \cellcolor{green!100}0.0
	 & \cellcolor{green!100}0.0\\
\hline
\multicolumn{1}{|c|}{\multirow{2}{*}{\geneticAlgo}} & $\pValue$
	 & \cellcolor{green!100}\num{1.8e-15}
	 & \cellcolor{green!98}\num{1.1e-09}
	 & 
	 & \cellcolor{green!100}\num{1.8e-15}
	 & \cellcolor{green!100}\num{1.8e-15}\\
\multicolumn{1}{|c|}{} & $\effectSize$
	 & \cellcolor{green!100}0.0
	 & \cellcolor{green!98}0.01
	 & 
	 & \cellcolor{green!100}0.0
	 & \cellcolor{green!100}0.0\\
\hline
\multicolumn{1}{|c|}{\multirow{2}{*}{\mosa}} & $\pValue$
	 & \cellcolor{green!100}\num{1.8e-15}
	 & \cellcolor{red!100}\num{1.8e-15}
	 & \cellcolor{red!100}\num{1.8e-15}
	 & 
	 & \cellcolor{red!100}\num{1.8e-15}\\
\multicolumn{1}{|c|}{} & $\effectSize$
	 & \cellcolor{green!100}0.0
	 & \cellcolor{red!100}1.0
	 & \cellcolor{red!100}1.0
	 & 
	 & \cellcolor{red!100}1.0\\
\hline
\multicolumn{1}{|c|}{\multirow{2}{*}{\nsgaThree}} & $\pValue$
	 & \cellcolor{green!100}\num{1.8e-15}
	 & \cellcolor{red!100}\num{1.8e-15}
	 & \cellcolor{red!100}\num{1.8e-15}
	 & \cellcolor{green!100}\num{1.8e-15}
	 & \\
\multicolumn{1}{|c|}{} & $\effectSize$
	 & \cellcolor{green!100}0.0
	 & \cellcolor{red!100}1.0
	 & \cellcolor{red!100}1.0
	 & \cellcolor{green!100}0.0
	 & \\
\hline
\end{tabular}
\end{table}

\begin{table}[t]
\centering
\setlength\tabcolsep{3.5pt}
\footnotesize
\caption{Comparison of genetic algorithms for \joomla (400 s).}
\label{tab:joomla:algos400}
\begin{tabular}{cc|c|c|c|c|c|}
\hline
\multicolumn{2}{|c|}{costs} & \algoRandom & \algoGreedy & \geneticAlgo & \mosa & \nsgaThree \\
\hline
\multicolumn{1}{|c|}{\multirow{2}{*}{\algoRandom}} & $\pValue$
	 & 
	 & \cellcolor{red!100}\num{1.8e-15}
	 & \cellcolor{red!100}\num{1.8e-15}
	 & \cellcolor{red!100}\num{1.8e-15}
	 & \cellcolor{red!100}\num{4.7e-10}\\
\multicolumn{1}{|c|}{} & $\effectSize$
	 & 
	 & \cellcolor{red!100}1.0
	 & \cellcolor{red!100}1.0
	 & \cellcolor{red!100}1.0
	 & \cellcolor{red!100}1.0\\
\hline
\multicolumn{1}{|c|}{\multirow{2}{*}{\algoGreedy}} & $\pValue$
	 & \cellcolor{green!100}\num{1.8e-15}
	 & 
	 & \cellcolor{red!32}\num{3.2e-02}
	 & \cellcolor{green!100}\num{1.8e-15}
	 & \cellcolor{green!70}\num{1.3e-05}\\
\multicolumn{1}{|c|}{} & $\effectSize$
	 & \cellcolor{green!100}0.0
	 & 
	 & \cellcolor{red!32}0.66
	 & \cellcolor{green!100}0.0
	 & \cellcolor{green!70}0.15\\
\hline
\multicolumn{1}{|c|}{\multirow{2}{*}{\geneticAlgo}} & $\pValue$
	 & \cellcolor{green!100}\num{1.8e-15}
	 & \cellcolor{green!32}\num{3.2e-02}
	 & 
	 & \cellcolor{green!100}\num{1.8e-15}
	 & \cellcolor{green!78}\num{9.4e-07}\\
\multicolumn{1}{|c|}{} & $\effectSize$
	 & \cellcolor{green!100}0.0
	 & \cellcolor{green!32}0.34
	 & 
	 & \cellcolor{green!100}0.0
	 & \cellcolor{green!78}0.11\\
\hline
\multicolumn{1}{|c|}{\multirow{2}{*}{\mosa}} & $\pValue$
	 & \cellcolor{green!100}\num{1.8e-15}
	 & \cellcolor{red!100}\num{1.8e-15}
	 & \cellcolor{red!100}\num{1.8e-15}
	 & 
	 & \cellcolor{red!69}\num{1.8e-05}\\
\multicolumn{1}{|c|}{} & $\effectSize$
	 & \cellcolor{green!100}0.0
	 & \cellcolor{red!100}1.0
	 & \cellcolor{red!100}1.0
	 & 
	 & \cellcolor{red!69}0.84\\
\hline
\multicolumn{1}{|c|}{\multirow{2}{*}{\nsgaThree}} & $\pValue$
	 & \cellcolor{green!100}\num{4.7e-10}
	 & \cellcolor{red!70}\num{1.3e-05}
	 & \cellcolor{red!78}\num{9.4e-07}
	 & \cellcolor{green!69}\num{1.8e-05}
	 & \\
\multicolumn{1}{|c|}{} & $\effectSize$
	 & \cellcolor{green!100}0.0
	 & \cellcolor{red!70}0.85
	 & \cellcolor{red!78}0.89
	 & \cellcolor{green!69}0.16
	 & \\
\hline
\end{tabular}
\end{table}

\begin{table}[t]
\centering
\setlength\tabcolsep{3.5pt}
\footnotesize
\caption{Comparison of genetic algorithms for \joomla (600 s).}
\label{tab:joomla:algos600}
\begin{tabular}{cc|c|c|c|c|c|}
\hline
\multicolumn{2}{|c|}{costs} & \algoRandom & \algoGreedy & \geneticAlgo & \mosa & \nsgaThree \\
\hline
\multicolumn{1}{|c|}{\multirow{2}{*}{\algoRandom}} & $\pValue$
	 & 
	 & \cellcolor{red!100}\num{1.8e-15}
	 & \cellcolor{red!100}\num{1.8e-15}
	 & \cellcolor{red!100}\num{1.8e-15}
	 & \cellcolor{red!100}\num{1.8e-15}\\
\multicolumn{1}{|c|}{} & $\effectSize$
	 & 
	 & \cellcolor{red!100}1.0
	 & \cellcolor{red!100}1.0
	 & \cellcolor{red!100}1.0
	 & \cellcolor{red!100}1.0\\
\hline
\multicolumn{1}{|c|}{\multirow{2}{*}{\algoGreedy}} & $\pValue$
	 & \cellcolor{green!100}\num{1.8e-15}
	 & 
	 & \cellcolor{red!32}\num{3.2e-02}
	 & \cellcolor{green!100}\num{1.8e-15}
	 & \cellcolor{red!32}\num{3.2e-02}\\
\multicolumn{1}{|c|}{} & $\effectSize$
	 & \cellcolor{green!100}0.0
	 & 
	 & \cellcolor{red!32}0.66
	 & \cellcolor{green!100}0.0
	 & \cellcolor{red!32}0.66\\
\hline
\multicolumn{1}{|c|}{\multirow{2}{*}{\geneticAlgo}} & $\pValue$
	 & \cellcolor{green!100}\num{1.8e-15}
	 & \cellcolor{green!32}\num{3.2e-02}
	 & 
	 & \cellcolor{green!100}\num{1.8e-15}
	 & \num{1.0e+00}\\
\multicolumn{1}{|c|}{} & $\effectSize$
	 & \cellcolor{green!100}0.0
	 & \cellcolor{green!32}0.34
	 & 
	 & \cellcolor{green!100}0.0
	 & 0.5\\
\hline
\multicolumn{1}{|c|}{\multirow{2}{*}{\mosa}} & $\pValue$
	 & \cellcolor{green!100}\num{1.8e-15}
	 & \cellcolor{red!100}\num{1.8e-15}
	 & \cellcolor{red!100}\num{1.8e-15}
	 & 
	 & \cellcolor{red!100}\num{1.8e-15}\\
\multicolumn{1}{|c|}{} & $\effectSize$
	 & \cellcolor{green!100}0.0
	 & \cellcolor{red!100}1.0
	 & \cellcolor{red!100}1.0
	 & 
	 & \cellcolor{red!100}1.0\\
\hline
\multicolumn{1}{|c|}{\multirow{2}{*}{\nsgaThree}} & $\pValue$
	 & \cellcolor{green!100}\num{1.8e-15}
	 & \cellcolor{green!32}\num{3.2e-02}
	 & \num{1.0e+00}
	 & \cellcolor{green!100}\num{1.8e-15}
	 & \\
\multicolumn{1}{|c|}{} & $\effectSize$
	 & \cellcolor{green!100}0.0
	 & \cellcolor{green!32}0.34
	 & 0.5
	 & \cellcolor{green!100}0.0
	 & \\
\hline
\end{tabular}
\end{table}

\TSE{2.2, 2.3}{The results for \jenkins with a time budget of 600 seconds are detailed in \Cref{tab:jenkins:algos600}.
The small p-values and effect sizes observed in the \geneticAlgo row indicates that \textbf{\geneticAlgo obtained minimized input sets with significantly smaller costs than all the alternative approaches}.
Moreover, the effect size for all baselines but the greedy algorithm is $0$, indicating that the minimized input sets obtained by \geneticAlgo consistently have smaller costs across all 50 runs. 
For the greedy algorithm, the effect size is $0.01$ because it performs as well as \geneticAlgo for a few runs (6 out of 50 runs).
For \joomla, the results with a time budget of 400 and 600 seconds are respectively detailed in \Cref{tab:joomla:algos400,tab:joomla:algos600}.
For all time budgets, the small p-values and the effect sizes of $0$ indicate that \textbf{\geneticAlgo performed better than random search and \mosa in all 50 runs}, but the results are more nuanced for the greedy algorithm and \nsgaThree.
For all time budgets, p-values indicate that \textbf{\geneticAlgo performed significantly better than the greedy algorithm}.
The latter managed to find the optimal solution for most runs (41 out of 50 runs), yielding an effect size of $0.34$, which is still in favor of \geneticAlgo.
We conjecture this is because the greedy algorithm does not consider input block as individual objectives and hence, as opposed to \geneticAlgo, does not take into account relevant information when selecting inputs (\Cref{sec:aim:mao}).
For a 400-second time budget, the p-value indicates that \textbf{\geneticAlgo performed significantly better than \nsgaThree}, but \nsgaThree managed to find the optimal solution for some runs, hence the effect size of $0.11$, which is still in favor of \geneticAlgo.
For a 600-second time budget, the p-value is $1.0$ and the effect size is $0.5$, indicating that both approaches find the same results for all 50 runs, i.e., the optimal solutions.
We conjecture that \nsgaThree manages to achieve the same results as \geneticAlgo for \joomla but not \jenkins because the former's reduced input set has fewer redundant inputs to be removed compared to \jenkins, making the problem computationally simpler.
Since \geneticAlgo was the only search algorithm able to consistently find the optimal solution for \jenkins for every time budget and since, for \joomla, \geneticAlgo finds the optimal solution almost three orders of magnitude faster than the only baseline, i.e., \nsgaThree, that manages to obtain the same  results, we conclude that \textbf{\geneticAlgo outperforms all baselines when accounting for both minimized input set cost and execution time}.}

% \TSE{2.2}{\YM{Previous version:} %\textbf{Nevertheless, the greedy algorithm finds approximations that are relatively close to those of \geneticAlgo and it is unclear whether the difference is significant in practice.}
%We speculate this small difference is due to the small size of the considered datasets, containing small connected components (\Cref{defi:overlapGraph}) where the approximation made by the greedy algorithm---based on the cost effectiveness of one input at a time and not on the optimal combination of several inputs to cover their connected component---does not make a large difference in the final cost.
%To assess this difference with larger input sets, we conducted another experiment…}

\TSE{2.2}{
\textbf{Nevertheless, the greedy algorithm finds approximations that are relatively close to those of \geneticAlgo.}
To determine the significance of the difference in practice, we estimate how this difference in cost translates into a difference in execution time, based on our execution time results (\Cref{sec:aim:results:magnitudeOfReduction}).
For \jenkins, it took 6119 minutes to execute a minimized input set of cost 119089 actions.
Since the greedy algorithms obtained minimized inputs sets with on average 611.74 more actions and up to 1095 actions, this translates into a difference of 31.43 minutes on average to execute the MRs, up to 56.3 minutes in the worst case.
For \joomla, the minimized input set of cost 582181 took 3675 minutes, hence given the difference between the greedy algorithm and \geneticAlgo of 949.32 actions on average with a maximum of 5274 actions, this translates into a difference of 6 minutes on average, up to 33.29 minutes in the worst case. However, relatively to the total execution time, these differences may not have a practical impact.
\textbf{Therefore, we conclude, from our case studies, that the greedy algorithm, despite its limitations, is a good alternative.}
}

\section{Threats to Validity}
\label{sec:aim:results:threatsToValidity}

In this section, we discuss internal, conclusion, construct, and external validity according to conventional practices~\cite{WRH+12}.

\subsection{Internal Validity}

A potential internal threat concerns inadequate data pre-processing, which may adversely impact our results. Indeed, clustering relies on the computed similarity among the pre-processed outputs and inputs. To address this potential concern, we have conducted a manual investigation of the quality of the clusters obtained without pre-processing.
This led us to remove, from the textual content extracted from each Web page, all the content that was shared by many Web pages, like system version, date, or (when present) the menu of the Web page.

For RQ1 on vulnerability detection, one potential threat we face is missing inputs that would be able to exercise a vulnerability or incorrectly considering that an input is able to exercise a vulnerability.
To ensure our list of inputs triggering vulnerabilities is complete, one author inspected all the \mstWi execution logs to look for failures.

\subsection{Conclusion Validity}

For RQ2, we rely on a non-parametric test (i.e., Mann-Whitney-Wilcoxon test) to evaluate the statistical and practical significance of differences in results, computing $\pValue$  and Vargha and Delaney's $\vdaLong$ metric for effect size.
Moreover, to deal with the randomness inherent to search algorithms, all the configurations and baselines were executed over 50 runs.

Randomness may also arise from (1) the workload of the machines employed to conduct experiments, potentially slowing down the performance of \mstWi, \aim, and the case study subjects, and (2) the presence of other users interacting with the software under test, which can impact both execution time and system outputs. To address these concerns, we conducted experiments in dedicated environments, ensuring that the study subjects were exclusively utilized by \aim.

\subsection{Construct Validity}

The constructs considered in our work are vulnerability detection effectiveness and input set reduction effectiveness.
Vulnerability detection effectiveness is measured in terms of vulnerability detection rate. Reduction effectiveness is measured in terms of MR execution time,  size and cost of the minimized input set, and \aim execution time for each configuration.
As it is expensive to execute all 18 configurations on the MRs, we consider the size of the input set and its cost to select the most efficient configuration. The cost of the input set has been defined in \Cref{defi:cost} and shown to be linearly correlated with MR execution time, thus enabling us to evaluate the efficiency of the results.

Finally, we executed the minimized input set obtained from the best configuration on the MRs and compared the obtained execution time, plus the \aim execution time required to minimize the initial input set, with the MRs execution time obtained with the initial input set.
%for both systems under test.
Execution time is a direct measure, allowing us to evaluate whether, for systems akin to our case study subjects, \aim should be adopted for making vulnerability testing more efficient and scalable.

\subsection{External Validity}

One threat to the generalizability of our results stems from the benchmark that we used. It includes 160 inputs for \jenkins and 148 inputs for \joomla. Furthermore, we considered the list of vulnerabilities in \jenkins and \joomla that were successfully triggered with MST-wi.
%, which is nine vulnerabilities for \jenkins and three vulnerabilities for \joomla.
However, even if in this study we used \mstWi to collect our data, the \aim approach does not depend on a particular data collector, and using or implementing another data collector would enable the use of our approach with other frameworks.
Moreover, even if we relied on previously obtained MRs to be sure they detect vulnerabilities in the considered Web systems, \aim is a general approach for metamorphic security testing which does not depend on the considered MRs.
Finally, in \Cref{sec:aim:aim:benchmarks}, we highlighted that the different input/output interfaces provided by \jenkins and \joomla, along with the diverse types of vulnerabilities they contain, is in support of the generalizability of our results.
Furthermore, the \aim approach can be generalized to other Web systems, if the data collection and pre-processing components are updated accordingly.
Nevertheless, further studies involving systems with known vulnerabilities are needed.

\section{Related Work}
\label{sec:aim:related}

MT enables the execution of a SUT with a potentially infinite set of inputs thus being more effective than testing techniques requiring the manual specification of either test inputs or oracles. However, in MT, the manual definition of metamorphic relations (MRs) is an expensive activity because it requires that engineers first acquire enough information on the subject under test and then analyze the testing problem to identify MRs. For this reason, in the past, researchers focused on both the definition of methodologies supporting the identification of MRs~\cite{METRIC,METRICp} and the development of techniques for the automated generation of MRs, based on meta-heuristic search~\cite{AutoMR,Terragni:MT:generation} and natural language processing~\cite{BLASI:MeMo}, and targeting query-based systems~\cite{SeguraQueryBased} and Cyber-Physical Systems~\cite{Terragni:MT:generation}.

However, source inputs also impact the effectiveness and performance of MT; indeed, MRs generate follow-up inputs from source inputs and both are executed by the SUT.
Consequently, the research community has recently shown increasing interest towards investigating the impact of source inputs on MT. We summarize the most relevant works in the following paragraphs. 
Note that all these studies focus on general fault detection, while we focus on metamorphic security testing for Web systems.
However, our approach could also be applied to fault detection while the approaches below could also be applied to security testing.
We therefore compare these approaches without considering their difference in application.
However, we excluded from our survey those approaches that study the effect of source and follow-up inputs on the metamorphic testing of systems that largely differ from ours (i.e.,  sequence alignment programs~\cite{Tang:MET:2017}, system validation~\cite{systemValidation}, and deep neural networks~\cite{MT:DNN:2020}). 
In the following paragraphs, we group the surveyed works into three categories: input generation techniques, input selection techniques, and feedback-directed metamorphic testing.

\emph{Input generation techniques} for MT use white-box approaches based on knowledge of the source code (mainly, for statement or branch coverage), while we use a black-box approach based on input and output information (\Cref{sec:aim:doubleCLustering}).
For instance, a study~\cite{SK18} leveraged the evolutionary search approach EvoSuite~\cite{FA11} to evolve whole input sets in order to obtain inputs that lead to more branch coverage or to different results on the mutated and non-mutated versions of the source code.
Another example study~\cite{SLF+22} leveraged symbolic execution to collect constraints of program branches covered by execution paths, then solved these constraints to generate the corresponding source inputs.
Finally, the execution of the generated inputs on the SUT was prioritized based on their contribution regarding uncovered statements.
In this case, both generation and prioritization phases were white-box approaches based on branch coverage.
Note that, while our approach on input set minimization could be seen as similar to input prioritization, both studies focused on increasing coverage, while we focused on reducing cost while maintaining full coverage.

\emph{Input selection techniques} share the same objective of our work (i.e., reducing the number of source inputs while maximizing MT effectiveness).
Because of its simplicity, random testing (RT) is a common strategy for test suite minimization~\cite{WAG15,ZAY19} that has been used in MT~\cite{HWHY21}.
%A 2016 survey reported that 57\% of MT work employed RT to generate source inputs and 34\% selected source inputs from already existing test suite~\cite{SFSRC16}.
RT was enhanced with Adaptive Random Testing (ART), a technique for obtaining source inputs spread across the input domain with the aim of finding failures with fewer number of inputs than RT.
As input selection technique for MT, ART outperforms RT in terms of fault detection~\cite{BCK+16}, which (as in the following studies) was evaluated using the F-measure, i.e., the number of inputs necessary to reveal the first failure.
In the AIM approach, our action clustering step (\Cref{sec:aim:actionClustering}) bears similarities with ART, since we partition inputs based on action parameters which are relevant for our SUT.
But, instead of assuming that close inputs lead to close outputs, we directly used SUT outputs during our output clustering step (\Cref{sec:aim:outputClustering}), since they are inexpensive to obtain compared to executing MRs.
Finally, instead of only counting the number of inputs as in the F-measure (or the size of the input set, in the context of input generation), we considered the cost of each source input as the number of executed actions as surrogate measure, which is tailored to reducing MR execution time in the context of Web systems.
Instead of focusing only on distances between source inputs as in ART, another study~\cite{HWHY21} also investigated distances with follow-up inputs, which is an improvement since usually there are more follow-up inputs than source inputs.
This led to the Metamorphic testing-based adaptive random testing (MT-ART) technique, which performed better than other ART algorithms regarding test effectiveness, test efficiency, and test coverage (considering statement, block, and branch coverage).
Unfortunately, in the AIM approach, we could not consider follow-up inputs to drive the input selection, since executing MRs to generate these follow-up inputs would defeat our purpose of reducing MR execution time.
%is too expensive in our context.
%Since our approach aims at reducing MR execution time, executing MRs would defeat our purpose.

Finally, while studies on MT usually focus either on the identification of effective MRs or on input generation/selection, a recent study proposed \emph{feedback-directed metamorphic testing} (FDMT)~\cite{SDLC23} to determine the next test to perform (both in terms of source input and MR), based on previous test results.
They proposed adaptive partition testing (APT) to dynamically select source inputs, based on input categories that lead to fault detection, and a diversity-oriented strategy for MR selection (DOMR) to select an MR generating follow-up inputs that are as different as possible from the already obtained ones.
While this approach is promising in general, it is not adapted to our case, where we consider a fixed set of MRs, MR selection being considered outside the scope of this paper.
Moreover, since we aim to reduce MR execution time, we cannot  execute them and use execution information to guide source input or MR selection during testing.
Finally, in our problem definition (\Cref{sec:aim:framework}), we do not consider source inputs independently from each other, which is why we reduced (\Cref{sec:aim:search}) then minimized (\Cref{sec:aim:genetic}) the cost of the input set as a whole.

\section{Conclusion and  Future Work}
\label{sec:aim:conclusion}

As demonstrated in our previous work~\cite{BPGB23}, metamorphic testing alleviates the oracle problem for the automated security testing of Web systems.
However, metamorphic testing has shown to be a time-consuming approach.
Our approach (\aim ) aims to reduce the cost of metamorphic security testing by minimizing the initial input set while preserving its capability at exercising vulnerabilities. 
Our contributions include
1) a clustering-based black box approach that identifies similar inputs based on their security properties,
2) \impro, an approach to reduce the search space as much as possible, then divide it into smaller independent parts,
3) \geneticAlgo, a novel genetic algorithm which is able to efficiently select diverse inputs while minimizing their total cost,
and 4) a testing framework automatically performing input set minimization.

We considered 18 different configurations for \aim and we evaluated our approach on two open-source Web systems, \jenkins and \joomla, in terms of vulnerability detection rate and magnitude of the input set reduction.
Our empirical results show that the best configuration for \aim is \BagDbscanKmeans: Bag distance, \dbscan to cluster the outputs, and \kmeans to cluster the inputs.
The results show that our approach can automatically reduce MRs execution time by $84\%$ for \jenkins and $82\%$ for \joomla while preserving full vulnerability detection.
Across 50 runs, the \BagDbscanKmeans configuration consistently detected all vulnerabilities in \jenkins and \joomla. 
We also compared \aim with four baselines common in security testing.
Notably, none of the baselines reached full vulnerability coverage.
Among them, \ArtKmeans (ART baseline using \kmeans) emerged as the closest to achieving full vulnerability coverage.
All \aim configurations with full vulnerability coverage outperformed this baseline in terms of minimized input set size and cost, demonstrating the effectiveness of our approach in reducing the cost of metamorphic security testing.

\TSE{2.2}{Furthermore, we compared the effectiveness of \geneticAlgo, in terms of minimized input set cost and execution time, with four other search algorithms.
The results on \jenkins showed that \geneticAlgo obtained minimized input sets with significantly lower costs than all the alternative approaches. The only baseline that could find the optimal solutions on \joomla was \nsgaThree, though \geneticAlgo did so almost three orders of magnitude faster. 
Among the considered alternative search algorithms, greedy was the only algorithm that consistently found results close to those of \geneticAlgo, on both \jenkins and \joomla.
Therefore, we conclude that \geneticAlgo outperforms all baselines in terms of minimized input set cost and execution time, while the greedy algorithm, despite its theoretical limitations, remains a viable alternative in practice.}

 As part of future work, we intend to develop a test case prioritization technique that facilitates earlier vulnerability detection by prioritizing the inputs in the minimized input set that are most likely to detect vulnerabilities.

\ifCLASSOPTIONcompsoc
  \section*{Acknowledgments}
\else
  \section*{Acknowledgment}
\fi

This work is supported by the H2020 COSMOS European project, grant agreement No. 957254, NSERC of Canada under the Discovery and CRC programs, and the Science Foundation Ireland grant 13/RC/2094-2. It is part of a collaborative research program between the University of Ottawa's Nanda laboratory and the SnT centre at the University of Luxembourg. 

\phantomsection
%\addcontentsline{toc}{section}{\bibname}
%\bibliographystyle{ACM-Reference-Format}
%\citestyle{acmnumeric}
\bibliographystyle{IEEEtran}
\bibliography{./Bibliography/bibliography.bib, ./Bibliography/tools.bib}

%\appendix
%\clearpage

%\input{./Appendices/dump.tex}

\begin{IEEEbiography}[{\includegraphics[width=1in,height=1.25in,clip,keepaspectratio]{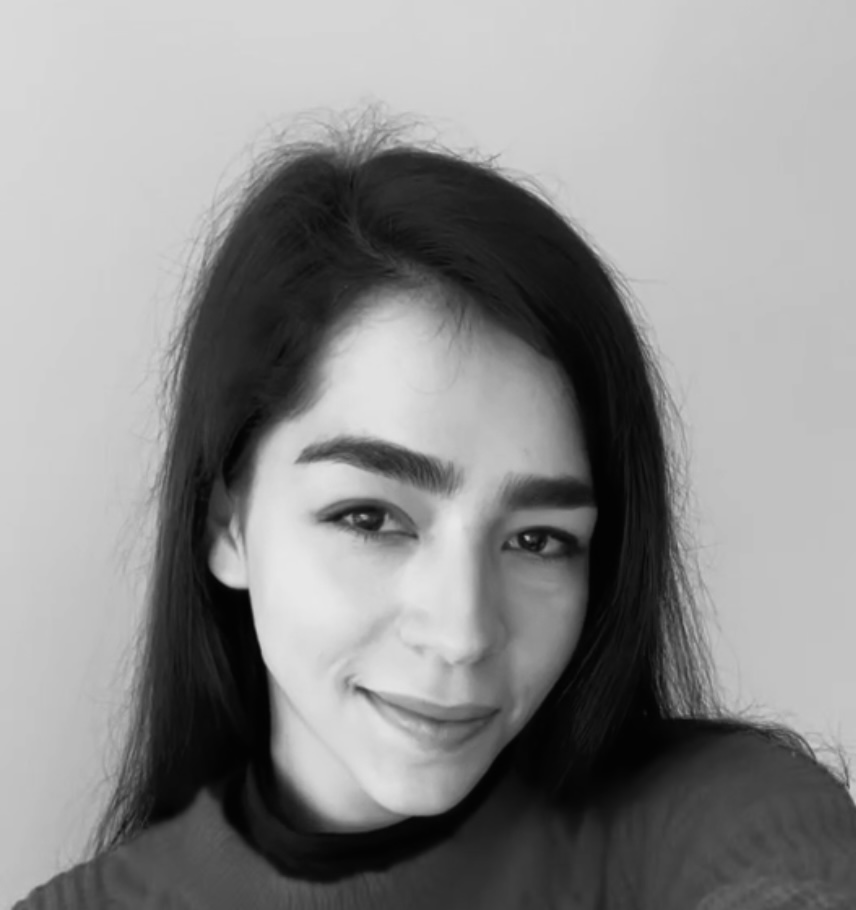}}]{Nazanin Bayati Chaleshtari} is a member of Nanda Lab and is currently working toward the PhD degree in the School of EECS, University of Ottawa. She gained valuable practical experience from her work with BlackBerry's security research and development lab in Canada. Throughout her academic career, she has been the recipient of several academic awards, including a PhD admission scholarship, an international doctoral scholarship from the University of Ottawa, and an honourable award for being an outstanding student during her master’s degree at the Iran University of Science and Technology. She was also ranked the best student among all computer engineering students at the Iran University of Science and Technology in 2019.
Her research interests include automated software testing concerning security testing,  applied data science and empirical software engineering.

\end{IEEEbiography}

\begin{IEEEbiography}[{\includegraphics[width=1in,height=1.25in,clip,keepaspectratio]{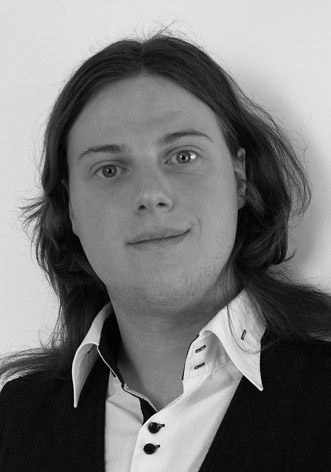}}]{Yoann Marquer} 
is a post-doctoral researcher at the Interdisciplinary Centre for Security, Reliability and Trust (SnT).
He obtained a ministerial scholarship for his PhD in Computer Science, obtained in 2015 from the University of Paris-Est Cr\'eteil, after wor\-king on the algorithmic completeness and implicit complexity of imperative programming languages.
He then worked with academic (universities and the Inria national institute) and industrial partners in several security-related, EU-founded projects.
His research interests concern non-functional properties of computation, especially security, including novel security metrics and countermeasures as well as source code testing and analysis to detect security vulnerabilities and refactoring to make it more secure.
\end{IEEEbiography}

\begin{IEEEbiography}[{\includegraphics[width=1in,height=1.25in,clip,keepaspectratio]{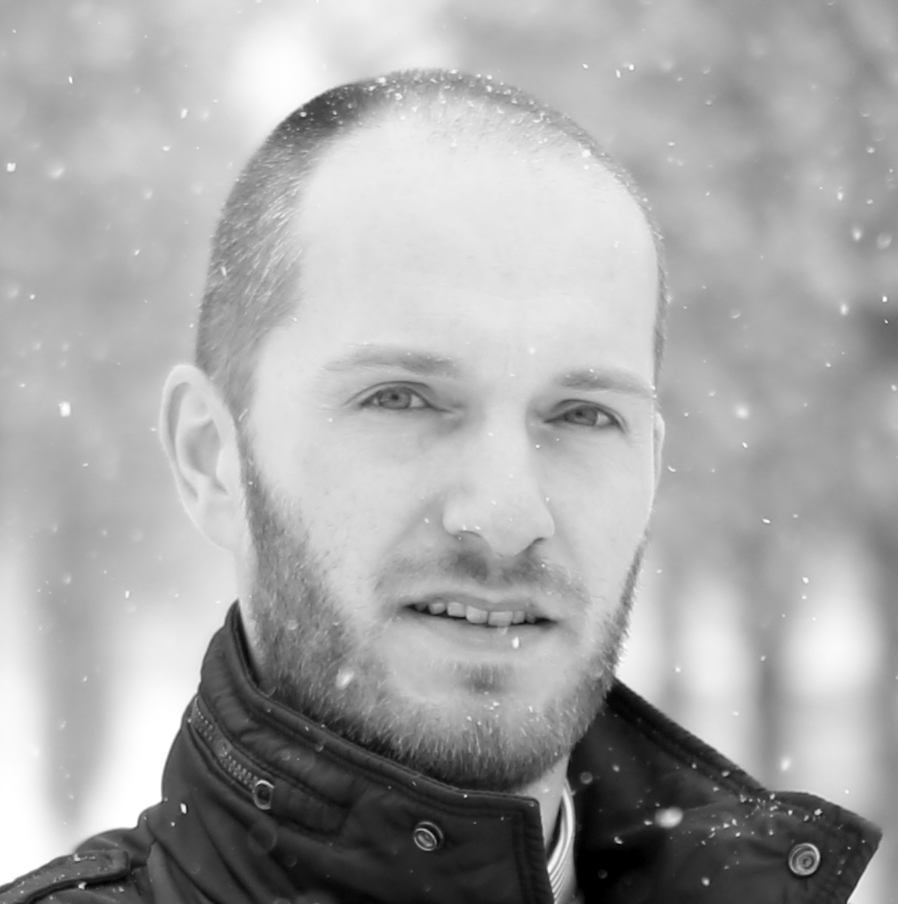}}]{Fabrizio Pastore}
is Chief Scientist II at the Interdisciplinary Centre for Security, Reliability and Trust (SnT), University of Luxembourg. He obtained his PhD in Computer Science in 2010 from the University of Milano - Bicocca.
His research interests concern automated software testing, including security testing and testing of AI-based systems; his work relies on the integrated analysis of different types of artefacts (e.g., requirements,  models, source code, and execution traces). He is active in several industry partnerships and national, ESA, and EU-funded research projects.
\end{IEEEbiography}
%\newpage
\begin{IEEEbiography}[{\includegraphics[width=1in,height=1.25in,clip,keepaspectratio]{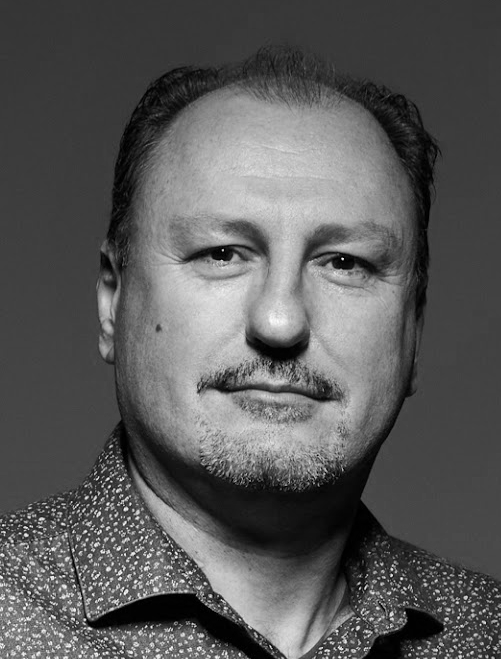}}]{Lionel C. Briand} is professor of software engineering and has shared appointments between (1) The University of Ottawa, Canada, and (2) The Lero SFI Centre---the national Irish centre for software research---hosted by the University of Limerick, Ireland. In collaboration with colleagues, for over 30 years, he has run many collaborative research projects with companies in the automotive, satellite, aerospace, energy, financial, and legal domains. Lionel has held various engineering, academic, and leading positions in seven countries.  He currently holds a Canada Research Chair (Tier 1) on "Intelligent Software Dependability and Compliance" and is the director of Lero, the national Irish centre for software research. Lionel was elevated to the grades of IEEE Fellow and ACM Fellow for his work on software testing and verification. Further, he was granted the IEEE Computer Society Harlan Mills award, the ACM SIGSOFT outstanding research award, and the IEEE Reliability Society engineer-of-the-year award. He also received an ERC Advanced grant in 2016 on modelling and testing cyber-physical systems, the most prestigious individual research award in the European Union and was elected a fellow of the Academy of Science, Royal Society of Canada in 2023. His research interests include: Testing and verification, trustworthy AI, search-based software engineering, model-driven development, requirements engineering, and empirical software engineering. More details can be found at: http://www.lbriand.info. 
\end{IEEEbiography}

\newpage



\section*{Supplementary material}

# AIM: Automated Input Set Minimization for Metamorphic Security Testing

Nazanin Bayati Chaleshtari, Yoann Marquer, Fabrizio Pastore, *Member, IEEE,* and Lionel C. Briand, *Fellow, IEEE*

## APPENDIX

We prove in this appendix theoretical results mentioned in the main paper and which demonstrates the correctness of the AIM approach. This includes Theorem 1 (Appendix C), presented in Section III-C on the objective functions, Proposition 3 and Theorem 2 (Appendix D) presented in Section VII-E on local dominance, Proposition 8 and Theorem 3 (Appendix E) presented in Section VII-F on dividing the problem, and desirable properties of roofer (Theorem 4) and miser (Theorem 5) populations (Appendix F) presented in Section VIII-F on the MOCCO population update. Appendices A and B present intermediate results.

### A. Input Coverage and Cost

**Lemma 1** (Non-Empty Coverage). *For every input $in$, we have $Cover(in) \neq \varnothing$.*

*Proof.* We consider only inputs containing at least one action (§VI-B). Let $act_1 = action(in, 1)$ be the first action of $in$. $act_1$ has an output in $outCl_1 = OutputClass(in, 1)$ (§VI-B). $act_1$ is in the action set $actSet_1 = ActionSet(outCl_1)$ (§VI-C). After clustering of this action set, let $bl_1 = Subclass(in, 1) = ActionSubclass(act_1, actSet_1)$ be the input block of the action $act_1$ executed in $in$ (§VI-C). Therefore, we have $bl_1 \in Cover(in)$. □

**Lemma 2** (Cost is Positive). *For every input $in$, we have $cost(in) > 0$. For every input set $I$, we have $cost(I) \geq 0$, and $cost(I) = 0$ if and only if $I = \varnothing$.*

*Proof.* Since we removed inputs with no cost (§III-A), for each input $in$, $cost(in) > 0$. The cost of an input set is the sum of the cost of its inputs, hence $cost(I) \geq 0$. In particular, if $I = \varnothing$, then $cost(I) = 0$. Finally, because the cost is positive, the only case where $cost(I) = 0$ is when $I = \varnothing$. □

### B. Redundancy

In this section, we prove that our characterization of redundancy is sound regarding the coverage of an input set

N. Bayati Chaleshtari and L. Briand are with the School of Electrical and Computer Engineering of University of Ottawa, Canada, Y. Marquer and F. Pastore are with the Interdisciplinary Centre for Security, Reliability, and Trust (SnT) of the University of Luxembourg, Luxembourg, and L. Briand is also with Lero SFI Centre for Software Research and University of Limerick, Ireland. Part of this work was done when L. Briand was affiliated with the Interdisciplinary Centre for Security, Reliability, and Trust (SnT) of the University of Luxembourg.
E-mail: n.bayati@uottawa.ca, yoann.marquer@uni.lu, fabrizio.pastore@uni.lu, lbriand@uottawa.ca
Manuscript received **Month DD**, 2024; revised **Month DD**, 2024.

(Proposition 1), then we introduce Lemma 4 which is useful to prove Lemma 6 in Appendix C and Proposition 8 in Appendix E.

First, if an input $in$ is in the considered input set $I$, then its redundancy is not negative. Indeed, there is always at least one input in $I$ that covers the input blocks covered by $in$, which is $in$ itself.

**Lemma 3** (Redundancy is Non-negative). *Let $I$ be an input set and $in$ be an input. If $in \in I$, then $redundancy(in, I) \geq 0$.*

*Proof.* Let $in$ be an input in $I$. Let $bl \in Cover(in)$ be any block covered by $in$. Because $bl \in Cover(in)$, we have that $in \in Inputs(bl)$ (§III-A). Moreover, $in \in I$, so $in \in Inputs(bl) \cap I$, thus we have $superpos(bl, I) \geq 1$ (§III-C). Therefore, for any $bl \in Cover(in)$ we have $superpos(bl, I) \geq 1$. So, $\min\{superpos(bl, I) \mid bl \in Cover(in)\} \geq 1$. By subtracting 1, we have $redundancy(in, I) \geq 0$ (§III-C). □

We prove that our characterization of redundancy is sound regarding the coverage of an input set. In other words, if an input is redundant in an input set, then it can be removed without reducing the coverage of the input set:

**Proposition 1** (Redundancy Soundness). *Let $I \subseteq I_{init}$ be an input set and $in \in I$. If $in \in Redundant(I)$, then $Cover(I \setminus \{in\}) = Cover(I)$.*

*Proof.* $Cover(I) = Cover(I \setminus \{in\}) \cup Cover(in)$ (§VI-C). We assume that $in$ is redundant in $I$. So, according to Lemma 3, we have $redundancy(in, I) > 0$ (§III-C). Thus, for each $bl \in Cover(in)$, we have that $superpos(bl, I) \geq 2$. Hence, according to our definition of superposition (§III-C), there are at least two inputs $in_1$ and $in_2$ in $I$ which also belong to $Inputs(bl)$. Because we have $bl \in Cover(in)$, we have that $in \in Inputs(bl)$ (§III-A). So, $in$ is one of these inputs. We assume this is the first one i.e., $in_1 = in$. Therefore, for each $bl \in Cover(in)$, there exists an input $in_2 \in Inputs(bl) \cap I$ which is not $in$. We denote $in_2 = in_{bl}$ this input. For each $bl \in Cover(in)$, we have $in_{bl} \in Inputs(bl)$. So, we have $bl \in Cover(in_{bl})$ (§III-A). Moreover, $in_{bl} \in I$ and is not $in$, so is in $I \setminus \{in\}$. Thus (§VI-C) we have $bl \in Cover(I \setminus \{in\})$. Therefore, $Cover(in) \subseteq Cover(I \setminus \{in\})$. Finally, $Cover(I) = Cover(I \setminus \{in\}) \cup Cover(in) = Cover(I \setminus \{in\})$. □

Finally, removing an input in an input set may update the redundancy of other inputs:

**Lemma 4** (Removing an Input). *Let $I$ be an input set and $in_2, in_1 \in I$ be two inputs in this input set.*

$$redundancy(in_1, I) - 1 \leq redundancy(in_1, I \setminus \{in_2\})$$
$$\leq redundancy(in_1, I)$$

*Proof.* Let $bl \in Cover(in_1)$. We denote $c_1 = |Inputs(bl) \cap I|$ and $c_2 = |Inputs(bl) \cap (I \setminus \{in_2\})|$. If $in_2 \in Inputs(bl)$ then $c_2 = c_1 - 1$. Otherwise, $c_2 = c_1$. Therefore, for every $bl \in Cover(in_1)$ we have $c_1 - 1 \leq c_2 \leq c_1$. This includes the input blocks minimizing the cardinality in the definition of redundancy (§III-C). Hence the result. □

### C. Valid Orders of Removal Steps

In this section, we first prove Lemma 5 that establishes that orders of removal steps contain inputs without repetition. Then, we introduce subsequences and prove Lemma 7, used in the rest of the section and in proofs of Appendix E. Finally, we prove Theorem 1 presented in §III-C.

**Lemma 5** (Order without Repetition). *Let $I$ be an input set. If $[in_1, \ldots, in_n] \in ValidOrders(I)$, then $in_1, \ldots, in_n$ are distinct.*

*Proof.* The proof is done by induction on $n$.

If $n = 0$, then $[in_1, \ldots, in_n] = [\,]$ is empty, hence there are no two identical inputs.

We now consider $[in_1, \ldots, in_n, in_{n+1}] \in ValidOrders(I)$. By induction hypothesis, $in_1, \ldots, in_n$ are distinct. According to the definition of valid order of removal steps (§III-C), we have $[in_1, \ldots, in_n, in_{n+1}]$ only if $in_{n+1} \in Redundant(I \setminus \{in_1, \ldots, in_n\})$.

Since $Redundant(I) = \{in \in I \mid redundancy(in, I) > 0\}$, we have $in_{n+1} \in I \setminus \{in_1, \ldots, in_n\}$.

Therefore, $in_{n+1}$ is distinct from the previous inputs $in_1, \ldots, in_n$. Hence the result, which concludes the induction step. □

**Lemma 6** (Redundant Inputs After Reduction). *Let $I$ be an input set. For each subset of inputs $\{in_1, \ldots, in_n\} \subseteq I$, we have $Redundant(I \setminus \{in_1, \ldots, in_n\}) \subseteq Redundant(I)$.*

*Proof.* The proof is done by induction on $n$.

If $n = 0$ then $I \setminus \{in_1, \ldots, in_n\} = I$, hence the result.

Otherwise, we assume by induction that $Redundant(I \setminus \{in_1, \ldots, in_n\}) \subseteq Redundant(I)$.

According to Lemma 4, redundancies can only decrease when performing a removal step. So, for every input $in_0 \in Redundant(I \setminus \{in_1, \ldots, in_n, in_{n+1}\})$, we have:

$$0 < redundancy(in_0, I \setminus \{in_1, \ldots, in_n, in_{n+1}\})$$
$$\leq redundancy(in_0, I \setminus \{in_1, \ldots, in_n\})$$

So, $Redundant(I \setminus \{in_1, \ldots, in_n, in_{n+1}\}) \subseteq Redundant(I \setminus \{in_1, \ldots, in_n\}) \subseteq Redundant(I)$. □

Since orders of removal steps contain inputs without repetition (Lemma 5), the index $index(in, \ell)$ of an input $in$ in a removal order $\ell$ containing $in$ is well defined. We leverage this to define subsequences.

**Definition 1** (Subsequences). *Let $[in_1, \ldots, in_n]$ and $[in'_1, \ldots, in'_m]$ be two orders of removal steps.*

We say that $[in'_1, \ldots, in'_m]$ is a *subsequence* of $[in_1, \ldots, in_n]$ if $[in'_1, \ldots, in'_m]$ can be obtained by removing some elements of $[in_1, \ldots, in_n]$, i.e., if $\{in \in [in'_1, \ldots, in'_m]\} \subseteq \{in \in [in_1, \ldots, in_n]\}$ and, for any two inputs $in_i, in_j \in [in'_1, \ldots, in'_m]$, if $index(in_i, [in_1, \ldots, in_n]) \leq index(in_j, [in_1, \ldots, in_n])$, then $index(in_i, [in'_1, \ldots, in'_m]) \leq index(in_j, [in'_1, \ldots, in'_m])$, where $index(in, \ell)$ is the index of input $in$ in $\ell$.

**Lemma 7.** *Let $I$ be an input set and $[in_1, \ldots, in_n], [in'_1, \ldots, in'_m]$ be two orders of removal steps in $I$. If $[in_1, \ldots, in_n] \in ValidOrders(I)$ and $[in'_1, \ldots, in'_m]$ is a subsequence of $[in_1, \ldots, in_n]$, then $[in'_1, \ldots, in'_m] \in ValidOrders(I)$.*

*Proof.* The proof is done by induction on $m$.

If $m = 0$ then $[in'_1, \ldots, in'_m] = [\,] \in ValidOrders(I)$.

Otherwise, we consider $[in'_1, \ldots, in'_m, in'_{m+1}]$ and we assume by induction that $[in'_1, \ldots, in'_m] \in ValidOrders(I)$.

Because $[in'_1, \ldots, in'_m, in'_{m+1}]$ is a subsequence of $[in_1, \ldots, in_n]$, there exists an index $i$ such that $in_i = in'_{m+1}$. By definition of valid removal steps (§III-C), we have $in_i \in Redundant(I \setminus \{in_1, \ldots, in_{i-1}\})$.

Moreover, $[in'_1, \ldots, in'_m, in'_{m+1}]$ is a subsequence of $[in_1, \ldots, in_i]$, so we have $\{in \in [in'_1, \ldots, in'_m]\} \subseteq \{in \in [in_1, \ldots, in_{i-1}]\}$. Hence, according to Lemma 6 we have $Redundant(I \setminus \{in_1, \ldots, in_{i-1}\}) \subseteq Redundant(I \setminus \{in'_1, \ldots, in'_m\})$. Therefore, $in'_{m+1} = in_i \in Redundant(I \setminus \{in'_1, \ldots, in'_m\})$.

$[in'_1, \ldots, in'_m] \in ValidOrders(I)$ and $in'_{m+1} \in Redundant(I \setminus \{in'_1, \ldots, in'_m\})$, so, according to our definition for valid orders of removal steps (§III-C), we have $[in'_1, \ldots, in'_m, in'_{m+1}] \in ValidOrders(I)$. □

Finally, we prove in Proposition 2 that inputs in a valid order of removal steps can be rearranged in any different order and the rearranged order of removal steps is also valid.

**Lemma 8** (Transposition of a Valid Order). *Let $I$ be an input set.*

*If $[in_1, \ldots, in_i, \ldots, in_j, \ldots, in_n] \in ValidOrders(I)$ and $1 \leq i < j \leq n$,*

*then $[in_1, \ldots, in_j, \ldots, in_i, \ldots, in_n] \in ValidOrders(I)$*

*where $in_i$ and $in_j$ were exchanged and the other inputs are left unchanged.*

*Proof.* The proof is done in four steps.

1) $[in_1, \ldots, in_{i-1}, in_j]$ is a subsequence of $[in_1, \ldots, in_{i-1}, in_i, \ldots, in_j, \ldots, in_n] \in ValidOrders(I)$. So, according to Lemma 7, we have $[in_1, \ldots, in_{i-1}, in_j] \in ValidOrders(I)$. Note that the first step even applies to the case $i = 1$.

2) If $j = i + 1$, then one can go directly to the third step. Otherwise, for each $i < k < j$, we denote:

$$I_{i+1} \stackrel{\text{def}}{=} I \setminus \{in_1, \ldots, in_{i-1}\}$$
$$I_{k+1} \stackrel{\text{def}}{=} I_k \setminus \{in_k\}$$

and, for the sake of conciseness, $I_k^i = I_k \setminus \{in_i\}$ and $I_k^j = I_k \setminus \{in_j\}$.

We now prove by induction on $i \leq k \leq j-1$ that:

$$[in_1, \ldots, in_{i-1}, in_j, in_{i+1}, \ldots, in_k] \in ValidOrders(I)$$

For the initialization $k = i$, we have from the first step:

$$[in_1, \ldots, in_{i-1}, in_j] \in ValidOrders(I)$$

We now assume by induction hypothesis on $i \leq k < j-1$ that:

$$[in_1, \ldots, in_{i-1}, in_j, in_{i+1}, \ldots, in_k] \in ValidOrders(I)$$

We first prove, for each $bl \in Cover(in_{k+1})$, that:

$$\left| Inputs(bl) \cap I_{k+1}^j \right| > 1$$

According to Lemma 5, $in_1, \ldots, in_i, \ldots, in_j, \ldots, in_n$ are distinct. We consider two cases.

a) We assume $bl \in Cover(in_j)$.

Because $I_j^i = I \setminus \{in_1, \ldots, in_{i-1}, in_i, in_{i+1}, \ldots, in_{j-1}\}$ and $I_{k+1}^j = I \setminus \{in_1, \ldots, in_{i-1}, in_{i+1}, \ldots, in_k, in_j\}$, we have $I_j^i \cup \{in_{k+1}\} \subseteq I_{k+1}^j \cup \{in_j\}$. Thus, for each $bl \in Cover(in_j) \cap Cover(in_{k+1})$:

$$\left| Inputs(bl) \cap I_j^i \right| + 1 \leq \left| Inputs(bl) \cap I_{k+1}^j \right| + 1$$

Because $[in_1, \ldots, in_i, \ldots, in_j, \ldots, in_n] \in ValidOrders(I)$, we have:

$$redundancy(in_j, I_j^i) > 0$$

So, for each $bl \in Cover(in_j)$, we have:

$$\left| Inputs(bl) \cap I_j^i \right| > 1$$

Thus, for each $bl \in Cover(in_j) \cap Cover(in_{k+1})$, we have:

$$\left| Inputs(bl) \cap I_{k+1}^j \right| > 1$$

b) We assume $bl \notin Cover(in_j)$.

For each $bl \in Cover(in_{k+1}) \setminus (Cover(in_i) \cup Cover(in_j))$, we have:

$$\left| Inputs(bl) \cap I_{k+1}^i \right| = \left| Inputs(bl) \cap I_{k+1}^j \right|$$

Moreover, for each $bl \in Cover(in_i) \setminus Cover(in_j)$, we have:

$$\left| Inputs(bl) \cap I_{k+1}^i \right| + 1 = \left| Inputs(bl) \cap I_{k+1}^j \right|$$

So, for each $bl \in Cover(in_{k+1}) \setminus Cover(in_j)$, we have:

$$\left| Inputs(bl) \cap I_{k+1}^i \right| \leq \left| Inputs(bl) \cap I_{k+1}^j \right|$$

Because $[in_1, \ldots, in_i, \ldots, in_j, \ldots, in_n] \in ValidOrders(I)$, we have:

$$redundancy(in_{k+1}, I_{k+1}^i) > 0$$

So, for each $bl \in Cover(in_{k+1})$, we have:

$$\left| Inputs(bl) \cap I_{k+1}^i \right| > 1$$

Thus, for each $bl \in Cover(in_{k+1}) \setminus Cover(in_j)$, we have:

$$\left| Inputs(bl) \cap I_{k+1}^j \right| > 1$$

So, we proved by case, for each $bl \in Cover(in_{k+1})$, that:

$$\left| Inputs(bl) \cap I_{k+1}^j \right| > 1$$

Thus, we have $redundancy(in_{k+1}, I_{k+1}^j) > 0$.

Moreover, according to Lemma 5, $in_1, \ldots, in_i, \ldots, in_j, \ldots, in_n$ are distinct, so $in_{k+1} \in I_{k+1}^j$. Hence, $in_{k+1} \in Redundant(I_{k+1}^j)$. By induction hypothesis, we have:

$$[in_1, \ldots, in_{i-1}, in_j, in_{i+1}, \ldots, in_k] \in ValidOrders(I)$$

So, according to the definition of valid removal steps (§III-C):

$$[in_1, \ldots, in_{i-1}, in_j, in_{i+1}, \ldots, in_k, in_{k+1}] \in ValidOrders(I)$$

which concludes the induction step.

Therefore, we proved by induction:

$$[in_1, \ldots, in_{i-1}, in_j, in_{i+1}, \ldots, in_{j-1}] \in ValidOrders(I)$$

3) We now prove that, for each $bl \in Cover(in_i)$, we have:

$$\left| Inputs(bl) \cap I_j^j \right| > 1$$

where $I_j^i$ and $I_j^j$ are defined in step two. We consider two cases.

a) We assume $bl \in Cover(in_j)$. Because $I_j^i \cup \{in_i\} = I_j^j \cup \{in_j\}$, for each $bl \in Cover(in_i) \cap Cover(in_j)$, we have:

$$\left| Inputs(bl) \cap I_j^i \right| + 1 = \left| Inputs(bl) \cap I_j^j \right| + 1$$

Because $[in_1, \ldots, in_i, \ldots, in_j, \ldots, in_n] \in ValidOrders(I)$, we have:

$$redundancy(in_j, I_j^i) > 0$$

So, for each $bl \in Cover(in_j)$, we have:

$$\left| Inputs(bl) \cap I_j^i \right| > 1$$

Thus, for each $bl \in Cover(in_i) \cap Cover(in_j)$, we have:

$$\left| Inputs(bl) \cap I_j^j \right| > 1$$

b) We assume $bl \notin Cover(in_j)$. In that case, for each $bl \in Cover(in_i) \setminus Cover(in_j)$, we have:

$$\left| Inputs(bl) \cap I_j^i \right| + 1 = \left| Inputs(bl) \cap I_j^j \right|$$

We consider two cases.

i) There exists $i+1 \leq k \leq j-1$ such that $bl \in Cover(in_k)$. In that case, let $k_{\max}$ be the largest. So, we have:

$$\left( Inputs(bl) \cap I_j^i \right) \cup \{in_{k_{\max}}\} = Inputs(bl) \cap I_{k_{\max}}^i$$

Hence:

$$\left| Inputs(bl) \cap I_j^i \right| + 1 = \left| Inputs(bl) \cap I_{k_{\max}}^i \right|$$

Thus:

$$\left| Inputs(bl) \cap I_j^j \right| = \left| Inputs(bl) \cap I_{k_{\max}}^i \right|$$

Because $[in_1, \ldots, in_i, \ldots, in_j, \ldots, in_n] \in ValidOrders(I)$, we have:

$$redundancy(in_{k_{\max}}, I_{k_{\max}}^i) > 0$$

So, because $bl \in Cover(in_{k_{\max}})$, we have:

$$\left|Inputs(bl) \cap I^i_{k_{\max}}\right| > 1$$

Finally:

$$\left|Inputs(bl) \cap I^j_j\right| > 1$$

ii) Otherwise, for each $i+1 \leq k \leq j-1$, we have $bl \notin Cover(in_k)$. Note that this is also the case if $j = i+1$. In that case, we have:

$$\left(Inputs(bl) \cap I^i_j\right) \cup \{in_i\} = Inputs(bl) \cap I_i$$

where $I_i = I \setminus \{in_1, \ldots, in_{i-1}\}$. So:

$$\left|Inputs(bl) \cap I^i_j\right| + 1 = |Inputs(bl) \cap I_i|$$

Thus:

$$\left|Inputs(bl) \cap I^j_j\right| = |Inputs(bl) \cap I_i|$$

Because $[in_1, \ldots, in_i, \ldots, in_j, \ldots, in_n] \in ValidOrders(I)$, we have:

$$redundancy(in_i, I_i) > 0$$

So, because $bl \in Cover(in_i)$, we have:

$$|Inputs(bl) \cap I_i| > 1$$

Finally:

$$\left|Inputs(bl) \cap I^j_j\right| > 1$$

This concludes the proof by case for $bl \in Cover(in_i) \setminus Cover(in_j)$. Therefore, we proved by case that, for each $bl \in Cover(in_i)$, we have:

$$\left|Inputs(bl) \cap I^j_j\right| > 1$$

Thus, we have $redundancy(in_i, I^j_j) > 0$.

Moreover, according to Lemma 5, $in_1, \ldots, in_i, \ldots, in_j, \ldots, in_n$ are distinct, so $in_i \in I^j_j$. Thus, $in_i \in Redundant(I^j_j)$.

Finally, we have from the second step (or the first one, if $j = i+1$):

$$[in_1, \ldots, in_{i-1}, in_j, in_{i+1}, \ldots, in_{j-1}] \in ValidOrders(I)$$

Therefore, according to the definition of valid orders of removal steps (§III-C):

$$[in_1, \ldots, in_{i-1}, in_j, in_{i+1}, \ldots, in_{j-1}, in_i] \in ValidOrders(I)$$

4) Finally, if $j = n$ then the proof is complete. Otherwise, we prove by induction on $j \leq k \leq n$ that:

$$[in_1, \ldots, in_j, \ldots, in_i, in_{j+1}, \ldots, in_k] \in ValidOrders(I)$$

For the initialization $k = j$, we have from the third step:

$$[in_1, \ldots, in_j, \ldots, in_i] \in ValidOrders(I)$$

We now assume by induction hypothesis on $j \leq k < n$ that:

$$[in_1, \ldots, in_j, \ldots, in_i, in_{j+1}, \ldots, in_k] \in ValidOrders(I)$$

Because $[in_1, \ldots, in_i, \ldots, in_j, \ldots, in_n] \in ValidOrders(I)$, we have:

$$in_{k+1} \in Redundant(I \setminus \{in_1, \ldots, in_i, \ldots, in_j, \ldots, in_k\})$$

Moreover, we have:

$$\{in_1, \ldots, in_i, \ldots, in_j, \ldots, in_k\}$$
$$= \{in_1, \ldots, in_j, \ldots, in_i, \ldots, in_k\}$$

So:

$$in_{k+1} \in Redundant(I \setminus \{in_1, \ldots, in_j, \ldots, in_i, \ldots, in_k\})$$

Thus, according to the definition of valid orders of removal steps (§III-C):

$$[in_1, \ldots, in_j, \ldots, in_i, in_{j+1}, \ldots, in_k, in_{k+1}]$$
$$\in ValidOrders(I)$$

which concludes the induction step.

Therefore, we proved by induction the lemma:

$$[in_1, \ldots, in_j, \ldots, in_i, in_{j+1}, \ldots, in_n] \in ValidOrders(I)$$

$\square$

The issue with our definition of the gain (§III-C) is that, to compute $gain(I)$ of an input set $I$, one has to try all the possible order of removal steps to determine which ones are valid. If $I$ contains $n$ inputs and we consider orders of $0 \leq k \leq n$ removal steps, there are $n \times \cdots \times (n-k+1) = \frac{!n}{!(n-k)}$ possibilities to investigate, which is usually large.

Thus, we present an optimization to reduce the cost of computing the gain. First, we prove in Proposition 2 that, while the order of inputs matters to determine if an order of removal steps is valid, it does not matter anymore once we know the order is valid. In other words, inputs in a valid order of removal steps may be rearranged in any different order, and the rearranged order of removal steps would be valid. Formally, if $[in_1, \ldots, in_n]$ is a valid order of removal steps, then $[in_{\sigma(1)}, \ldots, in_{\sigma(n)}]$ is also a valid order, when $\sigma$ denotes a permutation of the $n$ inputs, i.e., the same inputs but (potentially) in a different order:

**Proposition 2** (Permutation of a Valid Order)**.** *Let $I$ be an input set. If $[in_1, \ldots, in_n] \in ValidOrders(I)$ and $\sigma$ is a permutation on $n$ elements, then $[in_{\sigma(1)}, \ldots, in_{\sigma(n)}] \in ValidOrders(I)$.*

*Proof.* Every permutation of a finite set can be expressed as the product of transpositions [1] (p.60). Let $m$ be the number of such transpositions for $\sigma$. We prove the result by induction on $m$.

If $m = 0$, then $\sigma$ is the identity and we have by hypothesis:

$$[in_{\sigma(1)}, \ldots, in_{\sigma(n)}] \in ValidOrders(I)$$

If $\sigma = \tau_{m+1} \circ \tau_m \circ \cdots \circ \tau_1$, then we denote $\sigma' = \tau_m \circ \cdots \circ \tau_1$. By induction hypothesis, we have:

$$[in_{\sigma'(1)}, \ldots, in_{\sigma'(n)}] \in ValidOrders(I)$$

Then, by applying Lemma 8 to the transposition $(ij) = \tau_{m+1}$, we have:

$$[in_{\tau_{m+1} \circ \sigma'(1)}, \ldots, in_{\tau_{m+1} \circ \sigma'(n)}] \in ValidOrders(I)$$

Finally, because $\sigma = \tau_{m+1} \circ \sigma'$, we conclude the induction step. □

Since inputs in a valid order of removal steps are always distinct (Lemma 5) and their order does not make a difference (Proposition 2), such orders can simply be seen as sets even if, in practice, for IMPRO and MOCCO, orders of removal steps are implemented as lists. More precisely, two valid orders of removal steps are equivalent if they correspond to the same set of inputs. We leverage this property to reduce the number of orders of removal steps to consider, e.g., since $[in_1, in_2]$ is equivalent to $[in_2, in_1]$, we only have to check that $[in_1, in_2]$ is valid to know whether $[in_2, in_1]$ is valid or not. This means that, when inputs are provided in a given order $[in_1, \ldots, in_n]$, these tools need only to check for valid orders of removal steps in increasing order (hence without backtracking on previous inputs), i.e., $[in_{i_1}, \ldots, in_{i_m}]$, where $1 \leq i_1 < \cdots < i_m \leq n$. We call *canonical orders* such orders of removal steps. For instance, $[in_2, in_4, in_7]$ is in canonical order, but $[in_4, in_2, in_7]$ is not. Thus, to save time when checking the validity of orders of removal steps, we consider only canonical orders.

**Theorem 1** (Canonical Order). *Let $I = \{in_1, \ldots, in_n\}$ be an input set. There exists $[in_{i_1}, \ldots, in_{i_m}] \in ValidOrders(I)$ such that $1 \leq i_1 < \cdots < i_m \leq n$ and:*

$$\sum_{1 \leq j \leq m} cost(in_{i_j}) = gain(I)$$

*Proof.* Let $[in_{i'_1}, \ldots, in_{i'_m}] \in ValidOrders(I)$ be a valid order of removal steps with a maximal cumulative cost i.e., according to our definition of the gain (§III-C):

$$\sum_{1 \leq j \leq m} cost(in_{i'_j}) = gain(I)$$

If $m = 0$, then $[\,]$ satisfies the theorem for a gain $= 0$.

Otherwise, we assume $m > 0$. According to Lemma 5, $[in_{i'_1}, \ldots, in_{i'_m}]$ contains $m$ distinct inputs. Let $\sigma \in S_m$ be the permutation such that:

$$[in_{\sigma(i'_1)}, \ldots, in_{\sigma(i'_m)}] = [in_{i_1}, \ldots, in_{i_m}]$$

with $1 \leq i_1 < \cdots < i_m \leq n$.

According to Proposition 2, we have $[in_{i_1}, \ldots, in_{i_m}] \in ValidOrders(I)$.

Moreover, because a permutation of the elements of a sum does not change the value of the sum, we have:

$$\sum_{1 \leq j \leq m} cost(in_{i_j}) = \sum_{1 \leq j \leq m} cost(in_{i'_j}) = gain(I)$$

Hence the theorem. □

Theorem 1 allows us to focus on inputs in increasing order, instead of exploring all possible input orders. Thus, this optimization saves computation steps and makes computing the gain more tractable. More precisely, since there are $!k$ ways of ordering $k$ selected inputs, there are "only" $\frac{!n}{!k!(n-k)}$ remaining possibilities to investigate, instead of $\frac{!n}{!(n-k)}$.

### D. Local Dominance

In this section we prove, as presented in §VII-E, that the local dominance relation is local (Proposition 3), hence is faithful to its name, and that non locally-dominated inputs, as a whole, locally-dominate all the locally dominated inputs (Theorem 2). We start by the locality property.

**Proposition 3** (Local Dominance is Local). *Let $in_1 \in I_{search}$ be a remaining input and $S \subseteq I_{search}$ be a subset of the remaining inputs. If $in_1 \sqsubseteq S$ then $in_1 \sqsubseteq S \cap \{in_2 \in I_{search} \mid in_1 \sqcap in_2\}$.*

*Proof.* We assume $in_1 \sqsubseteq S$. So, by definition of local dominance (§VII-E), we have $in_1 \notin S$, $Cover(in_1) \subseteq Cover(S)$, and $cost(in_1) \geq cost(S)$. First, because $in_1 \notin S$, we have $in_1 \notin S \cap \{in_2 \in I_{search} \mid in_1 \sqcap in_2\}$.

Moreover, for every $in_2 \in I_{search}$, if there is no overlap between $in_1$ and $in_2$ then, according to the overlapping relation (§VII-E), we have $Cover(in_1) \cap Cover(in_2) = \varnothing$. Thus:

$$Cover(in_1) \cap Cover(S \cap \{in_2 \in I_{search} \mid \neg in_1 \sqcap in_2\}) = \varnothing$$

Hence, because $Cover(in_1) \subseteq Cover(S)$, we have:

$Cover(in_1)$
$= Cover(in_1) \cap Cover(S)$
$= Cover(in_1) \cap (Cover(S \cap \{in_2 \in I_{search} \mid in_1 \sqcap in_2\})$
$\quad \cup Cover(S \cap \{in_2 \in I_{search} \mid \neg in_1 \sqcap in_2\}))$
$= (Cover(in_1) \cap Cover(S \cap \{in_2 \in I_{search} \mid in_1 \sqcap in_2\}))$
$\quad \cup (Cover(in_1)$
$\quad\quad \cap Cover(S \cap \{in_2 \in I_{search} \mid \neg in_1 \sqcap in_2\}))$
$= (Cover(in_1) \cap Cover(S \cap \{in_2 \in I_{search} \mid in_1 \sqcap in_2\}))$

So, $Cover(in_1) \subseteq Cover(S \cap \{in_2 \in I_{search} \mid in_1 \sqcap in_2\})$.

Finally, $cost(S) \geq cost(S \cap \{in_2 \in I_{search} \mid in_1 \sqcap in_2\})$. Thus, because $cost(in_1) \geq cost(S)$, we have $cost(in_1) \geq cost(S \cap \{in_2 \in I_{search} \mid in_1 \sqcap in_2\})$.

Therefore, $in_1 \sqsubseteq S \cap \{in_2 \in I_{search} \mid in_1 \sqcap in_2\}$. □

We now prove that the local dominance relation is asymmetric (Proposition 5). First, because an input always covers some objectives, it can be locally dominated only by a non-empty input subset.

**Lemma 9** (Non-Empty Local Dominance). *Let $in \in I_{search}$ be an input and $S \subseteq I_{search}$ be a subset of the remaining inputs. If $in \sqsubseteq S$ then $S \neq \varnothing$.*

*Proof.* The proof is done by contradiction. If $S = \varnothing$ then $Cover(S) = \varnothing$. Because $in \sqsubseteq S$ we have $Cover(in) \subseteq Cover(S)$. So, $Cover(in) = \varnothing$, which contradicts Lemma 1. □

Second, when two inputs locally dominates each other, they are equivalent in the sense of the equivalence relation from §VII-D (two inputs are equivalent if they have the same coverage and cost), that we denote $\equiv$ in the following. But, because we removed duplicates in §VII-D, this case can happen only if the inputs are the same, which is excluded by the definition of local dominance (§VII-E).

**Lemma 10** (Local Dominance for Singletons is Asymmetric). *Let $in_1, in_2 \in I_{search}$ be two inputs. If $in_1 \sqsubseteq \{in_2\}$, then $in_2 \not\sqsubseteq \{in_1\}$.*

*Proof.* We assume by contradiction that $in_1 \sqsubseteq \{in_2\}$ and $in_2 \sqsubseteq \{in_1\}$.

Hence, we have: 1) $Cover(in_1) \subseteq Cover(in_2)$ and $Cover(in_2) \subseteq Cover(in_1)$, so $Cover(in_1) = Cover(in_2)$. 2) $cost(in_1) \geq cost(in_2)$ and $cost(in_2) \geq cost(in_1)$, so $cost(in_1) = cost(in_2)$. So, we have $in_1 \equiv in_2$.

Because we removed duplicates in §VII-D, we have $in_1 = in_2$. So, $in_1 \sqsubseteq \{in_1\}$, which contradicts in the definition of local dominance (§VII-E) that an input cannot locally dominates itself. □

Note that this result is also useful to prove transitivity (Proposition 7).

Third, because the cost of an input is positive, if an input is locally dominated by several inputs, then they have a strictly smaller cost.

**Lemma 11** (Cost Hierarchy). *Let $in_1 \in I_{search}$ be an input and $S \subseteq I_{search}$ be a subset of the remaining inputs. If $in_1 \sqsubseteq S$ and $|S| \geq 2$ then for every $in_2 \in S$ we have $cost(in_2) < cost(in_1)$.*

*Proof.* Because $in_1 \sqsubseteq S$ we have $cost(in_1) \geq cost(S)$. So, for every $in_2 \in S$ we have $cost(in_2) \leq cost(in_1)$. If $|S| \geq 2$, then the inequality is strict because, according to Lemma 2, the cost is positive $cost(in_2) > 0$. □

Finally, we prove as expected that the local dominance relation is asymmetric.

**Proposition 4** (Local Dominance for Subsets is Asymmetric). *Let $in_1, in_2 \in I_{search}$ be two inputs and $S_1, S_2 \subseteq I_{search}$ be two subsets of the remaining inputs. If $in_1 \sqsubseteq S_1$, $in_2 \in S_1$, and $in_2 \sqsubseteq S_2$, then $in_1 \notin S_2$.*

*Proof.* The proof is made by contradiction. We assume $in_1 \in S_2$ and prove a contradiction in different cases for $|S_1|$ and $|S_2|$.

$|S_1| = 0$ or $|S_2| = 0$ are not possible, because this would contradict Lemma 9.

If $|S_1| = 1$ and $|S_2| = 1$, then $S_1 = \{in_2\}$ and $S_2 = \{in_1\}$. Thus, $in_1 \sqsubseteq \{in_2\}$ and $in_2 \sqsubseteq \{in_1\}$, which contradicts Lemma 10.

If $|S_1| = 1$ then $S_1 = \{in_2\}$. Because $in_1 \sqsubseteq S_1$, we have $cost(in_1) \geq cost(in_2)$. Moreover, because $in_2 \sqsubseteq S_2$ and $in_1 \in S_2$, if $|S_2| \geq 2$ then according to Lemma 11 we have $cost(in_1) < cost(in_2)$, hence the contradiction.

Because $in_1 \sqsubseteq S_1$ and $in_2 \in S_1$, if $|S_1| \geq 2$ then according to Lemma 11 we have $cost(in_2) < cost(in_1)$. Because $in_2 \sqsubseteq S_2$ and $in_1 \in S_2$, if $|S_2| \geq 2$ then according to Lemma 11 we have $cost(in_1) < cost(in_2)$. Hence the contradiction $cost(in_1) < cost(in_1)$. □

Based on our definition of local dominance for subsets (§VII-E), we introduce the corresponding definition for inputs, in order to state more easily Corollary 1, then prove Theorem 2.

**Definition 2** (Local Dominance). The input $in_1 \in I_{search}$ *locally-dominates* the input $in_2 \in I_{search}$, denoted $in_1 \hookrightarrow in_2$, if there exists a subset $S \subseteq I_{search}$ such that $in_1 \in S$ and $in_2 \sqsubseteq S$.

**Proposition 5** (Local Dominance for Inputs is Asymmetric). *The $\hookrightarrow$ relation is asymmetric i.e., for every $in_1, in_2 \in I_{search}$, if $in_2 \hookrightarrow in_1$, then $in_1 \not\hookrightarrow in_2$.*

*Proof.* If $in_2 \hookrightarrow in_1$ then there exists $S_1 \subseteq I_{search}$ such that $in_2 \in S_1$ and $in_1 \sqsubseteq S_1$. The proof is done by contradiction, assuming $in_1 \hookrightarrow in_2$. Hence, there exists $S_2 \subseteq I_{search}$ such that $in_1 \in S_2$ and $in_2 \sqsubseteq S_2$ According to Proposition 4 we have $in_1 \notin S_2$, hence the contradiction. □

We now prove that the local dominance relation is transitive (Proposition 7).

**Proposition 6** (Local Dominance for Subsets is Transitive). *Let $in_1, in_2 \in I_{search}$ be two inputs and $S_1, S_2 \subseteq I_{search}$ be two subsets of the remaining inputs.*

*If $in_1 \sqsubseteq S_1$, $in_2 \in S_1$, and $in_2 \sqsubseteq S_2$, then $in_1 \sqsubseteq (S_1 \setminus \{in_2\}) \cup S_2$.*

*Proof.* Because $Cover(in_1) \subseteq Cover(S_1)$, $in_2 \in S_1$, and $Cover(in_2) \subseteq Cover(S_2)$, we have:

$$Cover(in_1) \subseteq Cover(S_1 \setminus \{in_2\}) \cup Cover(S_2)$$
$$= Cover((S_1 \setminus \{in_2\}) \cup S_2)$$

Moreover, because $cost(in_1) \geq cost(S_1)$, $in_2 \in S_1$, and $cost(in_2) \geq cost(S_2)$, we have:

$$cost(in_1) \geq cost(S_1 \setminus \{in_2\}) + cost(S_2)$$
$$\geq cost((S_1 \setminus \{in_2\}) \cup S_2)$$

Finally, because $in_1 \sqsubseteq S_1$ we have $in_1 \notin S_1$. We prove $in_1 \notin (S_1 \setminus \{in_2\}) \cup S_2$ by contradiction. If $in_1 \in (S_1 \setminus \{in_2\}) \cup S_2$ then, because $in_1 \notin S_1$, we have $in_1 \in S_2$. In that case, we have

$cost(in_1)$
$\geq cost(S_1 \setminus \{in_2\}) + cost(S_2)$
$= cost(S_1 \setminus \{in_2\}) + cost(S_2 \setminus \{in_1\}) + cost(in_1)$
$\geq cost(in_1)$

Hence, by subtracting $cost(in_1)$, we have:

$$cost(S_1 \setminus \{in_2\}) + cost(S_2 \setminus \{in_1\}) = 0$$

Thus, according to Lemma 2, we have $S_1 = \{in_2\}$ and $S_2 = \{in_1\}$. Therefore $in_1 \sqsubseteq \{in_2\}$ and $in_2 \sqsubseteq \{in_1\}$, which contradicts Lemma 10. □

**Proposition 7** (Local Dominance for Inputs is Transitive). *The $\hookrightarrow$ relation is transitive i.e., for every $in_1, in_2, in_3 \in I_{search}$, if $in_3 \hookrightarrow in_2$ and $in_2 \hookrightarrow in_1$, then $in_3 \hookrightarrow in_1$.*

*Proof.* If $in_3 \hookrightarrow in_2$ and $in_2 \hookrightarrow in_1$, then there exists $S_1, S_2 \subseteq I_{search}$ such that $in_2 \in S_1$, $in_1 \sqsubseteq S_1$, $in_3 \in S_2$, and $in_2 \sqsubseteq S_2$. According to Proposition 6, we have $in_1 \sqsubseteq (S_1 \setminus \{in_2\}) \cup S_2$. Hence, because $in_3 \in S_2 \subseteq (S_1 \setminus \{in_2\}) \cup S_2$ we have $in_3 \hookrightarrow in_1$. □

We finally prove that the local dominance relation is acyclic.

**Corollary 1** (Local Dominance is Acyclic). *There exists no cycle $in_1 \hookrightarrow \ldots \hookrightarrow in_n \hookrightarrow in_1$ such that $in_1, \ldots, in_n \in I_{search}$.*

*Proof.* The proof is done by contradiction, by assuming the existence of such a cycle $in_1 \hookrightarrow \ldots \hookrightarrow in_n \hookrightarrow in_1$. According to Proposition 7, we have by transitivity that $in_1 \hookrightarrow in_1$. Then, according to Proposition 5, we have by asymmetry $in_1 \not\hookrightarrow in_1$. Hence the contradiction. □

Finally, we use the transitivity (Proposition 7 and Proposition 6) and the acyclicity (Corollary 1) of the local-dominance relation to prove that non locally-dominated inputs, as a whole, locally-dominate all the locally dominated inputs.

**Theorem 2** (Local Dominance Hierarchy). *For every locally-dominated input $in \in I_{search}$, there exists a subset $S \subseteq I_{search}$ of not locally-dominated inputs such that $in \sqsubseteq S$.*

*Proof.* For each input $in_0 \in I_{search}$ we denote:

$$LocDoms(in_0) \stackrel{\text{def}}{=} \{in \in I_{search} \mid in \hookrightarrow in_0\}$$

the input set which locally dominate $in_0$.

Let $in_0 \in I_{search}$ be an input. We prove by induction on $LocDoms(in_0)$ that either $in_0$ is not locally dominated or there exists a subset $S \subseteq I_{search}$ of not locally-dominated inputs such that $in_0 \sqsubseteq S$.

For the initialization, we consider $LocDoms(in_0) = \varnothing$. In that case, $in_0$ is not locally dominated.

For the induction step, we consider $LocDoms(in_0) \neq \varnothing$, so $in_0$ is a locally dominated input. Hence, there exists $S_0 \subseteq I_{search}$ such that $in_0 \sqsubseteq S_0$. We consider two cases.

Either every input in $S_0$ is not locally dominated. In that case $S = S_0$ satisfies the inductive property.

Or there exists $n \geq 1$ input $in_1, \ldots, in_n$ in $S_0$ which are locally dominated. The rest of the proof is done in two steps.

First, let $1 \leq i \leq n$. We prove that $LocDoms(in_i) \subset LocDoms(in_0)$, with a strict inclusion.

Because $in_i \in S_0$ and $in_0 \sqsubseteq S_0$, we have $in_i \hookrightarrow in_0$. Hence, $in_i \in LocDoms(in_0)$.

Let $in \in LocDoms(in_i)$, so we have $in \hookrightarrow in_i$. So, according to Proposition 7, we have by transitivity $in \hookrightarrow in_0$, thus $in \in LocDoms(in_0)$. Hence, $LocDoms(in_i) \subseteq LocDoms(in_0)$.

According to Corollary 1, there is no cycle so $in_i \not\hookrightarrow in_i$. Hence, $in_i \notin LocDoms(in_i)$.

Finally, we have $LocDoms(in_i) \subseteq LocDoms(in_0)$ and $in_i \in LocDoms(in_0) \setminus LocDoms(in_i)$. Therefore $LocDoms(in_i) \subset LocDoms(in_0)$.

Because $LocDoms(in_i) \subset LocDoms(in_0)$, we can apply the induction hypothesis on $in_i$, which is locally dominated. Therefore, for each $1 \leq i \leq n$, there exists a subset $S_i \subseteq I_{search}$ of not locally-dominated inputs such that $in_i \sqsubseteq S_i$.

Second, we use these $S_i$ to define by induction on $0 \leq m < n$ the following input sets:

$$\begin{aligned} I_0 &= S_0 \\ I_{m+1} &= (I_m \setminus \{in_{m+1}\}) \cup S_{m+1} \end{aligned}$$

and we prove by induction on $0 \leq m \leq n$ that $in_0 \sqsubseteq I_m$ and that either, $m < n$ and $in_{m+1}, \ldots, in_n$ are the only locally dominated inputs in $I_m$, or $m = n$ and $I_m$ contains no locally dominated input.

For the initialization, we have $I_0 = S_0$ and the inductive property is satisfied because $in_0 \sqsubseteq S_0$ and $in_1, \ldots, in_n$ are the only locally dominated inputs in $S_0$.

For the induction step $m+1 \leq n$, we assume that $in_0 \sqsubseteq I_m$ and that $in_{m+1}, \ldots, in_n$ are the only locally dominated inputs in $I_m$.

Because $I_{m+1} = (I_m \setminus \{in_{m+1}\}) \cup S_{m+1}$ and $S_{m+1}$ contains no locally-dominated input, we have that either, $m+1 < n$ and $in_{m+2}, \ldots, in_n$ are the only locally dominated inputs in $I_{m+1}$, or $m+1 = n$ and $I_{m+1}$ contains no locally-dominated input.

Moreover, because $in_0 \sqsubseteq I_m$, $in_{m+1} \in I_m$, and $in_{m+1} \sqsubseteq S_{m+1}$, then according to Proposition 6, we have $in_0 \sqsubseteq (I_m \setminus \{in_{m+1}\}) \cup S_{m+1}$. Hence $in_0 \sqsubseteq I_{m+1}$, which concludes the induction step.

Therefore, $in_0 \sqsubseteq I_n$ and $I_n$ contains no locally-dominated input. Thus, $S = I_n$ satisfies the inductive property on $LocDoms(in_0)$. Hence the claim for not locally-dominated inputs. □

### E. Dividing the Problem

In this section, we prove that redundancy updates are local (Lemma 12), that reductions on different connected components can be performed independently (Proposition 8), and finally that the gain can be independently computed on each connected component (Theorem 3). The main purpose of the section is to justify we can divide our problem into subproblems (§VII-F). We start by the locality.

**Lemma 12** (Redundancy Updates are Local). *Let $I$ be an input set and let $in_1, in_2 \in I$.*

*If $redundancy(in_2, I \setminus \{in_1\}) \neq redundancy(in_2, I)$, then $in_1 \sqcap in_2$.*

*Proof.* The proof is done by contraposition. If $in_1$ and $in_2$ do not overlap, then $Cover(in_1) \cap Cover(in_2) = \varnothing$. So, for every $bl \in Cover(in_2)$, we have $in_1 \notin Inputs(bl)$, and thus $Inputs(bl) \cap (I \setminus \{in_1\}) = Inputs(bl) \cap I$. Therefore, $redundancy(in_2, I \setminus \{in_1\}) = redundancy(in_2, I)$. □

Then, we prove the independence of removal steps performed on different components.

**Lemma 13** (Independent Redundancies). *Let $I$ be an input set. For each connected component $C \in Comps(I)$, for each input $in_0 \in C$, and for each order of removal steps $[in_1, \ldots, in_n]$, if $in_1, \ldots, in_n \notin C$, then $redundancy(in_0, I \setminus \{in_1, \ldots, in_n\}) = redundancy(in_0, I)$.*

*Proof.* $Comps(I)$ are the connected components for the redundant inputs in $I$, hence they are disjoint. Therefore, for each $in_0 \in C$, because $in_1, \ldots, in_n \notin C$, we have that $in_0$ do not overlap with any of the $in_1, \ldots, in_n$. Therefore, according to Lemma 12, we have $redundancy(in_0, I) = redundancy(in_0, I \setminus \{in_1\}) = \cdots = redundancy(in_0, I \setminus \{in_1, \ldots, in_n\})$. □

**Proposition 8** (Independent Removal Steps).

Let $I$ be an input set and let $C_1, \ldots, C_c \in Comps(I)$ denote $c$ connected components. If, for each $0 \leq i \leq c$, there exists inputs $in_1^i, \ldots, in_{n_i}^i \in C_i$ such that $[in_1^i, \ldots, in_{n_i}^i] \in ValidOrders(I)$, then:

$$[in_1^1, \ldots, in_{n_1}^1] + \cdots + [in_1^c, \ldots, in_{n_c}^c] \in ValidOrders(I)$$

where $+$ denotes sequence concatenation.

*Proof.* The proof is done by induction on $c$.

If $c = 0$, then $[in_1^1, \ldots, in_{n_1}^1] + \cdots + [in_1^c, \ldots, in_{n_c}^c] = [] \in ValidOrders(I)$.

We assume by induction that $[in_1^1, \ldots, in_{n_1}^1] + \cdots + [in_1^c, \ldots, in_{n_c}^c] \in ValidOrders(I)$.

We denote $\ell$ this order of removal steps and for the sake of simplicity we also denote $S = \{in_1^1, \ldots, in_{n_1}^1, \ldots, in_1^c, \ldots, in_{n_c}^c\}$.

By induction, let $C_{c+1} \in Comps(I)$ be another connected component and let $in_1^{c+1}, \ldots, in_{n_{c+1}}^{c+1} \in C_{c+1}$ such that $[in_1^{c+1}, \ldots, in_{n_{c+1}}^{c+1}] \in ValidOrders(I)$.

We now prove that $\ell + [in_1^{c+1}, \ldots, in_{n_{c+1}}^{c+1}] \in ValidOrders(I)$.

This is done by proving by induction on $0 \leq j \leq n_{c+1}$ that:

$$\ell + [in_1^{c+1}, \ldots, in_j^{c+1}] \in ValidOrders(I)$$

and that, for each $in_0 \in C_{c+1} \setminus S_j$, we have:

$$redundancy(in_0, I \setminus (S \cup S_j)) = redundancy(in_0, I \setminus S_j)$$

where $S_j = \{in_1^{c+1}, \ldots, in_j^{c+1}\}$.

If $n_{c+1} = 0$, then $\ell + [in_1^{c+1}, \ldots, in_{n_{c+1}}^{c+1}] = \ell \in ValidOrders(I)$. Moreover, because $C_{c+1}$ is disjoint with $C_1, \ldots, C_c$, according to Lemma 13, performing the $\ell$ removal steps does not change the redundancies of inputs in $C_{c+1}$ i.e., for each $in_0 \in C_{c+1}$, we have $redundancy(in_0, I \setminus S) = redundancy(in_0, I)$.

We now assume by induction that $\ell + [in_1^{c+1}, \ldots, in_j^{c+1}] \in ValidOrders(I)$ and for each $in_0 \in C_{c+1} \setminus S_j$, we have $redundancy(in_0, I \setminus (S \cup S_j)) = redundancy(in_0, I \setminus S_j)$.

Because $[in_1^{c+1}, \ldots, in_{n_{c+1}}^{c+1}] \in ValidOrders(I)$, we have $in_{j+1}^{c+1} \in Redundant(I \setminus S_j)$. Moreover, $in_{j+1}^{c+1} \in C_{c+1}$ hence $in_{j+1}^{c+1} \notin S$. Therefore, using the induction hypothesis with $in_0 = in_{j+1}^{c+1}$, we have $in_{j+1}^{c+1} \in Redundant(I \setminus (S \cup S_j))$. Therefore, according to our definition of valid removal steps (§III-C), $\ell + [in_1^{c+1}, \ldots, in_{j+1}^{c+1}] \in ValidOrders(I)$.

We denote $S_{j+1} = S_j \cup \{in_{j+1}^{c+1}\}$. Let $in_0 \in C_{c+1} \setminus S_{j+1}$. To complete the induction step, we prove that:

$$redundancy(in_0, I \setminus (S \cup S_{j+1})) = redundancy(in_0, I \setminus S_{j+1})$$

We remind that, for each input set $X$, we have:

$$redundancy(in_0, X) = \min\{|Inputs(bl) \cap X| \mid bl \in Cover(in_0)\} - 1$$

We denote as critical the objectives contributing to the redundancy:

$$CritSubCls(in_0, X) \stackrel{def}{=} \arg\min\{|Inputs(bl) \cap X| \mid bl \in Cover(in_0)\}$$

Because $in_0 \in C_{c+1}$, we know that $in_0$ does not overlap with inputs in $S$ so, for each $bl \in Cover(in_0)$, we have $Inputs(bl) \cap (I \setminus S) = Inputs(bl) \cap I$. Thus, by removing inputs of $S_j$ from both sides, we have $Inputs(bl) \cap (I \setminus (S \cup S_j)) = Inputs(bl) \cap (I \setminus S_j)$. Therefore:

$$CritSubCls(in_0, I \setminus (S \cup S_j)) = CritSubCls(in_0, I \setminus S_j)$$

We consider two cases:

1) Either there exists $bl \in CritSubCls(in_0, I \setminus S_j)$ such that $in_{j+1}^{c+1} \in Inputs(bl)$. In that case:

$$\begin{aligned}|Inputs(bl) \cap (I \setminus (S \cup S_{j+1}))| \\ = |Inputs(bl) \cap (I \setminus (S \cup S_j))| - 1\end{aligned}$$

$$|Inputs(bl) \cap (I \setminus S_{j+1})| = |Inputs(bl) \cap (I \setminus S_j)| - 1$$

Because $bl$ is critical and a redundancy can decrease at most by 1 after a reduction (Lemma 4) so the cardinalities for other objectives cannot decrease below the previous redundancy minus one, we have:

$$\begin{aligned}redundancy(in_0, I \setminus (S \cup S_{j+1})) \\ = redundancy(in_0, I \setminus (S \cup S_j)) - 1\end{aligned}$$

$$redundancy(in_0, I \setminus S_{j+1}) = redundancy(in_0, I \setminus S_j) - 1$$

2) Or for each $bl \in CritSubCls(in_0, I \setminus S_j)$, we have $in_{j+1}^{c+1} \notin Inputs(bl)$. In that case:

$$\begin{aligned}|Inputs(bl) \cap (I \setminus (S \cup S_{j+1}))| \\ = |Inputs(bl) \cap (I \setminus (S \cup S_j))|\end{aligned}$$

$$|Inputs(bl) \cap (I \setminus S_{j+1})| = |Inputs(bl) \cap (I \setminus S_j)|$$

Because $bl$ is critical and a redundancy can decrease at most by 1 after a reduction (Lemma 4) so the cardinalities for non-critical objectives cannot decrease below the previous redundancy we have:

$$\begin{aligned}redundancy(in_0, I \setminus (S \cup S_{j+1})) \\ = redundancy(in_0, I \setminus (S \cup S_j))\end{aligned}$$

$$redundancy(in_0, I \setminus S_{j+1}) = redundancy(in_0, I \setminus S_j)$$

In both cases, by induction hypothesis $redundancy(in_0, I \setminus (S \cup S_j)) = redundancy(in_0, I \setminus S_j)$, hence we have $redundancy(in_0, I \setminus (S \cup S_{j+1})) = redundancy(in_0, I \setminus S_{j+1})$.

This completes the induction on $0 \leq j \leq n_{c+1}$. In particular, we proved that $\ell + [in_1^{c+1}, \ldots, in_{n_{c+1}}^{c+1}] \in ValidOrders(I)$, which completes the induction on $c$, hence the result. □

We now prove that the gain can be independently computed on each connected component (Theorem 3). To do so, we have to prove intermediate results, including Proposition 9.

**Lemma 14** (Redundancy within a connected component). *Let $I$ be an input set. For each connected component $C \in Comps(I)$, for each input $in_0 \in C$, and for each order of removal steps $[in_1, \ldots, in_n]$, if $in_1, \ldots, in_n \in C$, then:*

$$\begin{aligned}redundancy(in_0, C \setminus \{in_1, \ldots, in_n\}) \\ = redundancy(in_0, I \setminus \{in_1, \ldots, in_n\})\end{aligned}$$

*Proof.* We remind that, for each input set $X$, we have:

$$redundancy(in_0, X) = \min\{|Inputs(bl) \cap X| \mid bl \in Cover(in_0)\} - 1$$

For the sake of simplicity, we denote $X_C = C \setminus \{in_1, \ldots, in_n\}$ and $X_I = I \setminus \{in_1, \ldots, in_n\}$. Let $in_0 \in C$ and $bl \in Cover(in_0)$.

Because $C \in Comps(I)$, we have $C \subseteq I$, hence $X_C \subseteq X_I$. So $Inputs(bl) \cap X_C \subseteq Inputs(bl) \cap X_I$. Moreover, for each $in \in Inputs(bl) \cap X_I$ we have $bl \in Cover(in_0) \cap Cover(in)$, so $in_0 \sqcap in$, and thus $in \in X_C$.

Therefore, $Inputs(bl) \cap X_C = Inputs(bl) \cap X_I$.

Hence the result $redundancy(in_0, X_C) = redundancy(in_0, X_I)$. □

**Corollary 2** (Component Validity). *Let $I$ be an input set. For each connected component $C \in Comps(I)$ and for each order of removal steps $[in_1, \ldots, in_n]$, if $[in_1, \ldots, in_n] \in ValidOrders(C)$, then $[in_1, \ldots, in_n] \in ValidOrders(I)$.*

*Proof.* The proof is done by induction on $n$.

For the initialization $n = 0$, we have $[in_1, \ldots, in_n] = [] \in ValidOrders(I)$.

For the induction step, we consider $[in_1, \ldots, in_n, in_{n+1}] \in ValidOrders(C)$ and we assume by induction that $[in_1, \ldots, in_n] \in ValidOrders(I)$.

Because $[in_1, \ldots, in_n, in_{n+1}] \in ValidOrders(C)$, according to our definition of redundant inputs and removal steps (§III-C), we have $in_1, \ldots, in_n \in C$. Hence, according to Lemma 14, we have for each $in_0 \in C$:

$$redundancy(in_0, C \setminus \{in_1, \ldots, in_n\}) = redundancy(in_0, I \setminus \{in_1, \ldots, in_n\})$$

Thus, because $C \subseteq I$, according to the definition of redundancy (§III-C) we have:

$$Redundant(C \setminus \{in_1, \ldots, in_n\}) \subseteq Redundant(I \setminus \{in_1, \ldots, in_n\})$$

Because $[in_1, \ldots, in_n, in_{n+1}] \in ValidOrders(C)$, we have $in_{n+1} \in Redundant(C \setminus \{in_1, \ldots, in_n\})$ by definition of removal steps (§III-C). Thus, $in_{n+1} \in Redundant(I \setminus \{in_1, \ldots, in_n\})$.

By induction hypothesis $[in_1, \ldots, in_n] \in ValidOrders(I)$. Therefore $[in_1, \ldots, in_n, in_{n+1}] \in ValidOrders(I)$, which completes the induction step. □

**Proposition 9** (Reductions withing a connected component). *Let $I$ be an input set. For each connected component $C \in Comps(I)$ and for each order of removal steps $[in_1, \ldots, in_n]$, if $[in_1, \ldots, in_n] \in ValidOrders(I)$ and $in_1, \ldots, in_n \in C$, then $[in_1, \ldots, in_n] \in ValidOrders(C)$.*

*Proof.* The proof is done by induction on $n$.

If $n = 0$ then $[in_1, \ldots, in_n] = [] \in ValidOrders(C)$.

Otherwise, we consider $[in_1, \ldots, in_n, in_{n+1}]$ with $in_1, \ldots, in_n, in_{n+1} \in C$ and we assume by induction that $[in_1, \ldots, in_n] \in ValidOrders(C)$.

We assume $[in_1, \ldots, in_n, in_{n+1}] \in ValidOrders(I)$, so $in_{n+1} \in Redundant(I \setminus \{in_1, \ldots, in_n\})$.

According to Lemma 14 applied to inputs $in_0 \in C \setminus \{in_1, \ldots, in_n\}$ with redundancy $> 0$, we have:

$$Redundant(C \setminus \{in_1, \ldots, in_n\}) = Redundant(I \setminus \{in_1, \ldots, in_n\})$$

Hence, $in_{n+1} \in Redundant(C \setminus \{in_1, \ldots, in_n\})$.

Thus, because $[in_1, \ldots, in_n] \in ValidOrders(C)$, according to our definition of valid removal steps (§III-C), we have the result $[in_1, \ldots, in_n, in_{n+1}] \in ValidOrders(C)$. □

Finally, we conclude the section by proving the claim on Theorem 3, which is used to divide our problem into subproblems (§VII-F).

**Theorem 3** (Divide the Gain). *For each input set $I$:*

$$gain(I) = \sum_{C \in Comps(I)} gain(C)$$

*Proof.* Let $[in_1, \ldots, in_n] \in ValidOrders(I)$ be a valid order of removal steps such that its cumulative cost $\sum_{1 \leq i \leq n} cost(in_i) = gain(I)$ is maximal (§III-C), and let $Comps(I) = \{C_1, \ldots, C_c\}$ denote the connected components, without a particular order.

We denote $[in_1, \ldots, in_n]_{C_j}$ the largest sublist (Definition 1) of $[in_1, \ldots, in_n]$ containing only inputs in the connected component $C_j$. Because $[in_1, \ldots, in_n] \in ValidOrders(I)$, according to Lemma 7 we have $[in_1, \ldots, in_n]_{C_j} \in ValidOrders(I)$ as well. So, according to Proposition 8:

$$[in_1, \ldots, in_n]_{C_1} + \cdots + [in_1, \ldots, in_n]_{C_c} \in ValidOrders(I)$$

Because the connected components form a partition of the redundant inputs, the cumulative cost of this order of removal steps is the same as $[in_1, \ldots, in_n]$:

$$\sum_{1 \leq j \leq c} \sum_{in \in [in_1, \ldots, in_n]_{C_j}} cost(in) = \sum_{1 \leq i \leq n} cost(in_i) = gain(I)$$

We now consider any connected component $C_{j_1}$ and we prove that:

$$\sum_{in \in [in_1, \ldots, in_n]_{C_{j_1}}} cost(in) = gain(C_{j_1})$$

The proof is done in two steps.

First, note that because $[in_1, \ldots, in_n]_{C_{j_1}} \in ValidOrders(I)$ and contains only inputs in $C_{j_1}$, according to Proposition 9 we have $[in_1, \ldots, in_n]_{C_{j_1}} \in ValidOrders(C_{j_1})$ as well.

Second, we prove by contradiction that $[in_1, \ldots, in_n]_{C_{j_1}}$ has a maximal cumulative cost in $C_{j_1}$. We assume by contradiction that there exists a valid order of removal steps $[in'_1, \ldots, in'_{n'}] \in ValidOrders(C_{j_1})$ such that:

$$\sum_{in \in [in'_1, \ldots, in'_{n'}]} cost(in) > \sum_{in \in [in_1, \ldots, in_n]_{C_{j_1}}} cost(in)$$

Because $[in'_1, \ldots, in'_{n'}] \in ValidOrders(C_{j_1})$, according to Corollary 2 we have $[in'_1, \ldots, in'_{n'}] \in ValidOrders(I)$ as well. So, according to Proposition 8, we have:

$$[in_1, \ldots, in_n]_{C_1} + \cdots + [in_1, \ldots, in_n]_{C_{j_1-1}}$$

$$+[in'_1, \ldots, in'_{n'}] + [in_1, \ldots, in_n]_{C_{j_1+1}} + \ldots$$
$$+[in_1, \ldots, in_n]_{C_c} \in ValidOrders(I)$$

The cumulative cost of this valid order of removal steps is thus larger than the cumulative cost of $[in_1, \ldots, in_n]_{C_1} + \cdots + [in_1, \ldots, in_n]_{C_c}$:

$$\sum_{in \in [in'_1, \ldots, in'_{n'}]} cost(in) + \sum_{1 \leq j \leq c \land j \neq j_1} \sum_{in \in [in_1, \ldots, in_n]_{C_j}} cost(in)$$
$$> \sum_{1 \leq j \leq c} \sum_{in \in [in_1, \ldots, in_n]_{C_j}} cost(in) = gain(I)$$

which contradicts the maximality of $gain(I)$.

Hence, $[in_1, \ldots, in_n]_{C_{j_1}} \in ValidOrders(C_{j_1})$ has a maximal cumulative cost in $C_{j_1}$. So, according to the definition of the gain (§III-C), we have for any connected component $C_{j_1}$:

$$\sum_{in \in [in_1, \ldots, in_n]_{C_{j_1}}} cost(in) = gain(C_{j_1})$$

Therefore, we have the result:

$$gain(I) = \sum_{1 \leq j \leq c} \sum_{in \in [in_1, \ldots, in_n]_{C_j}} cost(in) = \sum_{1 \leq j \leq c} gain(C_j)$$

□

### F. Genetic Search

In this section, we prove desirable properties satisfied for each generation by roofers (Theorem 4) and misers (Theorem 5) during the genetic search (Section VIII). We start by roofers.

**Theorem 4** (Invariant of the Roofers). *For every generation $n$, we have:*

$$|Roofers(n)| = n_{size}$$
$$\min\{cost(I) \mid I \in Roofers(n)\}$$
$$= \min\{cost(I) \mid I \in \bigcup_{0 \leq m \leq n} Roofers(m)\}$$

*and for every $I \in Roofers(n)$, we have:*
- *$I$ is reduced (in the sense of §III-C)*
- *$Cover(I) = Coverage_{obj}(C)$*

*Proof.* There are $n_{size}$ individuals in the initial roofer population (§VIII-B). Moreover, the procedure detailed above to update the roofer population ensures that an offspring can only take the place of an existing roofer. Hence, the number of roofers does not change over generations.

In the initial population, a roofer $I$ is always replaced by its reduced counterpart $reduce(I)$ after adding an input (§VIII-B). Moreover, after mutation (§VIII-E) each offspring $I$ is replaced by its reduced counterpart $reduce(I)$ before determining if it is accepted in the roofer population or rejected. Hence, for each generation, each roofer is reduced.

Roofers in the initial population are built so that they cover all the objectives (§VIII-B). Moreover, in the above procedure, an offspring can be added to the roofer population only if it covers all the objectives. Hence, for each generation, each roofer covers all the objectives.

Finally, in the above procedure one can remove an individual from the roofer population only if a less or equally costly roofer is found. Hence, the minimal cost amongst the roofers can only remain the same or decrease over generations. Therefore, the minimal cost in the last generation is the minimal cost encountered so far during the search.    □

To ease the proof for misers (Theorem 5), we first prove in Lemma 16 that a miser $I_0$ can be removed between generation $n$ and generation $n+1$ only by a dominating miser $I$. Then, we prove in Corollary 3 that being dominated is carried from generation to generation.

**Lemma 15** (Transitivity of Pareto Dominance). *The $\succ$ relation (§II-E) is transitive i.e., for every input sets $I_1, I_2, I_3$, if $I_1 \succ I_2$ and $I_2 \succ I_3$, then $I_1 \succ I_3$.*

*Proof.* For each $0 \leq i \leq n$, we have $f_i(I_1) \leq f_i(I_2)$ and $f_i(I_2) \leq f_i(I_3)$, so $f_i(I_1) \leq f_i(I_3)$. Moreover, there exists $0 \leq i_1 \leq n$ such that $f_{i_1}(I_1) < f_{i_1}(I_2) \leq f_{i_1}(I_3)$ and there exists $0 \leq i_2 \leq n$ such that $f_{i_2}(I_1) \leq f_{i_2}(I_2) < f_{i_2}(I_3)$. $i_1$ and $i_2$ can be the same or distinct. In any case, we have $f_{i_1}(I_1) < f_{i_1}(I_3)$ and $f_{i_2}(I_1) < f_{i_2}(I_3)$.    □

**Lemma 16.** *For every generation $n$, if $I_0 \in Misers(n) \setminus Misers(n+1)$, then there exists $I \in Misers(n+1)$ such that $I \succ I_0$, where $\succ$ is the domination relation from (§II-E).*

*Proof.* In the miser population update (§VIII-F), $I_0$ can be removed from the miser population only if there exists a miser candidate $I_1$ such that $I_1 \succ I_0$. If $I_1$ is accepted in the population then $I = I_1$ satisfies the lemma.

Otherwise, $I_1$ is rejected only because there exists a miser $I_2$ such that $I_2 \succ I_1$. Hence, according to Lemma 15, we have by transitivity that $I_2 \succ I_0$. Note that there is at most two miser candidates per generation. If either $I_1$ is the only miser candidate or there exists a second miser candidate $I_3$ which does not dominate $I_2$, then $I_2$ is present in the next generation and $I = I_2$ satisfies the lemma.

Otherwise, there exists a second candidate $I_3$ such that $I_3 \succ I_2$. Hence, according to Lemma 15, we have by transitivity that $I_3 \succ I_0$. If $I_3$ is accepted in the population then $I = I_3$ satisfies the lemma.

Otherwise, $I_3$ is rejected only because there exists a miser $I_4$ such that $I_4 \succ I_3$. Hence, according to Lemma 15, we have by transitivity that $I_4 \succ I_0$. Because there is at most two miser candidates per generation, $I_4$ cannot be removed by another candidate. Therefore, $I_4$ is present in the next generation and $I = I_4$ satisfies the lemma.    □

**Corollary 3** (Dominance across Generations). *For every generation $n$ and for every $I_1 \in Misers(n)$, if there exists $0 \leq m \leq n$ and $I_2 \in Misers(m)$ such that $I_2 \succ I_1$, then there exists $I_3 \in Misers(n)$ such that $I_3 \succ I_1$.*

*Proof.* The proof is done by induction on $n - m$.

If $I_2 \in Misers(n)$ then $I_2 = I_3$ satisfies the corollary.

Otherwise, there exists a generation $m < g \leq n$ where $I_2$ was removed. In that case, according to Lemma 16, there exists $I_3 \in Misers(g)$ such that $I_3 \succ I_2$. Hence, according to Lemma 15, we have by transitivity that $I_3 \succ I_1$. Finally,

because $n - g < n - m$, we have the result using the induction hypothesis on $I_3$. □

We finally prove, as expected, the properties satisfied by misers on each generation.

**Theorem 5** (Invariant of the Misers). *For every generation $n$, for every $I_1 \in Misers(n)$, we have:*

- *$I_1$ is reduced (in the sense of §III-C)*
- *$Cover(I) \subset Coverage_{obj}(C)$ (the inclusion is strict)*
- *There exists no $I_2 \in \bigcup_{0 \leq m \leq n} Misers(m)$ such that $I_2 \succ I_1$.*

*Proof.* The initial miser population is empty (§VIII-B), hence the properties trivially hold.

After mutation (§VIII-E) each offspring $I$ is replaced by its reduced counterpart $reduce(I)$ before determining if it is accepted in the miser population or rejected. Hence, for each generation, each miser is reduced.

In the miser population update (§VIII-F), an offspring can be a candidate to the miser population only if it does not cover all the objectives.

Finally, the last property is proved for a miser $I_1 \in Misers(n)$. Let $n'$ be the first generation when $I_1$ was accepted, so we have $n' \leq n$ and $I_1 \in Misers(n')$.

The proof is done by contradiction, assuming that there existed a generation $0 \leq m \leq n$ and a miser $I_2 \in Misers(m)$ such that $I_2 \succ I_1$. Let $m'$ be the first generation when $I_2$ was accepted, so we have $m' \leq m$ and $I_2 \in Misers(m')$. We consider two cases.

If $m' \leq n'$ then, according to Corollary 3, there exists $I_3 \in Misers(n')$ such that $I_3 \succ I_1$. This contradicts the above procedure, because if $I_1$ was dominated it would not have been accepted in $Misers(n')$.

If $m' > n'$ then, because $I_2 \succ I_1$, according to the above procedure $I_1$ is removed so $I_1 \notin Misers(m')$. But $m' \leq m \leq n$ and $I_1 \in Misers(n)$, so $I_1$ was added between $m'$ and $n$. Let $g$ be the first generation when this occurred.

In that case we have $I_1 \in Misers(g)$, $0 \leq m' \leq g$, $I_2 \in Misers(m')$, and $I_2 \succ I_1$. So, according to Corollary 3, there exists $I_3 \in Misers(g)$ such that $I_3 \succ I_1$, which contradicts the fact that $I_1$ was accepted in $Misers(g)$. □



\end{document}